\pdfoutput=1
\documentclass[fleqn,usenatbib,usedcolumn]{mnras}

\usepackage{graphicx,amssymb,hyperref}
\usepackage{multirow}
\usepackage{color}

\usepackage[fleqn]{amsmath}
\usepackage{abbrev}
\usepackage{siunitx}
\usepackage{comment} 
\usepackage{gensymb}
\usepackage{booktabs,tabularx}
\usepackage{flushend}
\usepackage{xspace}
\usepackage{physics}
 \usepackage{xcolor}
 
\usepackage{nicefrac}
\usepackage{units}
\usepackage{verbatim}
\usepackage[shortlabels]{enumitem}
\usepackage{subcaption}
\setlist[itemize]{leftmargin=5pt}

\def\mnras{MNRAS}
\def\apj{ApJ}

\def\simlt{\lower.5ex\hbox{$\; \buildrel < \over \sim \;$}}
\def\simgt{\lower.5ex\hbox{$\; \buildrel > \over \sim \;$}}
\def\etal{{\it et al.}}

\def\D{\mathrm{d}}
\def\rmd{\mathrm{d}}

\def\kpc{\mathrm{\, kpc}}

\def\mpc{\mathrm{\, Mpc}}
\def\msun{\mathrm{\, M_\odot}}
\def\kms{\mathrm{\, km \, s^{-1}}}
\def\cmsg{\, \mathrm{cm^2 \, g^{-1}}}
\def\gyr{ \, \mathrm{Gyr}}

\def\sigTt{\sigma_{\tilde{T}}}

\def\ro{\rho_{0}}
\def\so{\sigma_{0}}

\defcitealias{2012ApJ...761....1W}{W12}
\defcitealias{2018arXiv181005649R}{R18}

\newcommand{\eagle}{\textsc{eagle}\xspace}

\newcommand{\subfind}{\textsc{subfind}\xspace}

\newcommand{\bahamas}{\textsc{bahamas}\xspace}

\def\gs{\mathrel{\raise1.16pt\hbox{$>$}\kern-7.0pt \lower3.06pt\hbox{{$\scriptstyle \sim$}}}}         
\def\ls{\mathrel{\raise1.16pt\hbox{$<$}\kern-7.0pt \lower3.06pt\hbox{{$\scriptstyle \sim$}}}}   

\newcommand{\vect}[1]{\boldsymbol{#1}}

\newcommand{\orcid}[1]{\href{https://orcid.org/#1}{\includegraphics[scale=0.08]{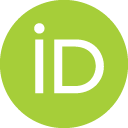}}}

\DeclareGraphicsExtensions{.pdf,.png}

\title[Isothermal SIDM modelling]{The surprising accuracy of isothermal Jeans modelling of self-interacting dark matter density profiles}

\author[A.\ Robertson \etal]
{\parbox{0.9\textwidth}{Andrew Robertson\thanks{e-mail: {\tt andrew.robertson@durham.ac.uk}}\orcid{0000-0002-0086-0524}$^1$, 
Richard Massey\orcid{0000-0002-6085-3780}$^1$, Vincent Eke\orcid{0000-0001-5416-8675}$^1$, Joop Schaye\orcid{0000-0002-0668-5560}$^2$ and Tom Theuns\orcid{0000-0002-3790-9520}$^1$}
\vspace{0.3cm}
\\$^1$Institute for Computational Cosmology, Durham University, South Road, Durham DH1 3LE, UK
\\$^2$Leiden Observatory, Leiden University, PO Box 9513, NL-2300 RA Leiden, the Netherlands
}

\begin{document}

\maketitle

\label{firstpage}

\begin{abstract}

Recent claims of observational evidence for self-interacting dark matter (SIDM) have relied on a semi-analytic method for predicting the density profiles of galaxies and galaxy clusters containing SIDM.
We present a thorough description of this method, known as isothermal Jeans modelling, and then test it with a large ensemble of haloes taken from cosmological simulations. Our simulations were run with cold and collisionless dark matter (CDM) as well as two different SIDM models, all with dark matter only variants as well as versions including baryons and relevant galaxy formation physics. Using a mix of different box sizes and resolutions, we study haloes with masses ranging from $\num{3e10}$ to $\num{3e15} \msun$. Overall, we find that the isothermal Jeans model provides as accurate a description of simulated SIDM density profiles as the Navarro-Frenk-White profile does of CDM halos. We can use the model predictions, compared with the simulated density profiles, to determine the input DM-DM scattering cross-sections used to run the simulations. This works especially well for large cross-sections, while with CDM our results tend to favour non-zero (albeit fairly small) cross-sections, driven by a bias against small cross-sections inherent to our adopted method of sampling the model parameter space. The model works across the whole halo mass range we study, although including baryons leads to DM profiles of intermediate-mass ($10^{12} - 10^{13} \msun$) haloes that do not depend strongly on the SIDM cross-section. The tightest constraints will therefore come from lower and higher mass haloes: dwarf galaxies and galaxy clusters.

\end{abstract}


\section{Introduction}

Uncovering the nature of dark matter (DM) is one of the major goals of science in the 21st century. The standard cosmological model, $\Lambda$-Cold Dark Matter (CDM), assumes that DM particles are collisionless, which is to say that the only DM interactions relevant for structure formation are gravitational. Self-interacting dark matter (SIDM) is an interesting alternative to CDM where DM particles can scatter with one another at astrophysically important rates. In regions of high density, primarily towards the centre of DM haloes, these interactions can transport heat through the DM halo, altering the halo structure.

SIDM was originally invoked in an astrophysical context as a way to address discrepancies between both the number and internal structure of observed dwarf galaxies, when compared with DM-only $\Lambda$CDM simulations \citep{Spergel:2000to}. Since then, it has become apparent that the inclusion of baryons into simulations can bring $\Lambda$CDM predictions into better agreement with observations \citep[e.g.][]{2014ApJ...786...87B, 2016MNRAS.457.1931S, 2016MNRAS.458.1559Z, 2017ApJ...850...97B}. Nevertheless, SIDM remains interesting because it is a viable alternative to CDM that can be tested with astrophysical observations \citep[for a review see][]{2018PhR...730....1T}, and because it provides a potential solution to the observed diversity of galaxy rotation curves (\citealt{2017PhRvL.119k1102K, 2017MNRAS.468.2283C, 2019PhRvX...9c1020R, 2019JCAP...12..010K, 2020PhRvL.124n1102S}, though see \citealt{2020MNRAS.495...58S}) and the anti-correlation between Milky Way satellites' pericentric distances and their central densities \citep{2019MNRAS.490..231K, 2020arXiv200702958C}. 

Attempts to measure or constrain the SIDM cross-section typically rely on either comparing the results of SIDM simulations directly with observations \citep[e.g.][]{2013MNRAS.430..105P, 2014MNRAS.437.2865K, 2015MNRAS.453...29E, 2016MNRAS.460.1399V, 2017MNRAS.465..569R, 2017MNRAS.469.1414K, 2018MNRAS.474..746B, 2018MNRAS.479..359S, 2019MNRAS.490.2117R, 2020JCAP...02..024B, 2020ApJ...896..112N, 2020arXiv200608596V}  
or use a semi-analytical model that can predict the effects of different SIDM cross-sections on the density profiles of DM haloes \citep{2014PhRvL.113b1302K, 2016PhRvL.116d1302K, 2017PhRvL.119k1102K, 2018NatAs...2..907V, 2019PhRvX...9c1020R, 2020JCAP...06..027K, 2020arXiv200612515S}. While simulations have been used to place upper-limits on the allowed SIDM cross-section \citep[e.g.][]{2001MNRAS.325..435M, 2008ApJ...679.1173R, 2013MNRAS.430...81R, 2013MNRAS.431L..20Z, 2019MNRAS.488.1572H, 2019MNRAS.488.3646R}, positive evidence for a non-zero cross-section has typically come from this semi-analytical model. This model has various advantages over direct comparison with simulations, including that its low computational cost allows a scan over SIDM parameter space, and that it can model specific systems -- with the baryon distribution inferred for an observed system, and the effects this has on the SIDM density profile, included by construction.

Evidence for a large DM--DM scattering cross-section would rule out many popular DM candidates, and would therefore alter the most promising regions of DM parameter space at which to target direct and indirect detection experiments \citep{2009PhRvD..80f3501Z, 2014PhRvD..89c5009K, 2014PhRvD..89k5017B, 2015PhRvD..91d3519K, 2015JCAP...10..055D}. This makes it crucially important to assess the efficacy of this semi-analytic model for SIDM density profiles.

The principal idea behind the semi-analytic model for SIDM density profiles is that in the inner regions of an SIDM halo, where the scattering rate is highest, DM self-interactions can keep the DM in thermal equilibrium. This means that the DM temperature (i.e. velocity dispersion) will be constant throughout the inner halo \citep{2014PhRvL.113b1302K}, which is why we refer to the method as ``\emph{isothermal} Jeans modelling'', with \emph{Jeans} reflecting the fact that the density profile in the isothermal region satisfies the Jeans equation. At large radii, the densities are substantially lower, leading to negligible rates of DM--DM scattering. The DM in the outskirts of the halo should therefore be unaffected by self-interactions and should be distributed as it would have been with CDM. The model assumes that there is a radius at which the behaviour abruptly transitions from collisional (i.e. isothermal) to fully collisionless, and that the role of the SIDM cross-section is to set this transition radius.

While this abrupt change in behaviour is clearly not exactly how SIDM affects a real halo, the density profiles predicted when making this assumption seem to agree well with those from $N$-body simulations with SIDM \citep[see the supplemental material of][]{2019PhRvX...9c1020R}. However, this sort of comparison has only been done for a limited number of haloes, and -- in all but a couple of cases \citep{2018MNRAS.476L..20R, 2020arXiv200612515S} -- has been done with DM-only simulations. The model has also been criticised on the basis that a number of the assumptions it makes (i.e. isotropic orbits in the isothermal region, and conservation of mass within the isothermal region) are not precisely borne out by SIDM simulations \citep{2018JCAP...12..038S}.

In this paper we address the question of how well the isothermal Jeans model describes the spherically-averaged density profiles of haloes taken from cosmological SIDM simulations, both DM-only and from simulations including baryons. Given that this model has been applied to observed systems across a wide range of mass scales, we take simulated haloes over five orders of magnitude in halo mass, ranging from dwarf galaxies to galaxy clusters. This is done by extracting haloes from simulations run 
with different box sizes and resolutions. We focus our attention on how well the isothermal Jeans model works in theory, rather than how well it works when applied to observational data. To this end, we compare its predicted density profiles directly with those of the simulated haloes, rather than generating the relevant observables from the simulations (stellar kinematics, gas rotation curves, strong and/or weak gravitational lensing, etc.) and fitting to those. 

This paper is structured as follows. In Section~\ref{sect:isothermal_modelling} we provide an overview of the isothermal Jeans model, including how to include the effects of baryons within the model. In Section~\ref{sect:sims} we describe the various SIDM (and CDM) simulations used throughout the paper, as well as how we extract relevant quantities from the simulations. In Section~\ref{sect:isothermal_fits} we describe how we fit the isothermal Jeans model to the density profiles of individual simulated haloes, before presenting the results of these fits to large ensembles of DM-only and hydrodynamical haloes in Sections~\ref{sect:DMO_results} and \ref{sect:hydro_results} respectively. In Section~\ref{sect:discussion} we discuss our results and provide an outlook on the use of the isothermal Jeans model, giving our conclusions in Section~\ref{sect:conclusions}.

All simulated density profiles used in this paper are taken from $z=0$ snapshots. The different simulation suites used in this paper assumed slightly different cosmologies from one another, but when applying the isothermal Jeans model we assumed a Planck 2013 cosmology throughout \citep[][and see Table~\ref{simulation_table}]{2014A&A...571A..16P}. This mainly enters our analysis in terms of the relationship between NFW halo masses and concentrations, and the corresponding scale densities and radii. Where not explicitly stated, $\log$ is $\log_{10}$, while we use $\ln$ for $\log_\mathrm{e}$.

\section{Overview of isothermal Jeans modelling}
\label{sect:isothermal_modelling}

The starting point for the model is a spherically symmetric \citet*[][hereafter NFW]{1997ApJ...490..493N} density profile,
\begin{equation}
\frac{\rho(r)}{\rho_\mathrm{crit}} = \frac{\delta_\mathrm{NFW}}{(r/r_\mathrm{s})(1+r/r_\mathrm{s})^2},
\label{eq:NFWrho}
\end{equation}
where $r_\mathrm{s}$ is the scale radius, $\delta_\mathrm{NFW}$ a dimensionless characteristic density, and $\rho_\mathrm{crit}=3H^2/8\pi G$ is the critical density. 
We define $r_{200}$ as the radius at which the mean enclosed density is 200 times $\rho_\mathrm{crit}$, and $M_{200}$ as the mass within $r_{200}$. The concentration parameter is defined as $c \equiv r_{200}/r_\mathrm{s}$ and can be related to the characteristic density by
\begin{equation}
\delta_\mathrm{NFW} = \frac{200}{3}\frac{c^3}{\ln (1+c)-c/(1+c)}.
\end{equation}

The model begins with an NFW density profile because it provides a good description of the density profiles of DM haloes in CDM-only simulations. The goal of the isothermal Jeans model is to take this profile, and predict how its inner regions are altered by DM self-interactions, as well as the presence of a baryonic mass component. The rest of this section describes how this is done, heavily inspired by previous work on the isothermal Jeans model, particularly \citet{2014PhRvL.113b1302K, 2016PhRvL.116d1302K} and \cite{2019PhRvX...9c1020R}.

\subsection{Finding the radius $r_{1}$}

Within the isothermal Jeans model, the SIDM halo is split into two regions. In one of these regions self-interactions are assumed to be frequent enough to keep the DM in thermal equilibrium, while in the other the effects of self-interactions are assumed to be negligible. The rate of scattering within an NFW halo decreases with increasing radius, and so the region where self-interactions maintain thermal equilibrium is in the centre of the halo where the scattering rate is highest. To determine at what radius the behaviour should switch, we find the radius, $r_1$, at which the local rate of scattering, multiplied by the age of the halo, is equal to one. Clearly this is simplistic, as in actuality there will not be a sharp transition in behaviour at this radius, but the validity of this assumption when translated into the predicted density profiles is one of the things we can test by comparing the model predictions with the density profiles of simulated systems. It is also not clear exactly what is meant by the `age' of a halo in a cosmology where structures grow hierarchically. For now we assume $t_\mathrm{age} = 7.5 \gyr$ for all haloes, but discuss this further in Section~\ref{sect:halo_age}.

The rate of scattering (per particle) as a function of radius is
\begin{equation}
\Gamma(r) = \frac{\sigma}{m} \, \langle v_\mathrm{pair}(r) \rangle  \rho(r) = \frac{\sigma}{m} \frac{4}{\sqrt{\pi}} \sigma_\mathrm{1D}(r) \rho(r), 
\label{eq:Gamma_r}
\end{equation}
where $\sigma/m$ is the SIDM cross-section divided by the DM particle mass, and is assumed here to be independent of velocity, $\langle v_\mathrm{pair}(r) \rangle$ is the mean pairwise velocity of particles at radius $r$, and the second equality comes from the fact that $\langle v_\mathrm{pair} \rangle = (4/\sqrt{\pi}) \, \sigma_\mathrm{1D}$ for a Maxwell-Boltzmann velocity distribution with a one-dimensional velocity dispersion of $\sigma_\mathrm{1D}$.

For an NFW halo with an isotropic velocity distribution, the one-dimensional velocity dispersion of particles is \citep{2001MNRAS.321..155L}:
\begin{equation}
\begin{split}
\sigma_{\mathrm{1D}}^{2}(x,c) 	&= \frac{1}{2} g(c) c \, x (1+x)^2 \frac{G M_{200}}{r_{200}} \left[ \pi^2 - \ln(x) - \frac{1}{x} \right. \\
				&- \frac{1}{(1+x)^2} - \frac{6}{1+x} + \left( 1 + \frac{1}{x^2} - \frac{4}{x} - \frac{2}{1+x} \right) \\
				&\times \left.\vphantom{\frac12} \ln(1+x) + 3 \ln^2(1+x) + 6 \, \mathrm{Li}_{2}(-x) \right],
\end{split}
\label{eq:NFWvdisp}
\end{equation}
where $x \equiv r/r_\mathrm{s}$, $g(c)\equiv[\ln (1+c) - c/(1+c)]^{-1}$,
and $\mathrm{Li}_{2}(y)$ is the dilogarithm (commonly referred to as Spence's function), defined by 
\begin{equation}
\mathrm{Li}_{2}(y) = \int_y^0 \frac{\ln (1-u)}{u} \, \rmd u.
\end{equation}

Putting equations \eqref{eq:NFWrho} and \eqref{eq:NFWvdisp} into equation \eqref{eq:Gamma_r} gives $\Gamma(r)$ for an NFW profile. Combining this with the halo age, $t_\mathrm{age}$, determines $r_1$. Outside of $r_1$, self-interactions are assumed to be unimportant, so the density profile will remain NFW, while inside of $r_1$ the DM will be in thermal equilibrium with a density profile that we now describe.

\subsection{Isothermal density profiles}
\label{sect:isothermal_profiles}

Inside $r_1$ frequent self-interactions are assumed to keep the DM in thermal equilibrium, and it therefore behaves like an isothermal ideal gas. The equation of state of an ideal gas, which links its density and pressure, is $p = \so^2 \, \rho$, where $\so$ is the 1D velocity dispersion. The temperature of the gas in this case is $k_\mathrm{B} T = m \so^2$, so the gas being isothermal implies that $\so$ is constant, independent of radius. 

Then, assuming the SIDM to be in hydrostatic equilibrium,\footnote{Where the inward force due to gravity is balanced by an outward force due to a pressure gradient.} and using the well-known result for the gravitational force from a spherically symmetric mass distribution, we find
\begin{equation}
\frac{\so^2}{\rho}  \dv{\rho}{r} = - \frac{G M_\mathrm{tot}(<r)}{r^2}.
\label{eq:hydrostatic}
\end{equation}
The total enclosed mass is the sum of the enclosed baryonic mass and the enclosed DM mass (i.e. $M_\mathrm{tot}(<r) = M_\mathrm{bar}(<r) + M(<r)$), with the enclosed DM mass related to the DM density by
\begin{equation}
\dv{M(<r)}{r} = 4 \pi r^2 \rho(r).
\label{eq:enclosed_mass}
\end{equation}
Equations~\eqref{eq:hydrostatic} and \eqref{eq:enclosed_mass} can be solved numerically\footnote{We use the \textsc{scipy} function scipy.integrate.odeint \citep{2020NatMe..17..261V}.} with appropriate boundary conditions.

For the DM-only case, we can make some headway towards understanding the solutions to these equations by taking the derivative of equation~\eqref{eq:hydrostatic}, and substituting in equation~\eqref{eq:enclosed_mass}
\begin{equation}
\dv{}{r} \left( r^2  \dv{\ln \rho}{r} \right) = - \frac{4 \pi G r^2 \rho}{\so^2}.
\label{eq:combo}
\end{equation}
Then defining $y = \ln(\rho / \rho_0)$, $r_0^2 = \so^2 / 4 \pi G \rho_0$ and $x = r / r_0$ one finds
\begin{equation}
\dv[2]{y}{x} + \frac{2}{x} \dv{y}{x} + \exp(y) = 0.
\label{eq:iso_dimensionless}
\end{equation}
This equation\footnote{Readers familiar with stellar structure may recognise this as the Lane-Emden equation with polytropic index $n \to \infty$, corresponding to an isothermal equation of state \citep{1939isss.book.....C}.}
 has different solutions for $y(x)$ depending on the boundary conditions imposed.\footnote{For example, a \emph{singular isothermal sphere} (which has $\rho \propto 1/r^2$) corresponds to $y = \ln (2/x^2)$, which leads to $\rho = 2 \ro r_0^2 / r^2$.} Given that simulated SIDM haloes have constant central density `cores', we impose that at $r=0$: $\rho = \ro$ and $\mathrm{d}\rho / \mathrm{d}r = 0$. Expressed in terms of $y$ these boundary conditions are that $y(0) = 0$ and $\mathrm{d}y / \mathrm{d}x|_{x=0} = 0$. These boundary conditions lead to a unique solution for $y(x)$, which means that the isothermal density can be written as
\begin{equation}
\rho(r) = \rho_0 \, f(r/r_0),
\label{eq:iso_density}
\end{equation}
where $f(x) = \exp(y)$. There are therefore two free parameters that describe the isothermal region of the halo: the central density, $\ro$, and a characteristic radius $r_0$. As $r_0$ is related to $\ro$ and $\so$, the two free parameters can also be thought of as $\ro$ and the isothermal velocity dispersion, $\so$.

\subsection{Matching criteria}
\label{sect:matching}

To determine the two parameters of the isothermal profile, $\ro$ and $\so$, requires two matching criteria. We match the profiles at $r_1$, requiring that the mass enclosed within $r_1$ and the density at $r_1$ be the same for the NFW profile and corresponding isothermal profile. We define $\rho_1 \equiv \rho_\mathrm{NFW}(r_1)$ and $M_1 \equiv M_\mathrm{NFW}(<r_1)$. The condition that the isothermal profile has $M(< r_1) = M_1$ is motivated by the fact that self-interactions re-distribute energy between particles, changing their radial distribution, but in a way that the total mass should remain constant. Requiring that  $\rho(r_1) = \rho_1$ then ensures that the density profile is continuous.

For a given $\rho_1$ and $M_1$ it is not immediately obvious which values of $\ro$ and $\so$ will satisfy our chosen matching criteria. In Appendix~\ref{App:matching} we demonstrate how the functional form of the isothermal density profile in the DM-only case (equation~\ref{eq:iso_density}) can be used to efficiently find $\ro$ and $\so$ from $r_1$, $\rho_1$ and $M_1$. This is useful in understanding whether or not there has to be an isothermal profile that matches (there does not, but this only happens when $r_1 \gg r_\mathrm{s}$) and whether there is a maximum of one solution (there can, rarely, be more), and the interested reader is encouraged to consult the appendix for more details. However, this method does not extend to the case including baryons, and so here we describe a more general iterative scheme for finding the matching isothermal profile.

\begin{figure*}
        \centering
        \includegraphics[width=\textwidth]{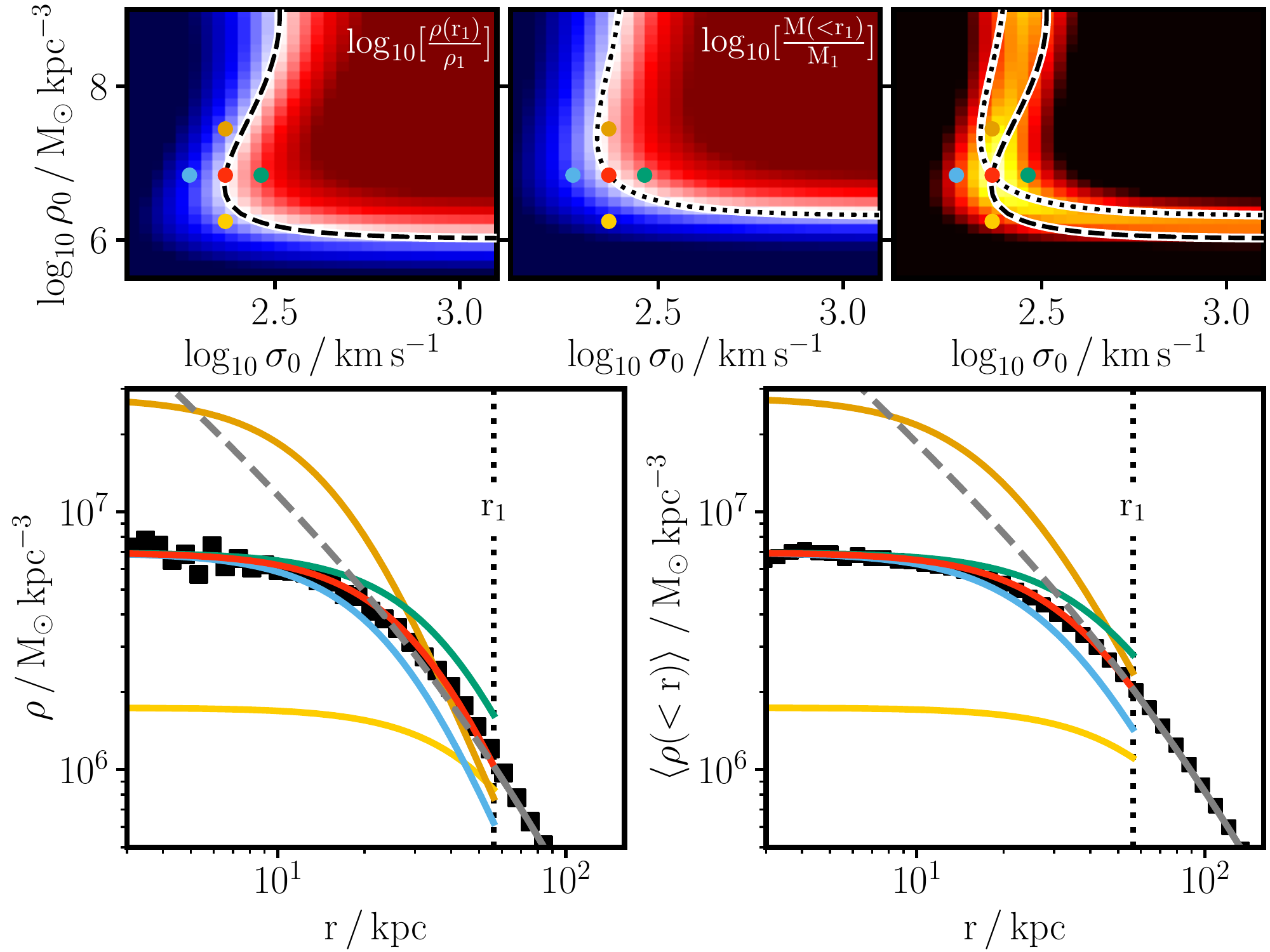}     
	\caption{A schematic illustration of finding the isothermal solution that matches onto an NFW profile at radius $r_1$, for a halo containing only DM. The bottom left panel shows the density as a function of radius, while the bottom right panel shows the mean enclosed density. The NFW profile shown in grey has $M_{200} = \num{1.9e13} \msun$ and $c=5.5$, which comes from fitting to the outskirts of the density profile of a DM-only halo simulated with $\sigma/m = 1 \cmsg$. The simulated density profile is plotted as the black squares. The radius, $r_1$, was calculated from the NFW profile, assuming the input cross-section of $1 \cmsg$ and a halo age of $7.5 \gyr$. The five different coloured lines, show five different isothermal density profiles, whose central densities, $\ro$, and velocity dispersions, $\so$, are marked in the top panels. The top left panel shows how the isothermal density profile's density at $r_1$ compares with that of the NFW profile, and the top centre panel shows the equivalent for the enclosed mass within $r_1$. Blue colours are where the isothermal density/mass is below the NFW one, while red colours are where it is above, the colours saturate at a difference of 0.5 dex. The density and mass matching criteria are satisfied along the black dashed and dotted lines respectively. These lines cross at the location of the red dot, which indicates that the red isothermal profile is the correct profile for matching onto the NFW at $r_1$. The top right panel shows $\sqrt{\left( \log_{10} \left[ \rho(r_1) / \rho_1 \right] \right)^2 + \left( \log_{10} \left[ M(<r_1) / M_1 \right] \right)^2}$, with the colour scale going from yellow (0) to black (0.5).}
\label{fig:matching}
\end{figure*}

Fig.~\ref{fig:matching} contains an illustration of how $\rho(r_1)$ and $M(< r_1)$ depend on $\ro$ and $\so$. In the lower panels we show $\rho(r)$ and $M(<r)$ (plotted as $\langle \rho(<r) \rangle = M(<r) / \frac{4}{3} \pi r^3$ to reduce the dynamic range on the y-axis) for five illustrative points in the $\ro - \so$ parameter space, including the point in $\ro - \so$ where both matching criteria are satisfied. The way in which $\rho(r_1)$ and $M(< r_1)$ vary as $\ro$ and $\so$ are varied is plotted in the top left and top centre panels of Fig.~\ref{fig:matching}. The top right panel then shows a combined `badness-of-fit' metric, $b \equiv\sqrt{\left( \log_{10} \left[ \rho(r_1) / \rho_1 \right] \right)^2 + \left( \log_{10} \left[ M(<r_1) / M_1 \right] \right)^2}$, which is 0 when the isothermal and NFW profiles correctly match at $r_1$.

Navigating the $\ro - \so$ parameter space to find the solution that meets our matching criteria could be done in a number of ways. For example, one could find the location where $b$ is minimised using a gradient descent algorithm or a similar optimisation method. The method that we found to work best, and that we used to find the solution for the case shown in Fig.~\ref{fig:matching}, is to use a root finding algorithm to find the zeroes of the vector $\vect{f}(\ro,\so) = ( \log_{10} \left[ \rho(r_1) / \rho_1 \right], \log_{10} \left[ M(<r_1) / M_1 \right])$.\footnote{Specifically, we use scipy.optimize.root with method=`hybr', which uses a modified version of the algorithm described in \citet{10.1093/comjnl/7.2.155}.} Requiring each component of $\vect{f}$ to have an absolute value less than $10^{-4}$, a solution could usually be found with fewer than 10 function evaluations, although this was dependent on a reasonable initial guess. Formulating such a guess is relatively straightforward in DM-only cases, and for individual systems, but is made more difficult in baryon-rich systems, or when trying to automate the isothermal Jeans modelling to run on haloes with a wide range of masses. A solution to this is to start with an isothermal solution, and ask what NFW profile can match onto it, rather than vice versa. We discuss this in Section~\ref{sect:inside_out}.

\subsection{Including baryons}
\label{sect:method_with_baryons}

The distribution of baryons within a DM halo can influence the distribution of the DM. For the case of collisionless DM, the way in which the DM responds to a baryon potential depends upon how the baryon distribution evolved to get to its present state. A baryon distribution that builds up gradually alters the distribution of DM particle orbits adiabatically, which means that particles' orbits will conserve quantities known as adiabatic invariants \citep[e.g.][]{1987gady.book.....B}. Gradual growth of the baryon potential typically contracts the DM halo \citep[e.g.][]{1984MNRAS.211..753B, 2004ApJ...616...16G}, making it more centrally concentrated. Rapid changes to the baryon potential, for example due to the expulsion of gas by supernovae explosions, lead to non-adiabatic changes to DM particles' orbits that can lower the central DM densities (e.g. \citealt*{1996MNRAS.283L..72N}; \citealt{2005MNRAS.356..107R}, and see \citealt{2014Natur.506..171P} for a review).

This picture is different with SIDM. As long as the timescale on which SIDM particles interact is shorter than that on which the gravitational potential due to the baryons varies, SIDM particles will be kept in equilibrium with the baryon potential as it is now. This means that we can include the effects of baryons into the isothermal Jeans model simply by including their contribution to $M_\mathrm{tot}(<r)$.\footnote{At $r_1$, the average time between interactions is (by definition) the age of the halo. This is a timescale on which the gravitational potential can significantly vary, violating the approximation that interactions maintain equilibrium. In this paper we demonstrate that this approximation works well for describing the effects of baryons on an SIDM halo, which is likely because the radii at which baryons make a significant contribution to the total enclosed mass are well within $r_1$ for even modest cross-sections.} 

\begin{figure*}
        \centering
        \includegraphics[width=\textwidth]{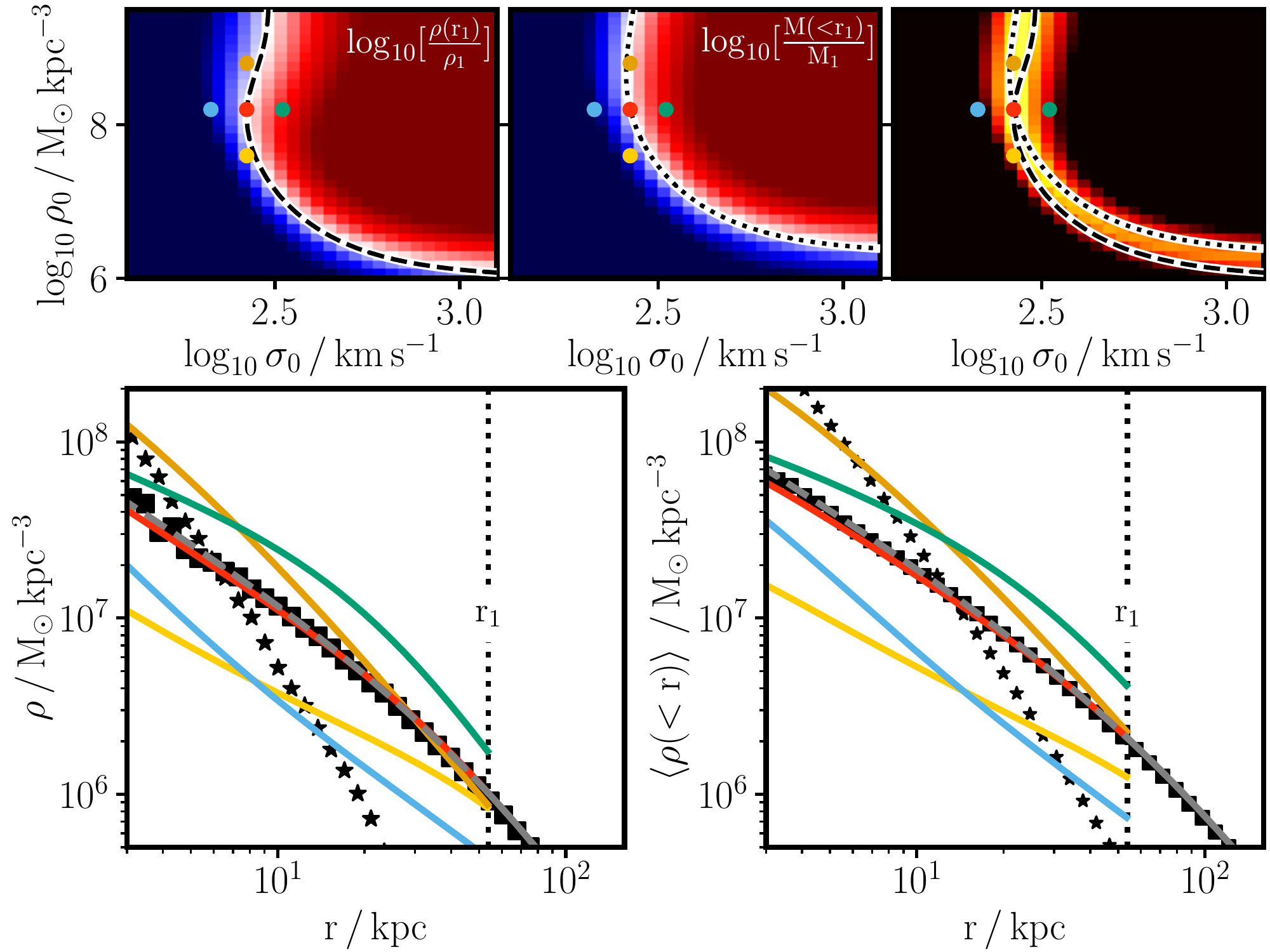}     
	\caption{The same as Fig.~\ref{fig:matching}, but for a case including baryons. The simulated halo is the same one shown in Fig.~\ref{fig:matching} (with $\sigma/m = 1 \cmsg$), but now with the addition of \eagle galaxy formation physics. The baryon density profile is plotted as black stars. The grey NFW profile has the same mass as in Fig.~\ref{fig:matching} ($M_{200} = \num{1.9e13} \msun$), but a slightly larger concentration of $c=6.5$ reflecting adiabatic contraction of the DM halo. The NFW profile is shifted down in density by a factor of $1 - f_\mathrm{bar} = 0.84$ to account for the fact that we are modelling only the DM as opposed to the total matter density. Note that the dynamic range on the $y$-axis is increased from Fig.~\ref{fig:matching} because the density profile is considerably steeper than in the DM-only case.}
\label{fig:matching_hydro}
\end{figure*}

In Fig.~\ref{fig:matching_hydro} we show an example of the isothermal Jeans model including baryons. The simulated halo is the same one shown in Fig.~\ref{fig:matching} but now from a simulation including gas and a model for galaxy formation. While DM dominates the total density at large radii, the inner $10 \kpc$ is baryon dominated. This has a dramatic effect on the simulated SIDM density profile, which no longer has the large constant density core seen in Fig.~\ref{fig:matching}, instead resembling an NFW profile over the radii shown.

The isothermal solution is calculated including $M_\mathrm{bar}(<r)$ for the simulated halo, which leads to a good match between the simulated SIDM profile and the isothermal prediction. We measured $M_\mathrm{bar}(<r)$ from the simulated halo within logarithmically spaced radii, and then interpolated the results so that our adopted ODE solver could find $M_\mathrm{bar}(<r)$ at arbitrary radii.

Including the effects of baryons into the isothermal Jeans model complicates the mapping from $r_1$, $\rho_1$ and $M_1$ to $\ro$ and $\so$, because this mapping now depends on the density profile of baryons, and so is different for each halo. Going from a DM-only case to one including baryons also leads to a subtlety about how our NFW profile is defined, because a fraction, $f_\mathrm{bar}$, of the mass in the Universe is no longer DM. For an NFW profile with mass and concentration, $M_{200}$ and $c$, we find the corresponding scale radius and characteristic density, $r_\mathrm{s}$ and $\delta_\mathrm{NFW}$. The DM density for this NFW profile is then calculated following equation~\eqref{eq:NFWrho}, but with the characteristic density scaled down by $1 - f_\mathrm{bar}$ to reflect the fact that we are only trying to model the DM component.

\section{SIDM simulations}
\label{sect:sims}

In order to thoroughly test the isothermal Jeans model for SIDM density profiles, we compare its predictions with a large number of simulated haloes, both from DM-only simulations, and from simulations including baryons. The isothermal Jeans model splits the halo into (1) a regime where the scattering rate is high and is assumed to fully thermalise the DM distribution, and (2) a regime where the scattering rate is low and is assumed not to affect the DM distribution. In contrast, the simulations can faithfully match the scattering physics in the transition regime where scattering is neither infrequent enough that it can be ignored, nor so frequent that the DM behaves like a collisional fluid with a short mean free path. All of our simulations are of cosmological boxes, with different box sizes and resolutions being used to study haloes of different mass. We describe these simulations below.

\subsection{Simulations suites}

For galaxy cluster scale haloes we use the \bahamas-SIDM simulations from \citet{2019MNRAS.488.3646R}, which use the \bahamas galaxy formation model described in \citet{2017MNRAS.465.2936M}. These have limited mass and spatial resolution, but a large box size, which allows us to study massive haloes. At intermediate halo masses, corresponding to Milky Way-like or massive elliptical galaxies, we use SIDM versions\footnote{The implementation of SIDM within \eagle was described in \citet{2018MNRAS.476L..20R}.} of the \eagle simulations \citep{2015MNRAS.446..521S, 2015MNRAS.450.1937C}. Our resolution and galaxy formation physics model was the same as for the `Reference' $100 \mpc$ \eagle box, but to reduce computational requirements we simulated smaller, $50 \mpc$, volumes. Finally, in order to study lower-mass galaxies, we ran small $12.5 \mpc$ boxes at approximately 25 times better mass resolution than our $50 \mpc$ simulations, using the initial conditions from \citep{2019MNRAS.488.2387B}. These also used the \eagle galaxy formation model, but with slightly adjusted (`Recal') parameters that better reproduce observed galaxy properties when running at higher resolution \citep[see][for more details of the Reference and Recal subgrid parameters]{2015MNRAS.446..521S}. Further specific details of the simulations are in Table~\ref{simulation_table}.

\subsection{Implementation of SIDM scattering}

The method used to simulate SIDM is shared by all of our simulation suites, and is described in \citet{2017MNRAS.465..569R}. It uses a Monte-Carlo approach to implement DM scattering, where at each time-step, particles search locally for neighbours, with random numbers drawn to see which nearby pairs scatter. The probability for a pair of particles to scatter depends on their relative velocity and the cross-section for scattering, which itself can be a function of the relative velocity. The search region around each particle is a sphere, with a radius equal to the Plummer-equivalent gravitational softening length. Our implementation can simulate anisotropic scattering cross-sections \citep{2017MNRAS.467.4719R}, which naturally arise when scattering cross-sections are velocity-dependent.

\begin{table*}
\centering
\caption{Box sizes and resolutions for the simulations used in this paper. The box sizes are comoving, and the gravitational softening lengths, $\epsilon_\mathrm{p}$, are proper Plummer-equivalent gravitational softening lengths \citep{2005MNRAS.364.1105S}, while $\epsilon_\mathrm{c}$ is a comoving softening length used at high redshift (the comoving softening length is used at redshifts where it is smaller than the proper one). The WMAP-9 cosmology has $\Omega_\mathrm{m}=0.2793$, $\Omega_\mathrm{b}=0.0463$, $\Omega_\mathrm{\Lambda}=0.7207$, $\sigma_8 = 0.812$, $n_\mathrm{s} = 0.972$ and $h = 0.700$ \citep{2013ApJS..208...19H} . The Planck 2013 cosmology has $\Omega_\mathrm{m}=0.307$, $\Omega_\mathrm{b}=0.04825$, $\Omega_\mathrm{\Lambda}=0.693$, $\sigma_8 = 0.8288$, $n_\mathrm{s} = 0.9611$ and $h = 0.6777$ \citep{2014A&A...571A..16P}. The WMAP-7 cosmology has  $\Omega_\mathrm{m}=0.272$, $\Omega_\mathrm{b}=0.0455$, $\Omega_\mathrm{\Lambda}=0.728$, $\sigma_8 = 0.81$, $n_\mathrm{s} = 0.967$ and $h = 0.704$ \citep{2011ApJS..192...18K}.}
\begin{tabular}{lllccccc}
\hline 
\\[-9pt]
Simulation  & Box size / Mpc & Cosmology   & $m_\mathrm{DM-only} / \msun$ & $m_\mathrm{DM} / \msun$ & $m_\mathrm{gas} / \msun$ & $\epsilon_\mathrm{p} / \kpc$ & $\epsilon_\mathrm{c} / \kpc$ \\[2pt] \hline 
\bahamas & $400 \, h^{-1}$          & WMAP-9   & $\num{6.6e9}$   & $\num{5.5e9}$ & $\num{1.1e9}$ & $5.7$ & 22.3\\
\eagle-50 & $50$          & Planck 2013  & $\num{1.2e7}$   & $\num{9.7e6}$ & $\num{1.8e6}$ & $0.7$ & 2.7\\
\eagle-12 & $12.5$          & WMAP-7  & $\num{4.8e5}$   & $\num{4.0e5}$ & $\num{8.1e4}$ & $0.23$ & 0.90
\end{tabular}
\label{simulation_table}
\end{table*}

\subsection{Simulated cross-sections}

In this paper we investigate three different DM models: collisionless CDM, a velocity-independent and isotropic cross-section of $1 \cmsg$ (SIDM1) and a velocity-dependent and anisotropic cross-section corresponding to DM particles scattering though a Yukawa potential (vdSIDM). Each of the three simulation suites described in Table~\ref{simulation_table} was run with these three DM models, both DM-only and including baryons. We will refer to simulations run with these cross-sections that include baryons as CDMb, SIDM1b and vdSIDMb. The differential cross-section that we simulate for vdSIDM is
\begin{equation}
\label{eq:yukawa_differential_cross-sect_alt}
\frac{\D \sigma}{\D \Omega} = \frac{\sigma_{T0}}{4 \pi \left( 1 + \frac{v^2}{w^2} \sin^2 \frac{\theta}{2} \right)^2 },
\end{equation}
with $\sigma_{T0} = 3.04 \cmsg$ and $w = 560 \kms$. These parameters were chosen to roughly reproduce the best-fitting cross-section in \citet{2016PhRvL.116d1302K}, which is claimed to successfully explain the density profiles of systems ranging from dwarf galaxies to galaxy clusters.

\begin{figure}
        \centering
        \includegraphics[width=\columnwidth]{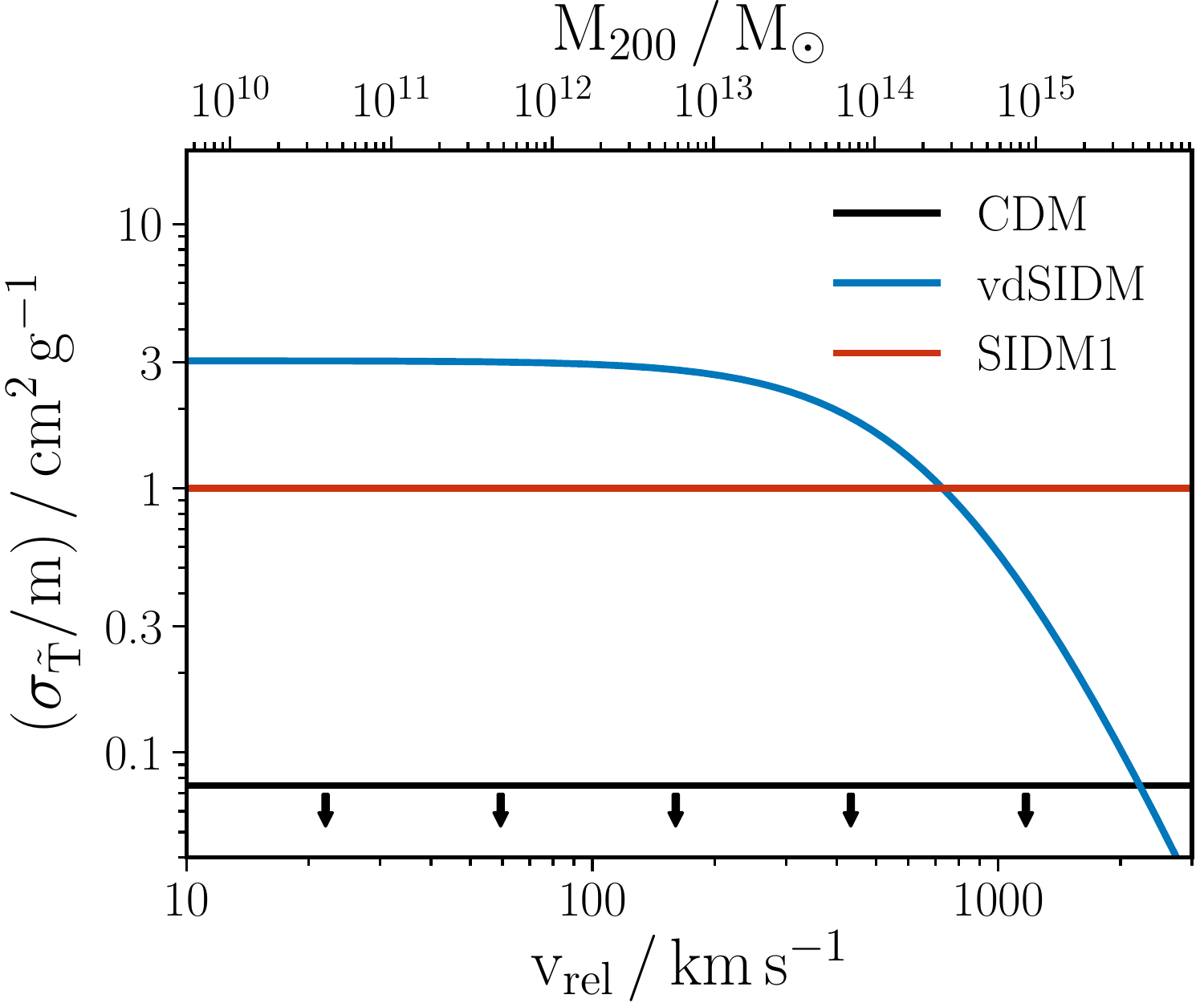}     
	\caption{The momentum-transfer cross-section as a function of velocity for the three particle models that we have simulated. The arrows below the CDM line reflect that CDM has zero cross-section, and therefore lies off the bottom of the plot. The mapping from a relative velocity between two DM particles (along the bottom) to a halo mass (along the top) is approximate, and is done assuming $v_\mathrm{rel} = \sqrt{G \, M_{200} / r_{200}}$.}
	\label{fig:cross_sects}
\end{figure}

In Fig.~\ref{fig:cross_sects} we plot the cross-section as a function of relative velocity for our three simulated DM models. Specifically, we plot the momentum transfer cross-section
\begin{equation}
\label{eq:sigT}
\sigma_{\tilde{T}} \equiv 2 \int (1 - |\cos \theta|) \frac{\D \sigma}{\D \Omega} \D \Omega ,
\end{equation}
which has been shown to be a more relevant quantity than the total cross-section for determining the rate at which cores form in isolated DM haloes \citep{2017MNRAS.467.4719R}. The $1 - |\cos \theta|$ term comes from weighting scatterings by the amount of momentum they transfer along the collision axis, taking into account that indistinguishable particles that scatter by $\theta > 90^{\circ}$ could be re-labelled such that the scattering was by less than $90^{\circ}$ \citep{2014MNRAS.437.2865K}. The factor of 2 means that for isotropic scattering $\sigma = \sigTt$. The momentum transfer cross-section for the differential cross-section that we implemented for vdSIDM (equation~\ref{eq:yukawa_differential_cross-sect_alt}) is
\begin{equation}
\label{eq:yukawa_sigmaTt}
\sigma_{\tilde{T}} = \sigma_{T0} \frac{4 w^4}{v^4} \left\{ 2 \ln \left(1 + \frac{v^2}{2 w^2} \right) - \ln \left(1 + \frac{v^2}{w^2} \right) \right\}.
\end{equation}

\subsection{Measuring density profiles}
\label{sect:measuring_rho_r}

For the tests that we wish to perform, the required information from each simulated halo is the DM density profile and (for cases including baryons) the baryonic enclosed-mass profile. We use only `centrals', i.e. we do not analyse galaxies that are satellites of something more massive. DM haloes are identified using the friends-of-friends algorithm \citep{1982ApJ...259..449P, 1985ApJ...292..371D}, and we define the centre of the halo as the position of the particle with the minimum gravitational potential energy. We then calculate the DM density profile by finding the mass in logarithmically-spaced spherical shells and dividing these masses by the volume of the relevant shell. For $M_\mathrm{bar}(<r)$ we add up the mass of the baryon particles (gas, stars and black holes\footnote{For two of the \bahamas haloes, one with CDM+baryons and one with vdSIDM+baryons, there were especially massive black hole particles ($\gtrsim 10^{11} \msun$) at the centre. The steep potential from these point masses made solving the coupled ODEs required to find isothermal density profiles challenging. For these two haloes we therefore softened the potential from the black hole particle using the gravitational softening length used when running the simulations. This could have been done for all haloes, but we only discovered this problem close to completing this work.}) within the same logarithmically-spaced radii as are used for the boundaries between the density shells. We used 100 radii ranging from $0.1 \kpc$ to $4 \mpc$. 

\section{Fitting the isothermal Jeans model to simulated density profiles}
\label{sect:isothermal_fits}

In Section~\ref{sect:isothermal_modelling} we described how the isothermal Jeans model can be used to generate a density profile that takes into account the effects of DM self-interactions, starting from an NFW profile (defined by $M_{200}$ and $c$) and an SIDM cross-section. In this section we will discuss fitting to the density profiles of simulated systems to extract a posterior distribution for the SIDM cross-section. The basic idea is to sample from the input parameters ($M_{200}$, $c$ and $\sigma/m$), generate the isothermal Jeans model density profile at each point in parameter space, and calculate a likelihood from comparing the model density profile with the measured density profile from the simulations. This procedure can then be wrapped in an MCMC sampler in order to generate samples of the input parameters drawn from their joint posterior distribution.

\subsection{Definition of a `good fit'}
\label{sect:good_fit}

In order to carry out the procedure outlined above, we need to define a likelihood function. When applying isothermal Jeans modelling to observed systems, the likelihood would take into account the uncertainties on measured quantities as well as any covariance between measurements. Here, we do not focus on any particular observational setup, and so instead must decide what constitutes a better or worse fit to a simulated density profile. To this end, we define our likelihood (up to a normalising constant) as $\mathcal{L} \propto \exp(-\chi^2 / 2)$ with
\begin{equation}
\label{eq:chi2}
\chi^2 = \sum_{i=1}^{N_\mathrm{bins}} \left( \frac{ \log_{10} \rho_\mathrm{sim} (r_i) - \log_{10} \rho_\mathrm{mod} (r_i) }{\delta \log_{10} \rho}\right)^2.
\end{equation}
We assume an uncorrelated error of 0.1 dex on $\log \rho$ (i.e. $\delta \log_{10} \rho = 0.1$), and the $r_i$ are taken from the same logarithmically-spaced radii at which the density profiles from the simulations were measured. We use all $r_i$ between $0.01 \, r_{200}$ and $r_{200}$, which leads to $N_\mathrm{bins} = 43$ or $44$ depending on the mass of the halo.  By assuming a constant error on $\log \rho$, and using logarithmically spaced radii, our notion of `goodness of fit' is essentially how similar in appearance the simulated and model density profiles are on a plot of $\log \rho$ against $\log r$ (e.g. in the bottom left panels of Figs.~\ref{fig:matching} and \ref{fig:matching_hydro}).

The reason that there is not a well defined value for the error on the density profile is that the differences between our simulated and isothermal-model density profiles are not random, but are systematic. Even in the absence of particle noise in the simulations, the density profiles of haloes would not be perfectly described by the isothermal Jeans model because the model makes several assumptions that are known not to be true. As examples, it assumes haloes are spherically symmetric and ignores substructure within the halo. This is no different from NFW profiles fit to CDM-only haloes. While the particle distributions from simulated CDM haloes are usually considered to be well-fit by NFW haloes, they are not well fit in the sense of being consistent with being precisely NFW except for some random error (for example Poisson noise on the number of particles in each radial bin). 

\subsection{Choice of model parameterisation}

A single isothermal Jeans model density profile is described by a number of parameters: $M_{200}$, $c$, $\sigma/m$, $r_1$, $\rho_0$ and $\sigma_0$, but only three of these are independent. So far we have discussed the isothermal Jeans model in terms of starting with an NFW profile, and calculating how this is affected by a given cross-section, making $M_{200}$, $c$ and $\sigma/m$ the natural parameters that describe a model density profile. However, we will find that parameterising the model in different ways can have benefits in terms of how quickly a likelihood can be evaluated.

\subsubsection{`Outside-in' fitting}

We refer to starting with the NFW parameters and then finding the matching isothermal profile for the inner halo as outside-in fitting. While this is a natural way to think about the physics of core-formation with SIDM, MCMC sampling of the  ($M_{200}$, $c$, $\sigma/m$) parameter space is problematic because finding the isothermal solution that matches onto an NFW profile is itself a process that requires iterating over parameters ($\ro$ and $\so$). Firstly, this means that running an MCMC chain is slow, because each likelihood evaluation requires multiple steps. Secondly, the iterative procedure for finding the isothermal profile that correctly matches the NFW (described in Section~\ref{sect:matching}) requires a reasonable initial guess for $\ro$ and $\so$ in order to converge on the correct solution, and sometimes there is no matching solution at all (see Appendix~\ref{App:matching}).

\subsubsection{`Inside-out' fitting}
\label{sect:inside_out}

A solution to the problem of iteratively finding the correct $\ro$ and $\so$ at a particular point in the sampled parameter space is to make $\ro$ and $\so$ (rather than $M_{200}$ and $c$) two of the parameters that are sampled by the MCMC sampler. In fact, it is convenient to make one more change to the sampled variables, changing from $\ro$ to the number of scatterings per particle in the centre of the halo
\begin{equation}
N_0 = \frac{\sigma}{m} \frac{4}{\sqrt{\pi}} \, \sigma_0 \, \rho_0 \, t_\mathrm{age}.
\end{equation}
This is convenient because isothermal Jeans modelling requires there to be a radius, $r_1$, at which $N(r_1)=1$. This cannot be achieved if $N_0 < 1$, and so a prior that $N_0 > 1$ limits us to isothermal solutions for which $r_1$ exists.

Instead of first considering the NFW profile that will become the outskirts of the model density profile, and then finding an isothermal profile that matches this NFW, the inside-out method starts from the isothermal profile in the inside and then find the NFW profile that matches onto this at $r_1$. This avoids any iteration, because the density profile and enclosed mass profile of an NFW halo are analytical, and these can be inverted to find the $M_{200}$ and $c$ that lead to $\rho_\mathrm{NFW}(r_1) = \rho_1$ and $M_\mathrm{NFW}(<r_1) = M_1$. Note that it is not always possible to find an NFW profile that matches a given $\rho_1$ and $M_1$. This happens when the isothermal region has a fairly constant density out to $r_1$, which produces values of $\rho_1$ and $M_1$ that cannot be matched by even the `flattest' region of an NFW halo (the $\rho \propto 1/r$ inner region). In particular, for a $\rho \propto 1/r$ density profile, $M(<r) = 2 \pi \rho(r) r^3$. So an isothermal profile that leads to $M_1 < 2 \pi \rho_1 r_1^3$ cannot be matched by an NFW profile. When doing inside-out fitting we assign a likelihood of zero to points in parameter space that do not match onto an NFW profile.

One subtlety that arises when switching from outside-in to inside-out isothermal Jeans modelling, is that previously $r_1$ was being calculated from the NFW profile (and $\sigma/m$). Starting from an isothermal profile defined by $N_0$ and $\so$ we need to know $r_1$ in order to find the matching NFW profile, this means that $r_1$ must be calculated from the inner (isothermal) profile. We do this following equation~\eqref{eq:Gamma_r}, where $\rho(r)$ is from the isothermal profile and $\sigma_\mathrm{1D}(r) = \so$. This is not the only way one could go about solving this problem. Instead, the sampled parameters could be $N_0$, $\so$ and $r_1$, from which the matching $M_{200}$ and $c$ could be found, and finally $\sigma/m$ could be determined from $M_{200}$, $c$ and $r_1$. This latter procedure would associate the same $\sigma/m$ with a model SIDM density profile as for our outside-in modelling. The disadvantage of this procedure is that the priors for our MCMC sampling will be defined on the parameters that are being sampled. Having $\sigma/m$ being one of these parameters is therefore good in that it allows us to choose our prior on the cross-section.

The extent to which the inside-out and outside-in procedures that we have described associate a different $\sigma/m$ with the same NFW + isothermal profile depends on how $\so$ compares with $\sigma_\mathrm{1D}^\mathrm{NFW}(r_1)$. If these agree then both procedures lead to the same $\sigma/m$, because the scattering rate is proportional to the product of $\sigma_\mathrm{1D}$, $\rho$ and $\sigma/m$, and the isothermal and NFW densities are equal at $r_1$ by definition. For DM-only haloes simulated with SIDM1 or vdSIDM, we find that the best-fitting isothermal Jeans models to well-resolved simulated haloes typically have $\sigma_\mathrm{1D}^\mathrm{NFW}(r_1) / \so$ in the range 1--1.3, which can increase up to 1.6 for CDM-only haloes.

The isothermal Jeans model is of course only approximate, with the radius $r_1$ dependent on the age of the halo (which does not have an unambiguous definition), and the somewhat arbitrary choice of one scattering per particle to separate the region strongly affected by self-interactions from that not affected at all. It is therefore not clear whether there are better or worse choices for the velocity dispersion used to calculate $r_1$, the definition of halo age, or the number of scatterings per particle at which the behaviour transitions from collisionless to fully collisional. Instead of worrying about these, we aim to state precisely what we have done and then show later that the results of fits to simulations do not lead to inferences on the cross-section that are obviously biased. Had we found that we typically under-predicted the cross-section in our fits by a factor of two, then this could be rectified by changing the transition radius from $r_1$ to $r_2$ (i.e. the radius at which two scatterings per particle have taken place) or by altering the definition of halo age such that the halo is only half as old as it was in our original fit. Given that these changes are perfectly degenerate, there is not a sense in which one is `best', rather fortuitously however, using one scattering per particle as the collisionless/collisional threshold, and a halo age somewhat shorter than the age of the Universe (we use $7.5 \gyr$), produces good results as we will soon discuss.


\subsubsection{Adopted priors}

The parameters that we sample are $N_0$, $\so$ and $\sigma/m$. At each sampled point in this parameter space we must calculate the corresponding $M_{200}$ and $c$ in order to find the density profile at $r > r_1$. We record the $M_{200}$ and $c$ values such that we can also express our posterior distribution in terms of these more familiar parameters. For the priors on the sampled parameters, we follow \citet{2019PhRvX...9c1020R} in using a flat prior on both the logarithm of $N_0$ and the logarithm of $\so$. We also use a flat prior on the logarithm of $\sigma/m$. Specifically, our priors are:

\noindent $\bullet$ $N_0$: Uniform prior on $\log N_0$ in the range $0 <   \log_{10} N_0 < 5$.

\noindent $\bullet$ $\so$: Uniform prior on $\log \so$ in the range $-1 <  \log_{10} \so / \kms < 3.5$.

\noindent $\bullet$ $\sigma/m$: Uniform prior on $\log \sigma/m$ in the range $-2 <  \log_{10} \sigma/m / \cmsg < 2$.

\noindent We do not adopt any prior on the concentration-mass relation, which is discussed in Section~\ref{sect:cM_relation}.

\subsubsection{`Effective' priors}

Our priors are defined in terms of the parameters being sampled, but our results are more familiar when presented in terms of $M_{200}$ and $c$. We can define an `effective prior' on the $(M_{200}, c, \sigma/m)$ parameter space, by sampling from our $(N_0, \so, \sigma/m)$ prior, and finding the corresponding points in $(M_{200}, c, \sigma/m)$. We do this using MCMC, setting the likelihood to a constant value when there is a valid NFW profile at that point in $(N_0, \so, \sigma/m)$, and setting it to zero when there is no matching NFW profile. For our adopted priors, the marginalised effective priors are shown in Fig.~\ref{fig:effective_priors} and are discussed in Appendix~\ref{App:priors}. In general the priors that we adopt on $N_0$, $\so$ and $\sigma/m$ lead to effective priors on $\log M_{200}$ and $\log c$ that are approximately uniform. There is however an effective-prior bias towards larger cross-sections (despite a uniform prior on $\log \sigma/m$), that increases at lower halo masses.

\subsection{MCMC fitting to example haloes}

\begin{figure}
        \centering
        \includegraphics[width=\columnwidth]{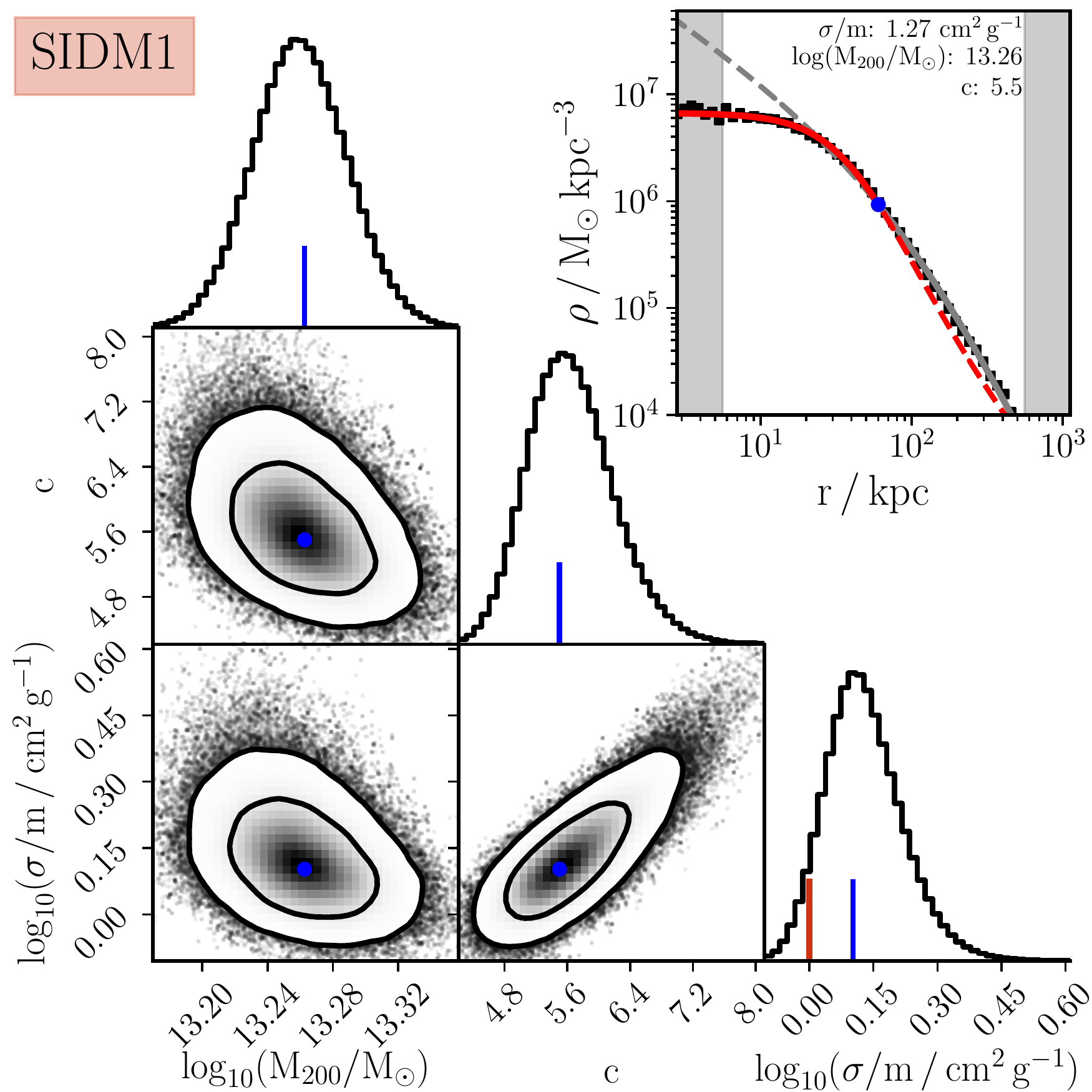}     
	\caption{An example of an isothermal Jeans model fit to a simulated SIDM1-only halo. The `corner plot' shows the marginalised posterior distributions on the parameters $M_{200}$, $c$ and $\sigma/m$, with the contours on the 2D plots enclosing 68\% and 95\% of the posterior probability. The top-right panel shows the simulated density profile as black squares, with the best-fitting isothermal profile shown by the red line and the corresponding NFW profile the grey line. The isothermal Jeans model prediction is that the density profile follows the isothermal solution inside of $r_1$ (the blue marker) and the NFW profile outside of $r_1$, which in this case provides a visually good fit to the simulated profile. The best-fit (maximum likelihood) parameter values are listed in the top-right panel and marked in blue on the corner plot. The adopted halo age is $7.5 \gyr$, and the input $\sigma/m$ of $1 \cmsg$ is marked in the bottom-right panel with the red vertical line.}
	\label{fig:DMO_MCMC}
\end{figure}

For a given simulated halo, we are now ready to calculate the posterior distribution on the $(M_{200}, c, \sigma/m)$ parameter space, using MCMC\footnote{We use the affine invariant Markov chain Monte Carlo ensemble sampler \texttt{emcee} \citep{2013PASP..125..306F}.} with the priors just described and the likelihood from Section~\ref{sect:good_fit}. We show an example of this in Fig.~\ref{fig:DMO_MCMC}, where the halo is the same one as in Fig.~\ref{fig:matching}, which is a DM-only halo from \eagle-50, simulated with $\sigma/m = 1 \cmsg$. The best-fitting (maximum likelihood) density profile is shown in the top-right of Fig.~\ref{fig:matching}, and is a very good fit to the simulated density profile. The inferred halo mass matches the true spherical-overdensity mass, and while the best-fit cross-section is slightly larger than the input cross-section ($1.27 \cmsg$ versus $1 \cmsg$), the posterior on the cross-section is consistent with the true value. It is worth recalling from Section~\ref{sect:good_fit} that there is some level of arbitrariness to the width of the posterior distribution (and therefore the marginalised posterior distributions), because they depend on our fairly arbitrary likelihood defined in equation~\eqref{eq:chi2}. Had we chosen a larger $\delta \log_{10} \rho$ in equation~\eqref{eq:chi2} then our posterior distributions would be broader, and had we used more radial bins they would be narrower.

\subsubsection{`Core collapse' solutions}

\begin{figure*}
        \centering
        \includegraphics[width=\columnwidth]{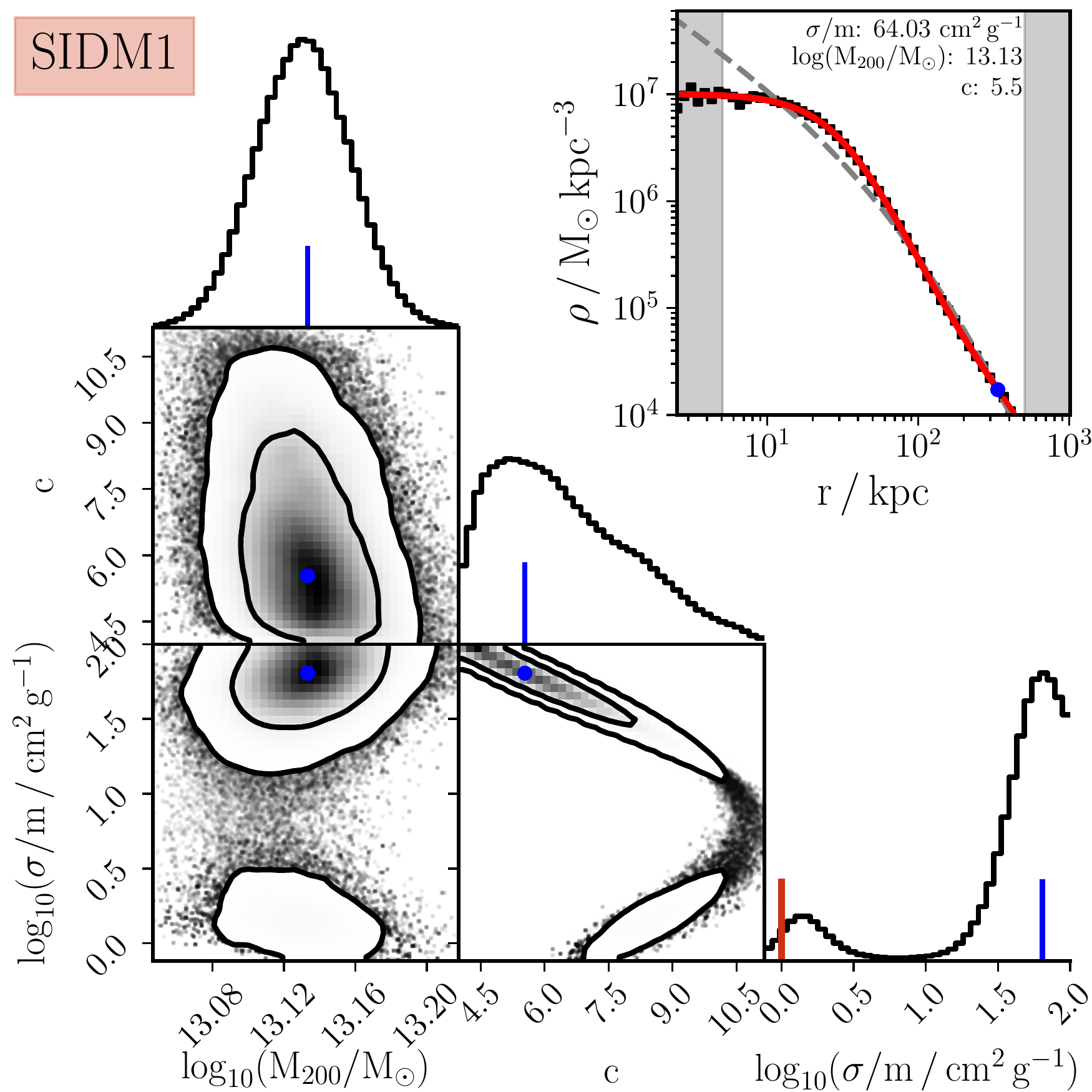}     
                \includegraphics[width=\columnwidth]{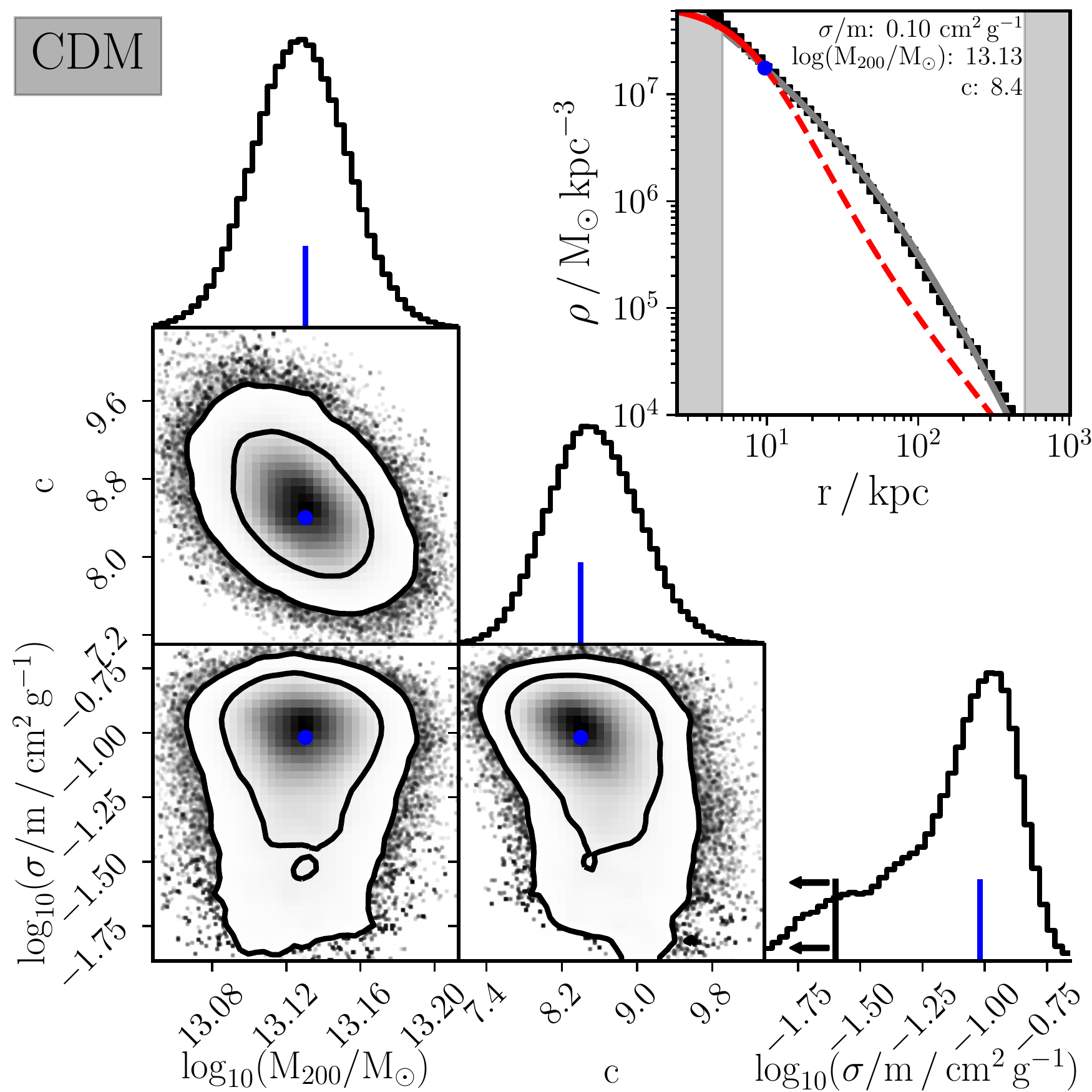}     
	\caption{Left: the same as Fig.~\ref{fig:DMO_MCMC} but for a slightly less massive halo, in which the SIDM density profile is well fit both by the `true' parameters of $\sigma/m=1 \cmsg$, $c \approx 8.5$, but also by a core collapsing solution with a large cross-section and low concentration. The maximum-likelihood solution is a core collapsing one, and the corresponding density profile is plotted alongside the simulated density profile in the top-right. Right: the same halo as in the left panel, but simulated with CDM. The halo mass agrees with the SIDM fits, and the concentration corresponds to the SIDM concentration in the $\sigma/m \approx 1 \cmsg$ peak in the SIDM1 posterior. While the input cross-section is zero, the bulk of the posterior probability is around $0.1 \cmsg$.}
	\label{fig:DMO_MCMC_core_collapse}
\end{figure*}

In the left panel of Fig.~\ref{fig:DMO_MCMC_core_collapse} we show a different halo, which is an example with a more complicated posterior distribution. In this case the marginalised posterior on the cross-section is bimodal, with one peak around the input cross-section of $1 \cmsg$ while the other is around $60 \cmsg$. This second solution corresponds to a halo undergoing `core collapse' \citep{2002PhRvL..88j1301B, 2019PhRvD.100f3007Z}, in that the isothermal Jeans model predicts this solution to become more centrally dense as the halo age is increased (or equivalently, as the cross-section is increased at fixed age). The `banana shaped' degeneracy between $\sigma/m$ and $c$ can then be explained as follows: at low $\sigma/m$, increasing the cross-section decreases the central density, and so the central density in the absence of self-interactions must be increased to compensate (hence an increase in $c$); at larger $\sigma/m$ the halo is undergoing core collapse, and larger cross-sections actually lead to larger central densities, as such, the concentration must now be decreased to maintain a similar density profile.

If we look at this same halo simulated with CDM (right panel of Fig.~\ref{fig:DMO_MCMC_core_collapse}) we see that it is well described by an NFW profile with $c \approx 8.5$. This corresponds to the value of $c$ in the posterior peak close to the input cross-section for the SIDM1 halo. Fitting to the SIDM1 simulated halo with knowledge of what this halo would have looked like in the absence of self-interactions, we could therefore identify the $c \approx 8.5$, $\sigma/m \approx 1 \cmsg$ peak as the truth (as opposed to the other peak at  $c \approx 5.5$, $\sigma/m \approx 60 \cmsg$) and make a correct inference on the cross-section. When dealing with observed systems, this could motivate a prior that halo concentrations roughly follow the concentration-mass relation, which we discuss further in Section~\ref{sect:cM_relation}.

Considering the core collapsing solution, work modelling SIDM as a fluid in which heat is transferred by thermal conduction \citep[e.g.][]{2020PhRvD.101f3009N} suggests that during core collapse the centre of the halo is no longer isothermal, but has a temperature that increases towards the centre of the halo. As such, the isothermal Jeans model we employ here probably does not provide a good description of the density profiles of core collapsing haloes. As we do not have simulated systems with cross-sections large enough for core collapse (ignoring the effects of baryons), we cannot comment further on the extent to which the isothermal Jeans model's description of core collapse is accurate, but we note here that the core collapsing density profiles as predicted by the isothermal Jeans model are sometimes good fits to simulated haloes in which the core is actually growing in time (with the isothermal density profile in the left panel of Fig.~\ref{fig:DMO_MCMC_core_collapse} a good example).

\subsubsection{Isothermal Jeans model fits to other haloes} 

While we have only shown a few example haloes in Figures~\ref{fig:DMO_MCMC} and~\ref{fig:DMO_MCMC_core_collapse}, similar corner plots are available online for our full sample of haloes.\footnote{\href{http://icc.dur.ac.uk/data/}{http://icc.dur.ac.uk/data/}} Further details about how we ran the MCMC, including a discussion of chain length, autocorrelation times, and the convergence of the posteriors, are contained in Appendix~\ref{App:MCMC}.

\section{Results with DM-only haloes}
\label{sect:DMO_results}

\begin{figure}
        \centering
        \includegraphics[width=\columnwidth]{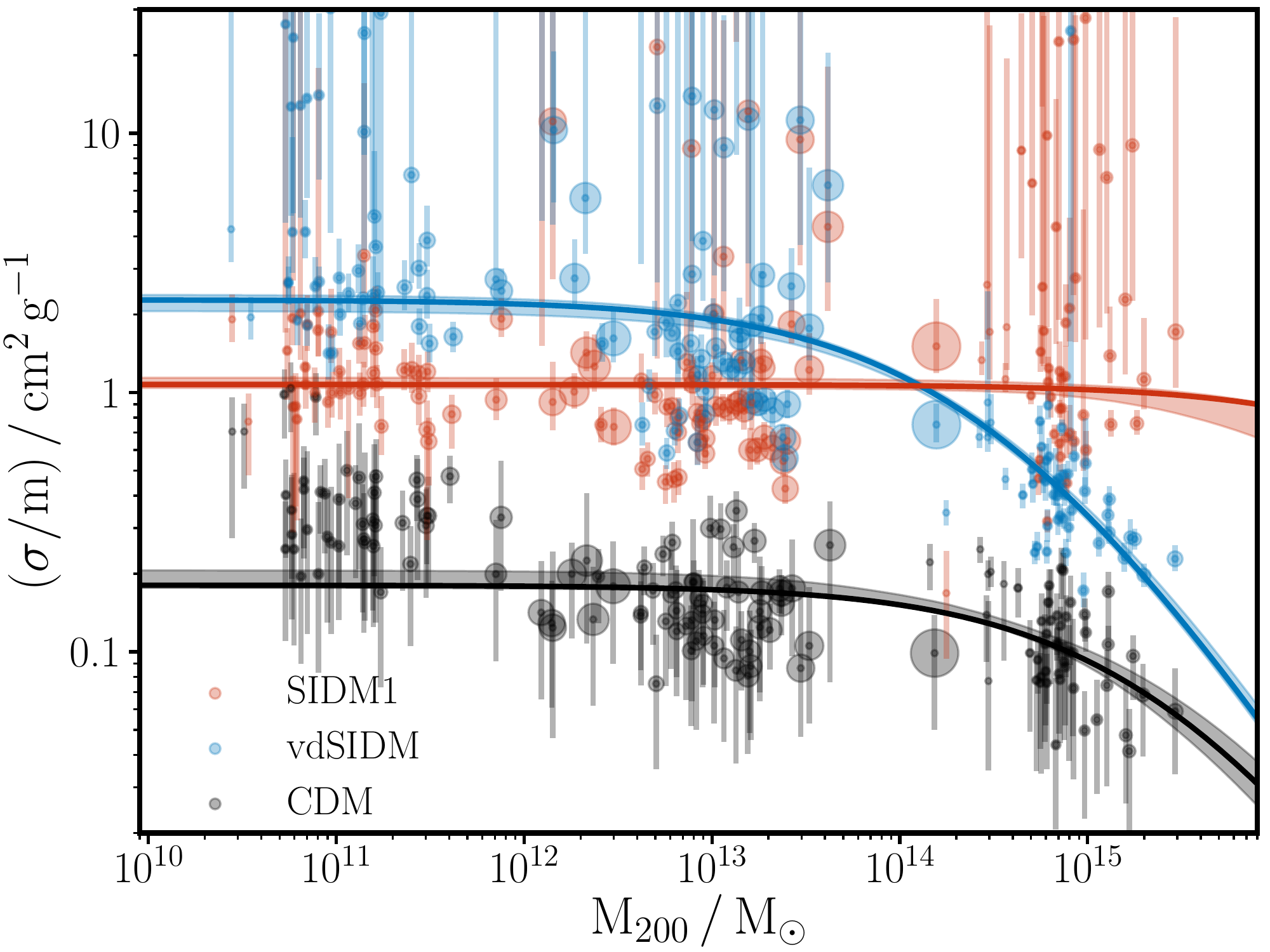}     
	\caption{The inferred masses and cross-sections from isothermal Jeans model fitting to our ensemble of DM-only haloes. Each marker area is proportional to the number of particles in the simulated halo (so at fixed mass, well-resolved haloes from a smaller volume simulation can be distinguished from haloes from a larger volume simulation). The markers are placed at the median $M_{200}$ and median $\sigma/m$ from the posterior distribution generated from each MCMC fit, with the error bars running from the 16th to 84th percentile of the $\sigma/m$ marginalised posterior. The lines show the maximum-likelihood velocity-dependent SIDM cross-section (see equation~\ref{eq:yukawa_sigmaTt_Mw}) fit to the distribution of points for each simulated cross-section, with the shaded regions around each line covering the 16th-84th percentiles of $\sigma/m(M_{200})$ from the velocity-dependent cross-section fits.}
	\label{fig:vdSIDM_fit_DMO}
\end{figure}

Having shown examples of fits to individual haloes, we now look at the results from ensembles of haloes from the different simulations described in Section~\ref{sect:sims}. We take the posterior distributions from isothermal Jeans model fits to the 50 most massive friends-of-friends haloes in each simulation and plot the median $\sigma/m$ from the posterior distributions as a function of the median $M_{200}$ in Fig.~\ref{fig:vdSIDM_fit_DMO}, with error bars on $\sigma/m$ extending from the 16th to 84th percentile of the marginalised posterior. The results are broadly consistent with what one would expect if the isothermal Jeans model is a good description of SIDM density profiles, with fits to SIDM1 haloes having cross-sections that scatter around $1 \cmsg$, CDM leading to cross-sections $\sigma/m \lesssim 0.2 \cmsg$ (except for at low masses, discussed in Section~\ref{sect:resolution}) and with vdSIDM leading to best-fit cross-sections $\approx 3 \cmsg$ at low halo masses, decreasing with increasing halo mass.

Each DM model has 150 simulated haloes spanning the mass range $\num{5e10} - \num{3e15} \msun$, and we fit a velocity-dependent SIDM model to the ensemble of haloes for each of the three models. At the particle physics level, the cross-section depends on the relative velocity between particles, while here we are considering it as a function of halo mass. Given that the typical velocities within a halo scale as $v_{200} = \sqrt{G M_{200} / r_{200}} \propto M_{200}^{1/3}$, we make an ansatz that the effective cross-section as a function of halo mass should look like equation~\eqref{eq:yukawa_sigmaTt}, but with $v/w$ replaced by $(M_{200} / M_w)^{1/3}$. The relative pairwise velocity below which the cross-section is approximately constant, $w$, is then replaced by a halo mass scale below which the cross-section is roughly constant, with the cross-section decreasing at higher halo masses. To be concrete, we fit the following functional form to the distribution of points in Fig.~\ref{fig:vdSIDM_fit_DMO}
\begin{equation}
\label{eq:yukawa_sigmaTt_Mw}
\sigma_{\tilde{T}}(M) = \sigma_{T0} \frac{4 M_w^{4/3}}{M^{4/3}} \left\{ 2 \ln \left(1 + \frac{M^{2/3}}{2 M_w^{2/3}} \right) - \ln \left(1 + \frac{M^{2/3}}{M_w^{2/3}} \right) \right\},
\end{equation}
where $M$ is $M_{200}$. Note that the vdSIDM differential cross-section can capture all three of our simulated cross-sections, not just the vdSIDM one. CDM is the case where $\sigma_{T0} = 0$, and SIDM1 corresponds to $\sigma_{T0}/m = 1 \cmsg$ and $w$ (or $M_w$) $\to \infty$. 

When fitting equation~\eqref{eq:yukawa_sigmaTt_Mw} to these points we use the posterior distribution generated from each isothermal-model fit to define the likelihood. Given that the masses are well constrained in the fits, we ignore the uncertainty on the mass of each system (using the median value). The MCMC isothermal-model fitting to each halo, $i$, produces a marginalised posterior probability density on the logarithm of the cross-section, $\mathrm{d}P_i / \mathrm{d} \log (\sigma/m)$ (with an example being the histogram plotted in the bottom-right of Fig.~\ref{fig:DMO_MCMC}). The likelihood for a given vdSIDM model (parameterised by $\sigma_{T0}$ and $M_w$) that we use when fitting a vdSIDM model to an ensemble of haloes is
\begin{equation}
\mathcal{L}(\sigma_{T0}, M_w) = \prod_{i=1}^{150} \dv{P_i}{\log (\sigma/m)} \left( \frac{\sigma_{\tilde{T}}}{m}(M_i; \sigma_{T0}, M_w ) \right),
\label{eq:vdSIMD_likelihood}
\end{equation}
which (in words) is the product over 150 haloes of the marginalised $\log \sigma/m$ posterior density, with each density evaluated at the $\sigma/m$ predicted by $\sigma_{T0}$ and $M_w$ at the mass of the halo in question.

In practice we don't actually have access to $\mathrm{d}P_i / \mathrm{d} \log (\sigma/m)$, instead having samples drawn from it. We therefore estimate this probability density (up to a constant) as the inverse of the distance (in $\log \sigma/m$) to the $n$th nearest posterior sample to $\sigma_{\tilde{T}}(M_i)/m$. With an infinite number of samples, this inverse distance tends to the (unnormalised) probability density, while with a finite number of samples it is an estimate of the mean probability density within a top-hat window centred on the model-predicted cross-section in a halo with mass $M_i$. We set $n$ such that the probability mass within the top-hat window is $1\%$ of the total. Specifically, our chains each contain $280 \, 000$ (non-independent) samples from the posterior, and we find the distance to the $2800$th nearest.

While we introduced this method as a way to estimate the probability density from samples drawn from it, it also serves to limit the influence of outliers. Taking SIDM1 haloes as an example and considering the case of fitting a model with a constant cross-section ($w \to \infty$), two of the 150 haloes have no posterior samples with $\sigma/m < 1 \cmsg$, while seven of them have no samples with $\sigma/m >1 \cmsg$. As such, if we obtained $\mathrm{d}P_i / \mathrm{d} \log (\sigma/m)$ by smoothing the distribution of samples with some small compact kernel, then $\mathcal{P}(\sigma_{T0}, M_w)$ would be zero for all values of $\sigma_0 / m$. Defining the probability density as proportional to the inverse of the distance to the $n$th nearest neighbour leads to a non-zero probability density at any value of the cross-section, circumventing this problem.\footnote{This is not an especially good way to deal with outliers, and the extent to which it penalises a set of model parameters for having outliers depends on the choice of $n$, but we find that our best-fitting vdSIDM models are relatively unaffected by a factor of 10 change in $n$ and so we believe this method is adequate for our current goal of comparing the sorts of cross-section one would infer from an ensemble of haloes to the true (input) cross-section.} 
 
In general we find that the best-fitting velocity-dependent SIDM cross-sections are good reflections of the input cross-sections used in the simulations, except for with CDM (see Section~\ref{sect:resolution}). Considering the fit to the SIDM1 haloes, the inferred SIDM model has  $M_w \gtrsim 10^{15} \msun$, meaning that the cross-section is correctly inferred to be velocity-independent over the range of halo masses studied. The normalisation of the cross-section, $\sigma_{T0} = (1.08 \pm ^{0.06}_{0.04}) \cmsg$, is slightly larger than the input, but with 150 haloes systematic errors become more important than random errors, and small systematic changes to the analysis (such as adopting an increased halo age of $8 \gyr$) would bring the inferred cross-section in line with the truth.

For our simulated vdSIDM model, the input velocity-scale for the cross-section is $w = 560 \kms$. The maximum-likelihood value of $M_w$ when fitting to the vdSIDM ensemble of haloes is $M_w = \num{1.3e14} \msun$, which would correspond to $w=730 \kms$ using the simple relationship between a halo mass and an effective velocity for DM interactions from Fig.~\ref{fig:cross_sects} ($v_\mathrm{rel} \approx \sqrt{G \, M_{200} / r_{200}}$). One could imagine that a better approach to mapping from a halo mass to an effective pairwise velocity would be to calculate the mean pairwise velocity for particles within the halo. Using the 1D velocity dispersion of DM particles, $\sigma_\mathrm{1D}$, as a function of halo mass from \citet{2013MNRAS.430.2638M}, combined with the fact that for a Maxwellian velocity distribution the mean pairwise velocity is $\langle v_\mathrm{rel} \rangle = 4/\sqrt{\pi} \, \sigma_\mathrm{1D}$, would lead to $M_w = \num{1.3e14} \msun$ mapping to a $\langle v_\mathrm{rel} \rangle$ of $1100 \kms$ -- further from the true value of $w$. This happens because the velocity dispersion in an NFW halo drops towards the centre of the halo, and it is the centre of the halo where interactions are important. As such, halo-wide estimates of the velocity dispersion over-predict the velocity at which SIDM interactions are typically taking place.

Note that it is important to properly account for the non-Gaussian marginalised posterior distributions for $\log \sigma/m$ (with a good example of this non-Gaussianity being in the left panel of Fig.~\ref{fig:DMO_MCMC_core_collapse}). As an alternative to the likelihood defined in equation~\eqref{eq:vdSIMD_likelihood}, we used the mean and variance of the $\log \sigma/m$ values in the MCMC chains to define a likelihood with a Gaussian term for each halo. The results were qualitatively similar to those using the full likelihood shown in Fig.~\ref{fig:vdSIDM_fit_DMO}, but the vdSIDM model parameters were typically further from their input values. Specifically, we found that when using a Gaussian likelihood the maximum-likelihood model for the CDM haloes had $\sigma_{T0} = 0.23 \cmsg$ (up from $0.18 \cmsg$), while for the vdSIDM haloes it was $\sigma_{T0} = 1.49 \cmsg$ (down from $2.28 \cmsg$ -- the true value is $3.04 \cmsg$). With SIDM1 the Gaussian likelihood and full likelihood lead to $\sigma_{T0} = 0.93$ and $1.07 \cmsg$ respectively.

\subsection{The cross-section as a function of velocity}
\label{sect:vrel_sigma}

\begin{figure}
        \centering
        \includegraphics[width=\columnwidth]{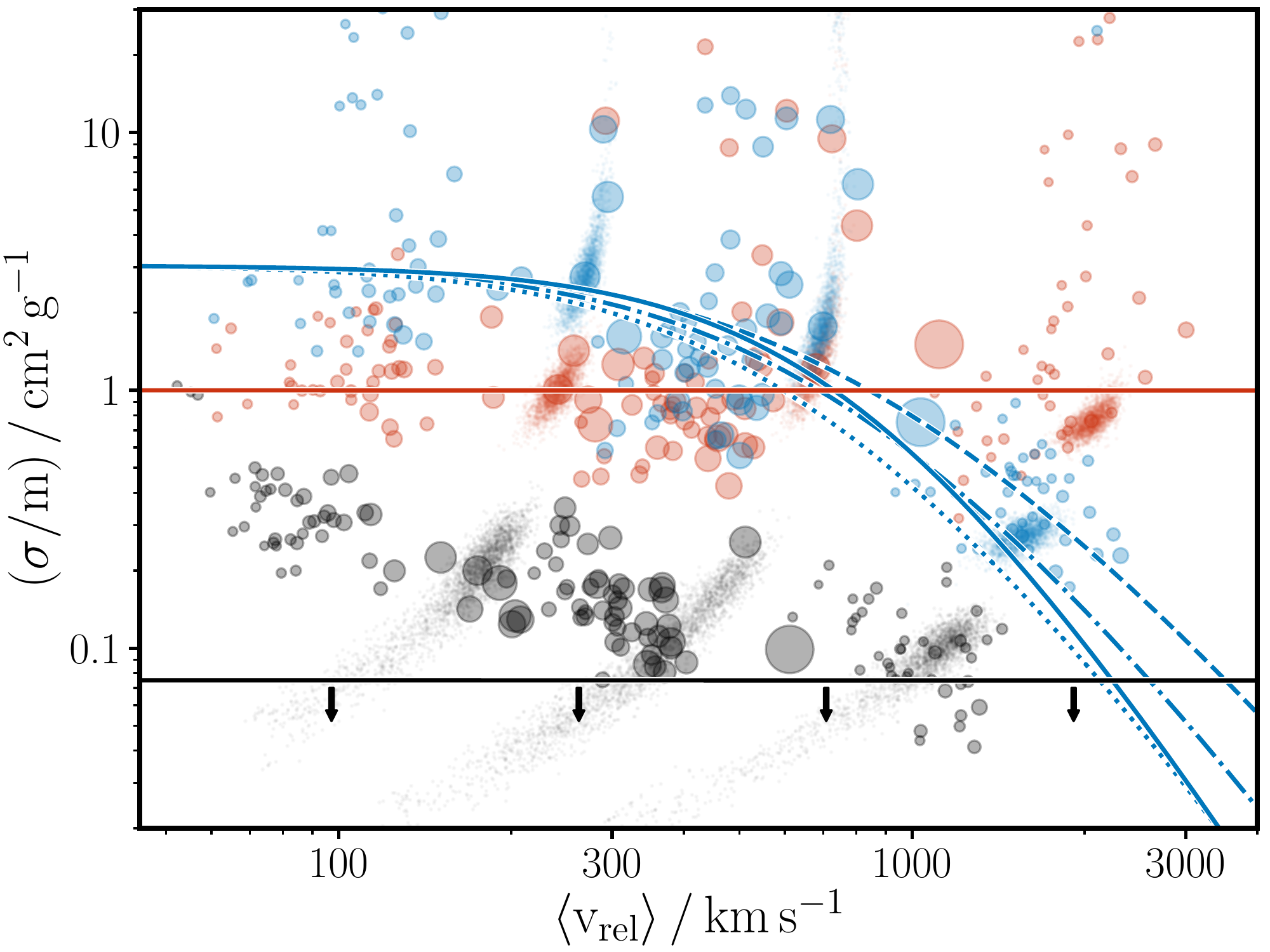}     
	\caption{The same as Fig.~\ref{fig:vdSIDM_fit_DMO} but now plotted as a function of $\langle v_\mathrm{rel} \rangle = 4/\sqrt{\pi} \, \sigma_0$. To highlight the degeneracy between $\sigma/m$ and $\sigma_0$ (and hence $\langle v_\mathrm{rel} \rangle$), we show -- as a cloud of points -- MCMC samples from fits to one halo from each simulation volume with each cross-section. This degeneracy makes fitting a velocity-dependent SIDM model challenging, so for comparison we show the input $\sigma_{\tilde{T}}(\langle v_\mathrm{rel} \rangle)/m$ as the solid lines (see Fig.~\ref{fig:cross_sects}). Owing to subtleties that arise with velocity-averaging of a velocity-dependent cross-section, for the vdSIDM model we plot three other lines in different line styles. The distinction between these lines is discussed in Section~\ref{sect:vrel_sigma} and elaborated upon further in Appendix~\ref{App:vel_av}.}
	\label{fig:sigma_vs_vrel}
\end{figure}

Given that interactions are taking place in the centre of the halo, and that -- within the isothermal Jeans model -- the velocity dispersion there is $\sigma_0$, another sensible approach would be to bypass an explicit mapping from halo masses to effective relative velocites altogether, and instead plot the results in terms of $\sigma/m$ against $\sigma_0$ (or $\langle v_\mathrm{rel} \rangle$). We show such a plot in Fig.~\ref{fig:sigma_vs_vrel}. There is a complication with fitting a vdSIDM cross-section to these however, in that there is a strong degeneracy between $\sigma/m$ and $\sigma_0$ in the isothermal Jeans model fits. At fixed halo mass, increasing the cross-section increases $r_1$, which increases the temperature of the isothermal region (because $\sigma_\mathrm{1D}$ increases with radius in the inner regions of an NFW profile). Fitting a vdSIDM model would therefore require a hierarchical model, in which one samples from the joint posterior of model values ($\sigma_{T0}$ and $w$) and the halo-specific parameters ($M_{200}$ and $c$ or $N_0$ and $\sigma_0$) of each halo, and then marginalises over the halo-specific parameters to get the posterior on $\sigma_{T0}$ and $w$. This is beyond the scope of this work, and so here we simply plot the input cross-sections on Fig.~\ref{fig:sigma_vs_vrel}, such that they can be visually compared with the median MCMC parameter values.

It is interesting to compare how the vdSIDM cross-sections inferred from isothermal Jeans model fits compare with the true input cross-section. The isothermal Jeans model assumes a constant (velocity-independent) cross-section, and in the region where scattering takes place the velocity distribution is assumed to be a Maxwell--Boltzmann distribution with a 1D velocity dispersion of $\sigma_0$. For a particular vdSIDM halo, one could imagine that the effective cross-section within the halo is simply the velocity-dependent cross-section $\sigma_{\tilde{T}}(v_\mathrm{pair})$, evaluated at the mean pairwise velocity in the isothermal region, $v_\mathrm{pair} = \langle v_\mathrm{rel} \rangle = 4/\sqrt{\pi} \, \sigma_0$ -- this is what is plotted as the solid line in Fig.~\ref{fig:sigma_vs_vrel}.

However, different pairs of particles in the isothermal region will have a wide range of relative velocities, and simply taking the cross-section at the mean pairwise velocity may not be an adequate reflection of the effects of scattering. Instead one could imagine that the mean value of the cross-section averaged over pairs of particles drawn from the Maxwell--Boltzmann distribution might give a better description. This is plotted as the dashed line in Fig.~\ref{fig:sigma_vs_vrel}, and does indeed seem to provide a better match to the $\sigma_0$--$\sigma/m$ values found from isothermal Jeans modelling of the vdSIDM haloes. More details about velocity averaging can be found in Appendix~\ref{App:vel_av}, which also describes the velocity averaging procedure used for the dotted and dot-dashed lines in Fig.~\ref{fig:sigma_vs_vrel}.

\subsection{Quality of fits}
\label{sect:quality_of_fit}

Aside from the question of whether one recovers the true input cross-section for a simulation when fitting the isothermal Jeans model to the simulated density profiles, another interesting question is how good these fits are. As we have already mentioned, the normalisation of our $\chi^2$ (equation~\ref{eq:chi2}) is fairly arbitrary, because the mismatch between our simulated and model density profiles is primarily systematic (i.e. the model not properly capturing the shape of the density profiles) rather than random (the deviations between the model and simulated density profiles are not driven by particle noise in the simulations for example). With the $\chi^2$ as we have previously defined it, the $\chi^2$ per degree of freedom for our best-fit density profiles are typically well below unity, reflecting the fact that the best-fitting density profiles fit to better than 0.1 dex, as can be seen in the examples in Figures~\ref{fig:DMO_MCMC} and  \ref{fig:DMO_MCMC_core_collapse}.

\begin{figure}
        \centering
        \includegraphics[width=\columnwidth]{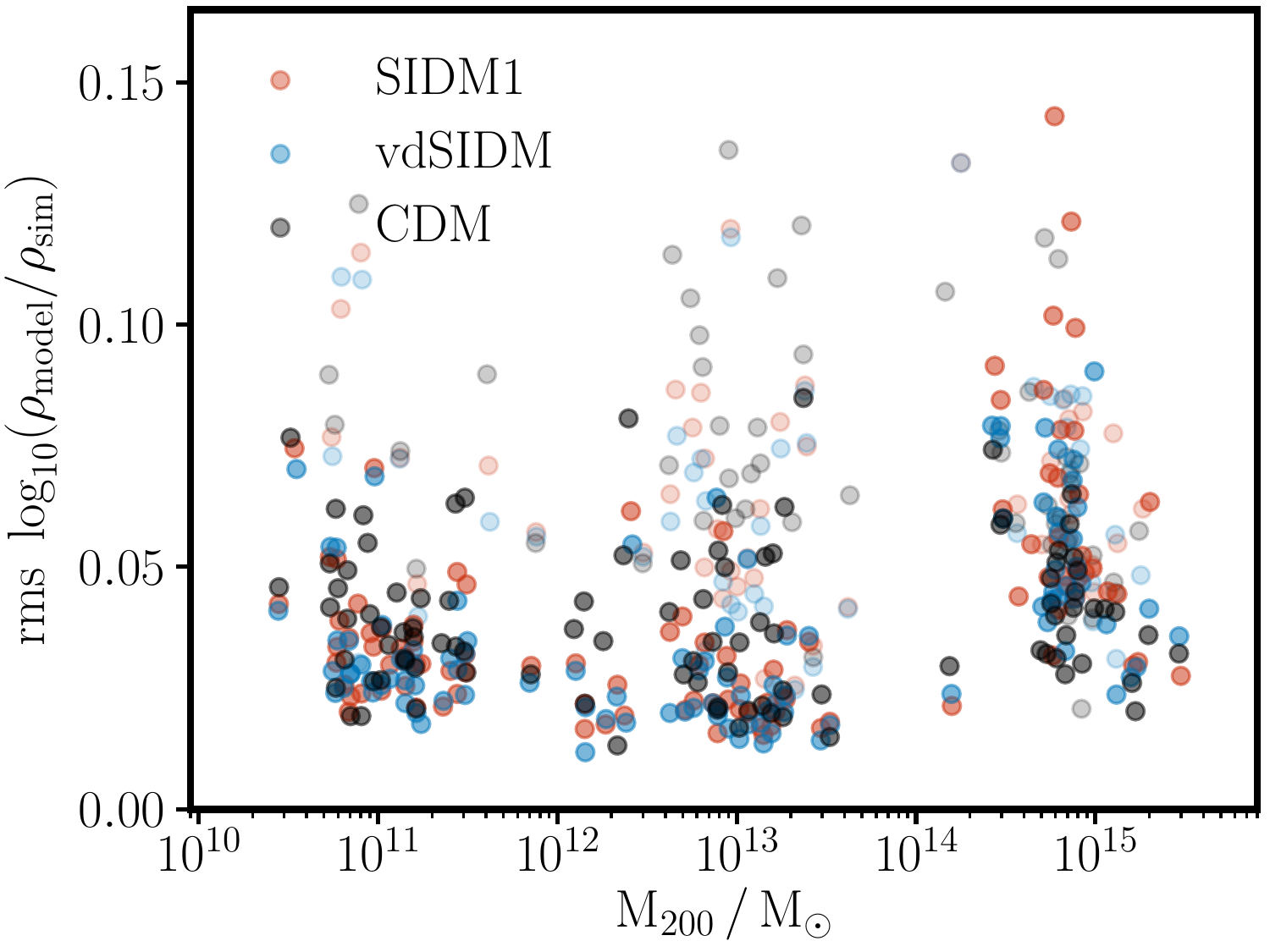}     
	\caption{The goodness-of-fit for the best-fitting isothermal Jeans model density profiles to the DM-only haloes in Fig.~\ref{fig:vdSIDM_fit_DMO}. The $y$-axis values are $\delta_\mathrm{rms}$ as defined in equation~\eqref{eq:rms}. The points are coloured according to the DM model, with faded points corresponding to haloes that are `unrelaxed' according to the \citet{2007MNRAS.381.1450N} criteria (see Section~\ref{sect:quality_of_fit}).}
	\label{fig:goodness_of_fit_DMO}
\end{figure}

To give a sense of how well the best-fitting density profiles match the simulated ones, we define the rms error on the logarithm of the density, $\delta_\mathrm{rms}$, by
\begin{equation}
\delta_\mathrm{rms}^2 = \frac{1}{N_\mathrm{bins}} \sum_{i=1}^{N_\mathrm{bins}} \left( \log_{10} \left[ \rho_\mathrm{sim} (r_i) / \rho_\mathrm{mod} (r_i) \right] \right)^2.
\label{eq:rms}
\end{equation}
We calculate this quantity for each of the haloes shown in Fig.~\ref{fig:vdSIDM_fit_DMO} and plot them in Fig.~\ref{fig:goodness_of_fit_DMO}. For haloes from \eagle-12 and \eagle-50 (primarily covering masses from $10^{11}$ to $10^{13} \msun$), the mean $\delta_\mathrm{rms}$ values are 0.041, 0.038 and 0.050 for SIDM1, vdSIDM and CDM respectively. These rise to 0.064, 0.059 and 0.053 for haloes from \bahamas. We note that this means that at low halo masses, our simulated haloes with larger cross-sections are better fit by the NFW + isothermal model, while at cluster scales this trend reverses and it is haloes simulated with smaller cross-sections whose density profiles can be better fit.

When fitting NFW profiles to CDM haloes, the concept of `relaxed' versus `unrelaxed' haloes is often invoked, as NFW profiles provide significantly better fits to relaxed haloes than unrelaxed ones. To investigate the effects of relaxedness on the quality of our fits, we use the relaxation criteria of \citet{2007MNRAS.381.1450N}, to determine if haloes are relaxed or not. The criteria for a halo to be relaxed are that the total mass within resolved substructures, as identified by the \subfind algorithm \citep{2001MNRAS.328..726S}, with centres within $r_{200}$ is less than 10\% of $M_{200}$; that the offset between the centre of mass of all particles within $r_{200}$ of the halo centre and the halo centre itself is less than 7\% of $r_{200}$;\footnote{As in Section~\ref{sect:measuring_rho_r} we define the centre of the halo as the position of the particle with the minimum gravitational potential energy.} and that the virial ratio, $2T/|U|$, is less than 1.35, where $T$ is the total kinetic energy of particles within $r_{200}$ and $U$ is their mutual gravitational potential energy.

Using these criteria, we find that around 80\% of the 50 most massive \eagle-12 DM-only haloes are relaxed, dropping to around 60\% for \eagle-50 and \bahamas. These fractions are roughly independent of DM model, although larger cross-sections do seem to produce a slightly larger fraction of relaxed galaxy clusters, with 33 of the 50 SIDM1 \bahamas haloes relaxed, while only 27 of the CDM ones are. As can be seen in Fig.~\ref{fig:goodness_of_fit_DMO}, the unrelaxed haloes are typically those with the largest $\delta_\mathrm{rms}$, and removing them leads to the bulk of \eagle-12 and \eagle-50 haloes fit to better than 0.05 dex, consistent with other work on the density profiles of CDM haloes and the goodness-of-fit of NFW profiles \citep[e.g.][]{2004MNRAS.349.1039N}. From Fig.~\ref{fig:goodness_of_fit_DMO} we conclude that the isothermal Jeans model fits the density profile of SIDM-only haloes at a similar level as the NFW profile fits CDM-only haloes,\footnote{As the isothermal Jeans model contains the NFW profile (in the limit of small $r_1$) our best-fit isothermal models to CDM density profiles could not be improved by fitting just an NFW profile.} except for in galaxy clusters where the four relaxed systems with the largest $\delta_\mathrm{rms}$ are all from SIDM1. In two of these four cases the corresponding CDM halo is deemed unrelaxed, so at some level these reflect the increased relaxed fraction with SIDM1. Another likely contributor is that in SIDM-only systems, the central density decreases with increasing halo mass. This leads to the inner regions of \bahamas-SIDM1 haloes having the fewest particles per radial bin of all simulated systems. Inspecting the simulated density profiles, there are indications of particle noise out to around $0.05 \, r_{200}$ which will increase the $\delta_\mathrm{rms}$ values.

\subsection{The concentration-mass relation}
\label{sect:cM_relation}

We have seen that the isothermal Jeans model provides a good description of SIDM density profiles, at a level similar to that with which NFW profiles describe CDM density profiles. It is then interesting to ask whether these good fits are achieved in the way the model envisaged (i.e. the $M_{200}$ and $c$ reflecting what the halo would look like in the absence of self-interactions, and then the isothermal region describing how self-interactions change the inner profile) or if, for example, a good fit can be achieved but only by using a very different value for $c$ than what would have been the case without self-interactions.

\begin{figure}
        \centering
        \includegraphics[width=\columnwidth]{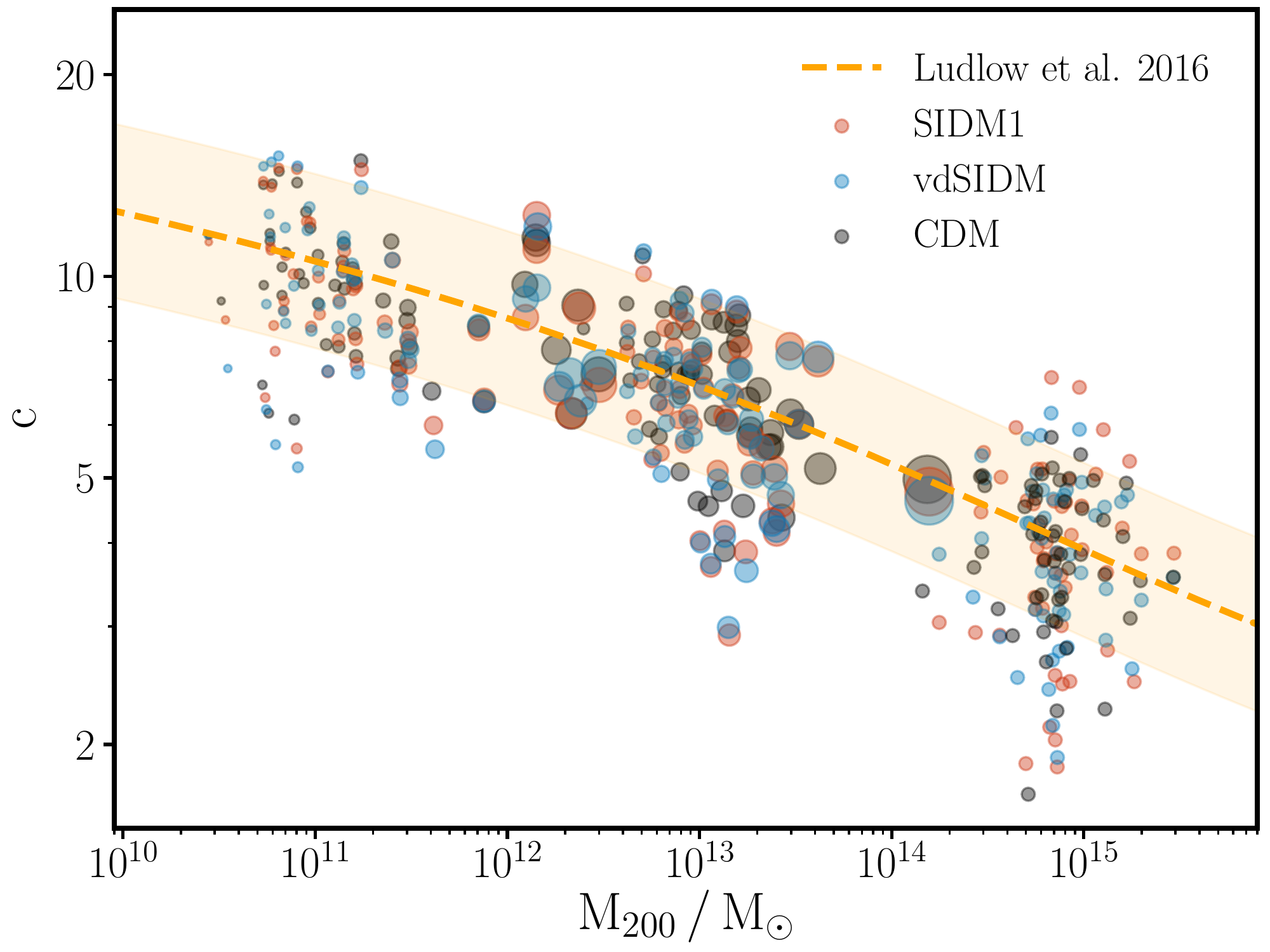}     
	\caption{The concentration-mass relation implied by our isothermal Jeans model fits. The $M_{200}$ and $c$ values are median values from the marginalised posterior distributions for each halo. The orange line shows the median concentration-mass relation from \citet{2016MNRAS.460.1214L}, with the shaded region showing the expected $1 \sigma$ scatter of 0.13 dex \citep{2014MNRAS.441.3359D}.}
	\label{fig:concentration_mass}
\end{figure}

We have already seen in Fig.~\ref{fig:DMO_MCMC_core_collapse} an example halo where the isothermal Jeans model provides a good fit when adopting the true cross-section and the concentration of the corresponding CDM halo. In Fig.~\ref{fig:concentration_mass} we plot the median posterior concentrations against the median posterior halo masses and show that it is generally true that the relationship between halo concentration and halo mass that one finds for isothermal Jeans model fits to SIDM simulated systems is in good agreement with the CDM concentration--mass relation, $c(M)$. This suggests that the isothermal Jeans model really is a reasonable approximation to the physics responsible for shaping the density profiles of SIDM haloes.

The fact that SIDM haloes modelled in the context of the isothermal Jeans model have concentrations that are broadly consistent with the CDM $c(M)$ relation, leads to two interesting questions. First, could our inference on the cross-section be improved by adopting a prior on the concentration-mass relation?  Second, should such a prior be adopted when dealing with observations of real systems?

To assess how a prior on $c(M)$ affects our results, we redid our analysis using a log-normal prior on $c(M)$ with a median relation from \citet{2016MNRAS.460.1214L}, and with a standard deviation of 0.13 dex \citep{2014MNRAS.441.3359D}.\footnote{We did not re-run our MCMC analyses. Instead we re-weighted each point in the chain with a weight of $w_i \propto p(c_i|M_i) \propto \exp \left[ -(\log c_i- \log c_\mathrm{L16}(M_i) )^2 / \, 2 \times 0.13^2 \right]$, where $c_i$ and $M_i$ are the concentration and mass of a point in the MCMC chain, $c_\mathrm{L16}(M)$ is the  \citet{2016MNRAS.460.1214L} concentration--mass relation at $z=0$, and 0.13 is the adopted scatter in $\log_{10} c$ at fixed halo mass.} For systems that are not well-fit by core-collapsing solutions, the concentrations are already well constrained by the likelihood, and adopting this prior makes little difference. For those systems well fit by core collapsing solutions (those in the top of Fig.~\ref{fig:vdSIDM_fit_DMO}, with a specific example being the SIDM1 halo from Fig.~\ref{fig:DMO_MCMC_core_collapse}) this prior can have a noticeable impact on the marginalised $\sigma/m$ posterior. However, these changes were not exclusively in the direction of improving the match between the $\sigma/m$ posterior and the input cross-section, with about half of the SIDM1 systems having a median $\sigma/m$ that actually moves away from the input value when adopting this prior. Core collapse solutions typically have lower concentrations than the `true' solutions (ones adopting the correct cross-section), and so the core collapse solutions of intrinsically high concentration systems can be a good match to the $c(M)$ relation. It is these haloes most likely to be well fit by core collapsing solutions, because core collapse sets in sooner in high concentration haloes \citep{2019PhRvL.123l1102E}. The halo in Fig.~\ref{fig:DMO_MCMC_core_collapse} is a good example of this, where the concentration-mass relation predicts $c \approx 6.5$ (so the CDM version of this halo with $c=8.4$ is a high-$c$ outlier), and a prior on $c(M)$ increases the probability associated with the high-$\sigma/m$ core-collapsing solutions at the expense of $\sigma/m \approx 1 \cmsg$ solutions.

The reason that our fits generally constrain the halo concentration quite tightly (without adopting a prior on $c(M)$) is that we fit to the density profile over a large range of radii. For observed systems this often may not be possible. For example, if using stellar kinematics or HI rotation curves to infer the DM density profile, one may only have measurements in the inner region of the halo. In such cases the halo concentration may be poorly constrained by the data, and it would therefore be sensible to adopt a prior on $c(M)$, as recently done in \citet{2019PhRvX...9c1020R} and \citet{2020arXiv200612515S}.

\subsection{Effects of halo age}
\label{sect:halo_age}

One aspect of the isothermal Jeans model that we have not yet given much attention to is the halo age. For inside-out matching, the scattering rate from equation~\eqref{eq:Gamma_r} leads to $r_1$ being defined by
\begin{equation}
\frac{\sigma}{m} \frac{4}{\sqrt{\pi}} \sigma_0 \rho_\mathrm{iso}(r_1) t_\mathrm{age} = 1.
\label{eq:N1_age}
\end{equation}
Setting the radius $r_1$ is the only place where the cross-section enters the isothermal Jeans model, and from equation~\eqref{eq:N1_age} it is therefore clear that the cross-section and age are perfectly degenerate, with their product being all that matters for the predicted density profile.

Our results thus far have assumed a common age for all haloes of $7.5 \gyr$. Here we investigate the impact of a physically motivated definition of halo age, to see if it can explain some of the scatter in the isothermal Jeans model cross-sections about the true (input) cross-sections. We focus on the \eagle-50 simulation with SIDM1 because we have merger trees available for \eagle-50 simulations (allowing us to track the growth of a halo through time) and because the SIDM1 simulations have a well-defined `correct' answer for the cross-section (CDM has zero cross-section and vdSIDM has an effective cross-section that we expect to vary with halo mass).

\begin{figure}
        \centering
        \includegraphics[width=\columnwidth]{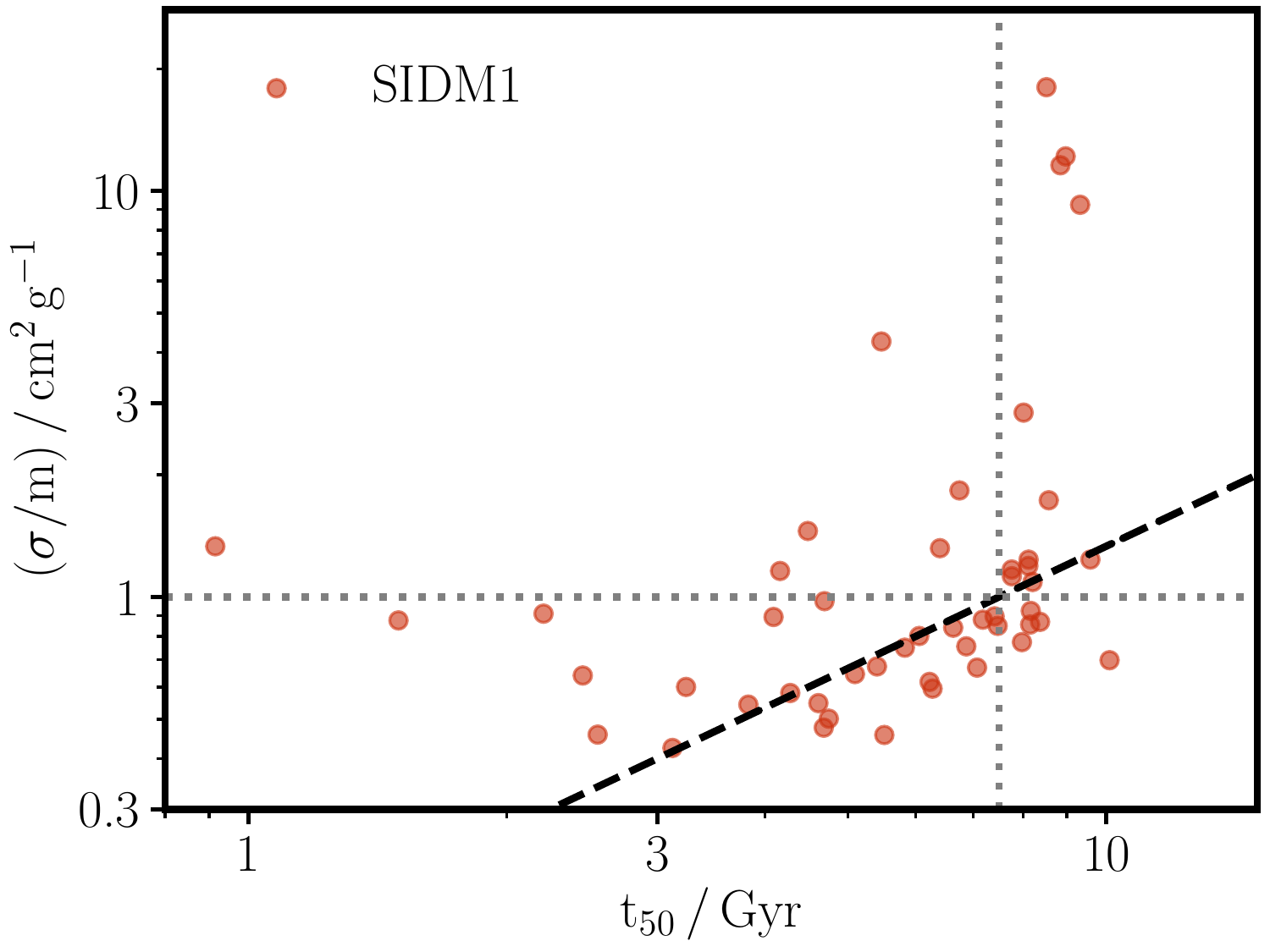}     
	\caption{The median posterior cross-section as a function of halo age (defined as the time since over 50\% of the halo's $z=0$ mass was first in a single halo), for the 50 SIDM1-only haloes from \eagle-50. The input cross-section of $1 \cmsg$ and the assumed halo age of $7.5 \gyr$ are marked with grey dotted lines. If differences between the true $\sigma/m$ and that inferred when fitting to the haloes density profiles were due solely to assuming a constant age -- when in fact the haloes have a range of ages -- the points would be expected to lie along the black dashed line.}
	\label{fig:formation_time}
\end{figure}

We follow \citet{1994MNRAS.271..676L} and define the halo age, $t_{50}$, as the time since the main progenitor contained at least 50\% of the present-day halo mass. In Fig.~\ref{fig:formation_time} we plot this halo age against the median posterior cross-section for each of our haloes. If variations in halo age were the sole driver of differences between the recovered $\sigma/m$ from isothermal Jeans modelling and the input $\sigma/m$ then the points in Fig.~\ref{fig:formation_time} would lie along the black dashed line. This is not what we find, although there is a slight trend for increasing inferred $\sigma/m$ with increasing $t_{50}$.

We experimented with other definitions of halo age, that required a lower fraction of the present day mass to be in the main progenitor (for example $t_{10}$ or $t_{3}$) but found that in none of these cases was there a clear linear relationship between the halo age and the cross-section inferred assuming a constant age. In fact, for definitions of halo age such as $t_3$ there is little spread in halo age, with the haloes in the mass range probed by \eagle-50 all being slightly younger than the age of the universe.\footnote{It is not that all haloes accumulated 3\% of their mass at the same time, but (as an example) the lookback times to $z=4$ and $z=10
$ differ by less than 10\%.} Note that it should probably not come as a surprise that there is not a simple definition of halo age for which the isothermal Jeans model then exactly works. In fact, the correlation between different age definitions is only weak \citep[see][for a comparison of $t_{50}$ and $t_{4}$]{2012MNRAS.422..185G}. Early-forming haloes in one definition can be late-forming by another, so formation histories are more complex than a single `age'. Combined with this, SIDM interactions in the smaller haloes that merge to form a large one have already affected the inner density and velocity structure of the DM halo, and so it is not only self-interactions after `formation' that are important.

Given that determining the age of a halo would be hard observationally, the fact that using the true age (for some definition of age) for the simulated systems does not lead to a large improvement on the inference of the cross-section suggests that observationally it is probably best just to assume some fixed age for haloes. The core collapsing solutions at large halo masses make it hard to be definitive, but there is an indication in Fig.~\ref{fig:vdSIDM_fit_DMO} that for SIDM1 haloes in which the cross-section is well constrained, the inferred cross-section decreases slightly with increasing halo mass. This trend would be consistent with the fact that in a $\Lambda$CDM (or $\Lambda$SIDM) cosmology, more massive haloes have formed more recently \citep[e.g.][]{1993MNRAS.262..627L}, and so adopting a halo age that decreases slightly with increasing halo mass may improve the inference on the cross-section.

\subsection{Effects of resolution}
\label{sect:resolution}

We end this section on isothermal fits to simulated DM-only systems with a discussion of the spatial resolution of our input simulations, and how this resolution could be affecting our results. In particular, in Fig.~\ref{fig:vdSIDM_fit_DMO} the CDM systems have median posterior cross-sections of order $0.1 \cmsg$ at cluster scales, which rises to around $0.4 \cmsg$ in $10^{11} \msun$ haloes. Given that the primary effect of numerical resolution on DM density profiles is to artificially decrease the central density of haloes \citep{2003MNRAS.338...14P}, one might imagine that the cross-sections returned by the isothermal Jeans model fit to CDM haloes are a result of the spatial resolution of the simulations.

However, when we consider the numerical parameters of our simulations and the radial range over which we fit to the density profiles, we do not expect resolution to be playing an important role. This is borne out in Fig.~\ref{fig:vdSIDM_fit_DMO} by the fact that the most massive \eagle-12 and \eagle-50 haloes are not obvious outliers with respect to the similarly-massive -- but much more poorly resolved -- haloes from \eagle-50 and \bahamas respectively. Rather than an effect of numerically-formed cores, we believe the driver for the non-zero cross-sections when fitting to CDM haloes is a combination of the minimum radius to which we fit the density profiles and a rather subtle effect of inside-out fitting and a resulting bias against small cross-sections. This is further discussed in Appendix~\ref{App:resolution}.

\section{Results with hydrodynamical haloes}
\label{sect:hydro_results}

As described in Section~\ref{sect:method_with_baryons}, the isothermal Jeans model can be readily extended to include the effects of the gravitational potential due to baryons. When dealing with observed systems, this usually involves taking observed optical images (for stars) or HI and/or CO data (for gas) and using these to infer the baryon profile. In this work we do not concern ourselves with the particulars of dealing with observations, instead focussing solely on how well the isothermal Jeans model describes SIDM density profiles when the distribution of baryons is known perfectly. To this end, we use the spherically averaged $M_\mathrm{bar}(<r)$ measured from the simulations in the isothermal Jeans model.

\begin{figure}
        \centering
        \includegraphics[width=\columnwidth]{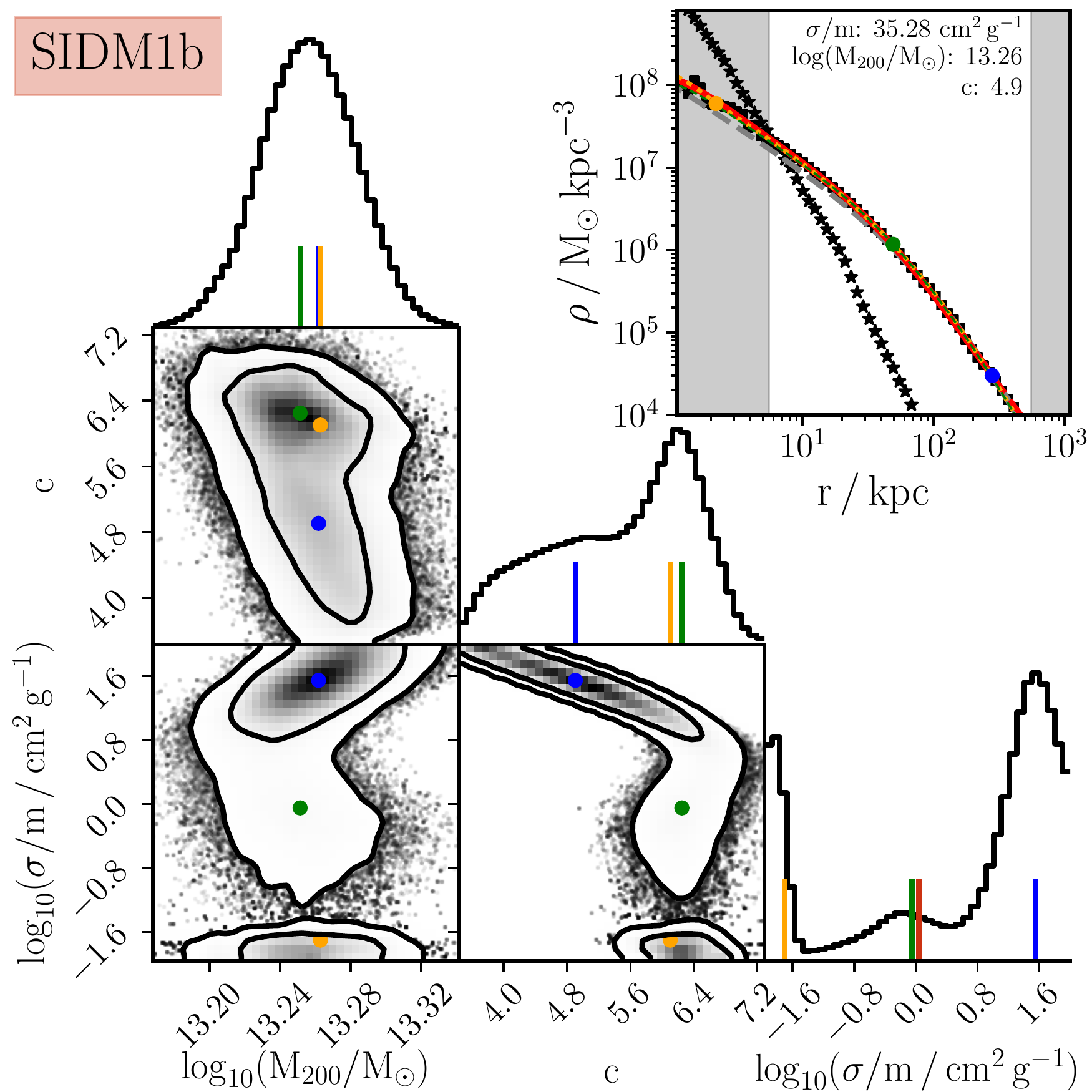}     
	\caption{The same as Fig.~\ref{fig:DMO_MCMC} but for the equivalent halo simulated with the addition of baryons (SIDM1b; also shown in Fig.~\ref{fig:matching_hydro}). Unlike the SIDM-only version of this halo, here the cross-section is poorly constrained by the isothermal Jeans model. Isothermal profiles in hydrostatic equilibrium with this halo's baryon potential are very close to NFW profiles, and a good fit to the simulated density profile can be achieved with either an NFW halo (i.e. with a low $\sigma/m$) or an entirely isothermal profile (a large $\sigma/m$). As well as the maximum likelihood point (blue dot), we mark two other points in the parameter space which correspond to the highest likelihood points in the chain with cross-sections of approximately $1 \cmsg$ (green; the true value) and $0.02 \cmsg$ (orange). The corresponding density profiles and $r_1$ values are shown in the top-right panel, although the density profiles are almost indistinguishable from the maximum likelihood one.}
	\label{fig:baryons_MCMC}
\end{figure}

In Fig.~\ref{fig:baryons_MCMC} we show an example halo, using the SIDM1b equivalent of the SIDM1-only halo shown in Fig.~\ref{fig:DMO_MCMC}. In the DM-only case, the isothermal Jeans model correctly identifies the true cross-section from this halo's density profile. When fitting to the simulated density profile that includes baryons, the marginalised $\sigma/m$ posterior is multi-modal, with reasonable fits to the SIDM density profile being achieved with NFW profiles ($r_1 < 0.01 r_{200}$) and also with almost entirely isothermal profiles ($r_1 \sim r_{200}$), corresponding to cross-sections spanning the full range of our prior. The profiles in these cases are virtually indistinguishable from one another, reflecting the fact that an isothermal species in hydrostatic equilibrium with this particular baryon potential can have a density profile very close to NFW. For this reason, it is hard to distinguish between large and small cross-sections, but this is at least reflected in a broad posterior (in other words, the isothermal Jeans model knows that it does not know the cross-section).

\begin{figure}
        \centering
        \includegraphics[width=\columnwidth]{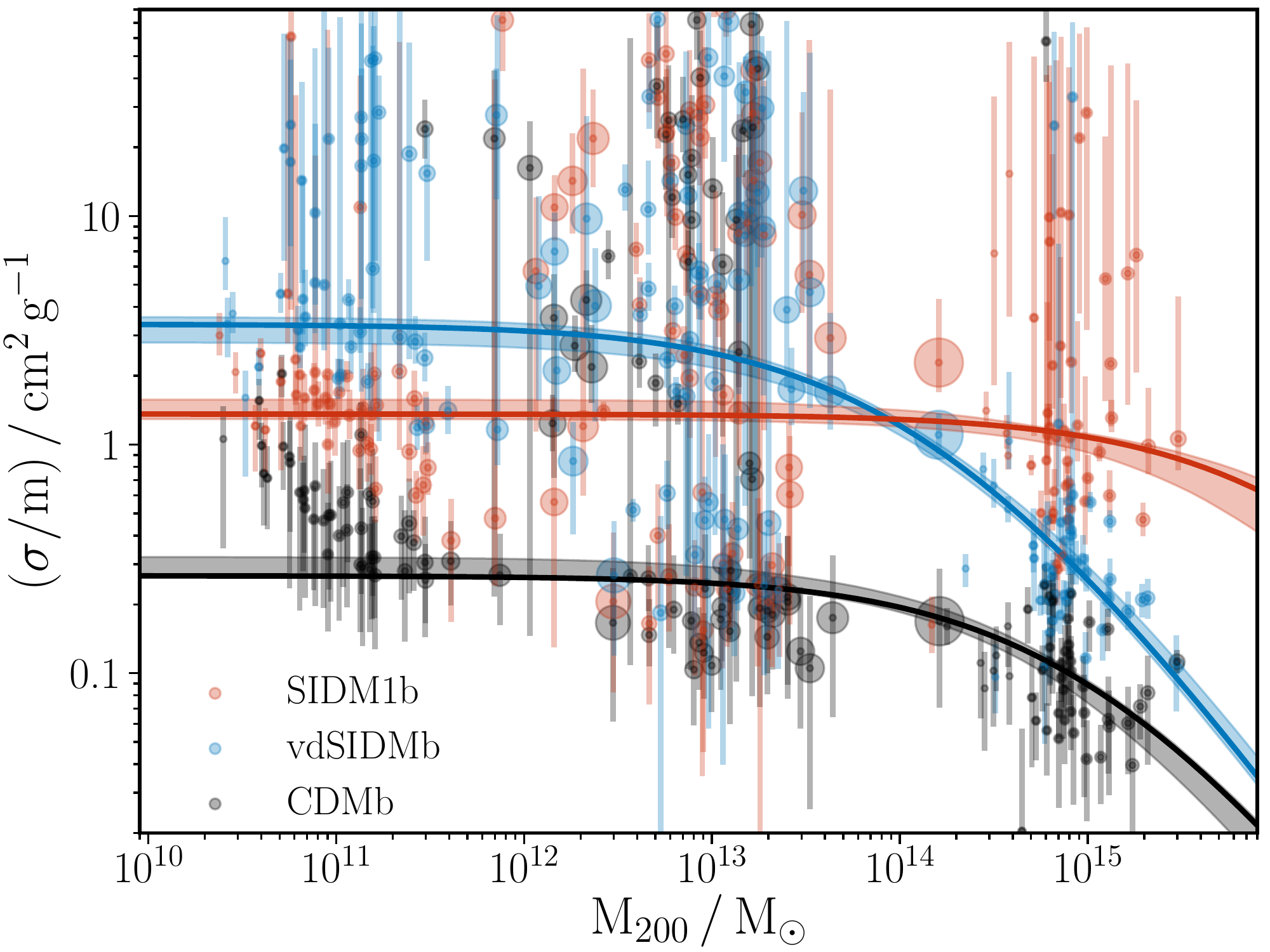}     
	\caption{The same as Fig.~\ref{fig:vdSIDM_fit_DMO} but for simulations that include baryons. Low-mass galaxies and galaxy clusters typically have well-constrained inferred cross-sections, consistent with those used in the respective simulations. In intermediate halo masses ($\sim 10^{13} \msun$) the isothermal Jeans model struggles to determine the true cross-section, although this is typically reflected by larger error bars. Fitting a vdSIDM model to each ensemble of haloes, the inferred cross-sections are slightly larger than when fitting to DM-only haloes with the same input cross-section, although still in reasonable agreement with the true values.}
	\label{fig:vdSIDM_fit_hydro}
\end{figure}

Looking at the $\sigma/m$ posteriors for all simulated systems including baryons in Fig.~\ref{fig:vdSIDM_fit_hydro} we see that broad posteriors (and significant scatter about the input cross-section) are typical for systems with $10^{12} \lesssim M_{200} / \msun \lesssim  \num{3e13}$. However, at both higher and lower halo masses the isothermal Jeans model can correctly discriminate between our different simulated DM models. The well-constrained cross-sections at high and low masses mean that -- when fitting to the ensemble of haloes -- the velocity-dependent SIDM model returned is a reasonable match to the input cross-section of the corresponding simulations.

\begin{figure*}
        \centering
        \includegraphics[width=\textwidth]{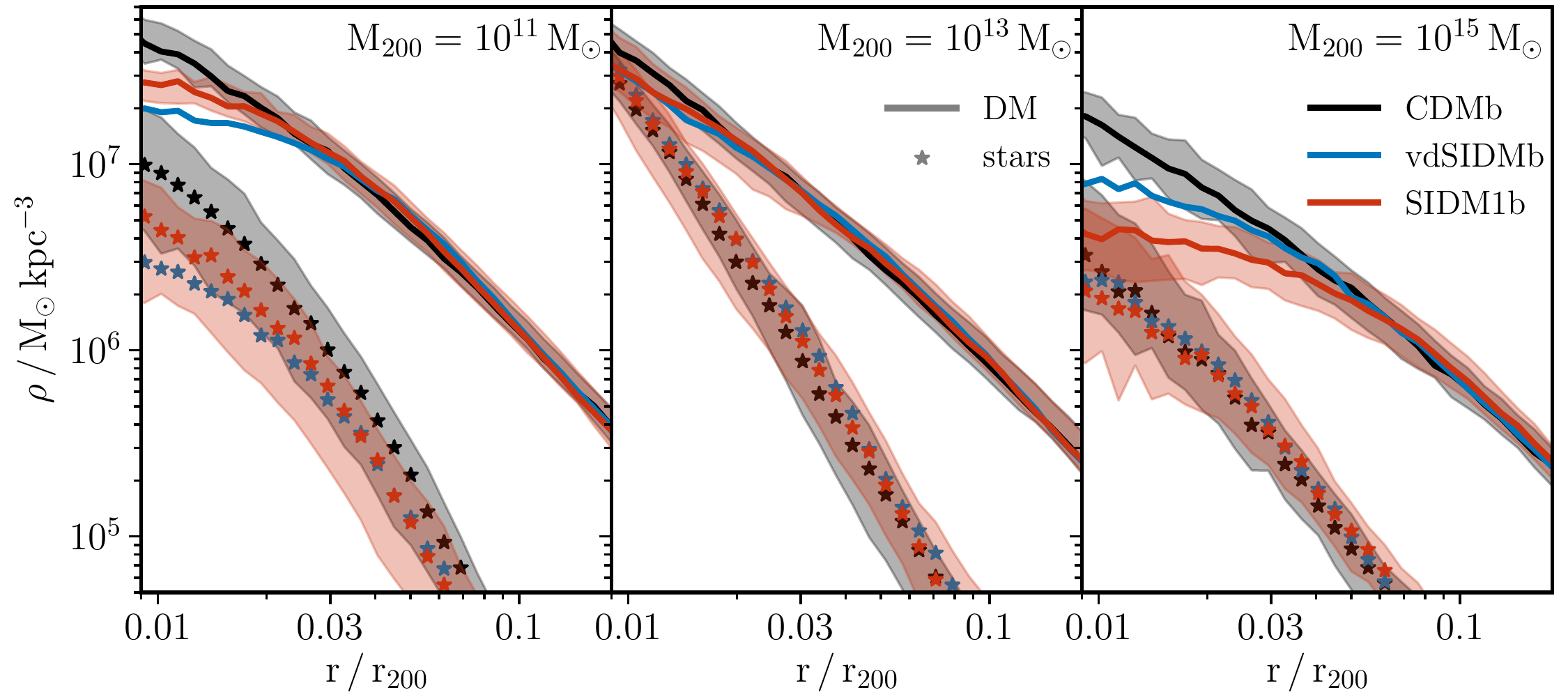}     
	\caption{Stacked density profiles of haloes from our hydrodynamical simulations. Each panel shows the median DM density (lines) and stellar density (stars) as a function of radius, with the shaded regions corresponding to the 16th-84th percentiles of the density at a given radius. From left to right the haloes increase in mass, with the first panel corresponding to haloes with $10.8 < \log_{10} M_{200} / \msun < 11.2$, and the second and third panels increasing the halo masses by successive factors of 100 (keeping a 0.4 dex bin width). The vdSIDMb shaded regions have been omitted for clarity.}
	\label{fig:stacked_hydro_profiles}
\end{figure*}

The fact that intermediate-mass haloes have similar density profiles with CDMb and SIDMb is well known, and we show this explicitly in Fig.~\ref{fig:stacked_hydro_profiles}, where we plot stacked density profiles from our hydrodynamical simulations. The first use of the isothermal Jeans model was to show that the core size expected for SIDM in the Milky Way is substantially smaller (when accounting for baryons) than the prediction from SIDM-only simulations \citep{2014PhRvL.113b1302K}. More recently, \citet{2019MNRAS.484.4563D} used hydrodynamical simulations to show that  intermediate-mass haloes simulated with SIDM could develop profiles that are similar to or cuspier than their CDM counterparts, and \citet{2020arXiv200606623B} showed that the maximal surface density (a quantity related to the density profile) of simulated haloes with $M_{200} \sim 10^{12} \msun$ are very similar between CDM+baryons and SIDM+baryons. The reason why baryons affect the SIDM density profiles of these intermediate-mass haloes the most, is because this is where the stellar mass fraction (i.e. $M_* / M_{200}$) peaks \citep[e.g.][]{2013MNRAS.428.3121M, 2013ApJ...762L..31B, 2015MNRAS.446..521S}, which is apparent in Fig.~\ref{fig:stacked_hydro_profiles} as it is intermediate-mass haloes that have the highest stellar densities at a fixed fraction of $r_{200}$.

\subsection{Quality of fits}

\begin{figure}
        \centering
        \includegraphics[width=\columnwidth]{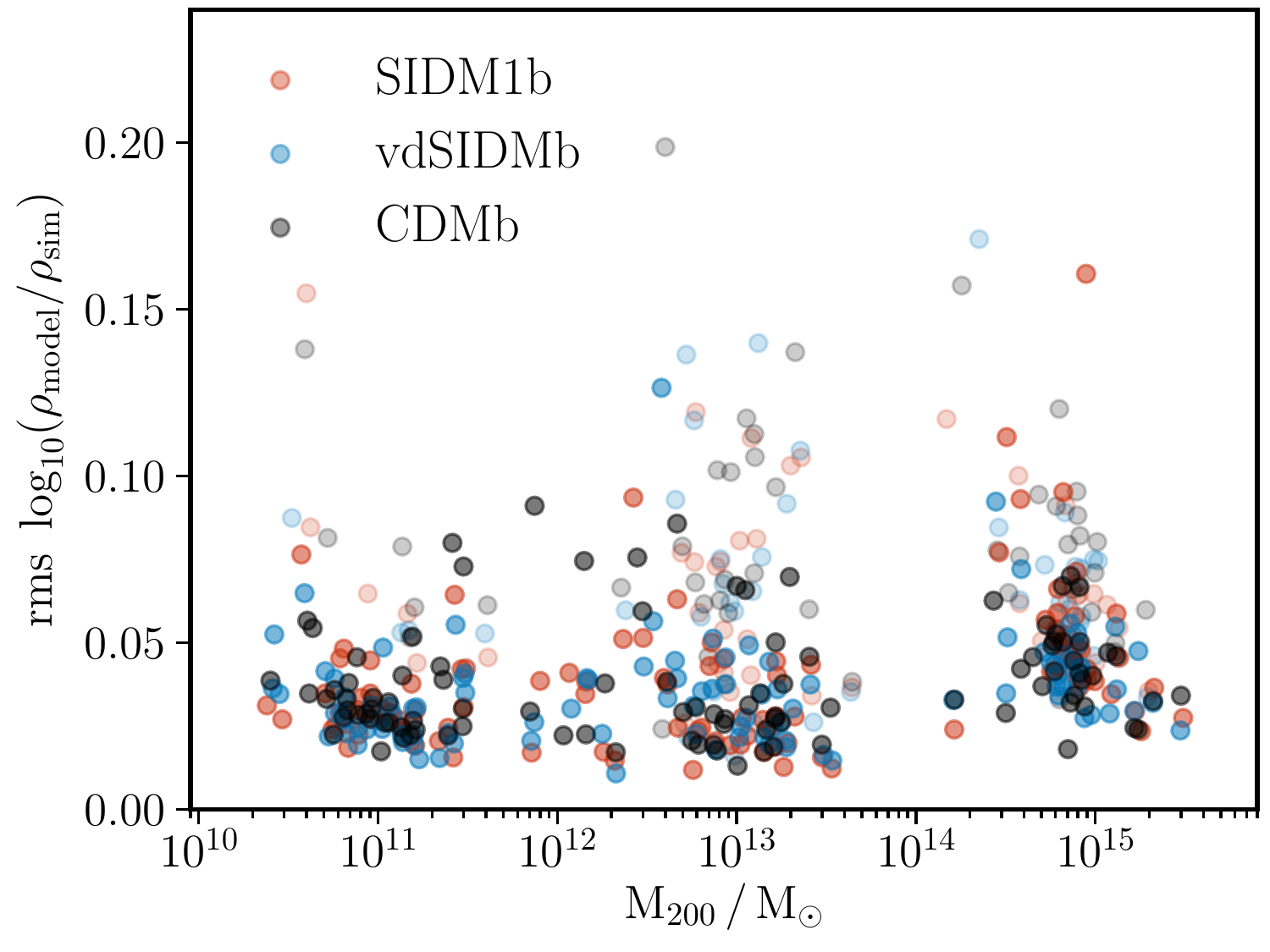}     
	\caption{The same as Fig.~\ref{fig:goodness_of_fit_DMO}, but for simulations including baryons. The quality of the isothermal Jeans model fits to density profiles from hydrodynamical simulations is generally similar as for the DM-only equivalents, although the worst fitting systems have larger $\delta_\mathrm{rms}$ with baryons than without. Note that the y-axis range here is different from in Fig.~\ref{fig:goodness_of_fit_DMO}.}
	\label{fig:goodness_of_fit_hydro}
\end{figure}

Despite the added physical complexity of systems containing baryons, the goodness-of-fit of the isothermal Jeans model density profiles are similar between the DM-only and hydrodynamical simulations. This demonstrates that the assumption made in isothermal Jeans modelling -- that the baryons impact the SIDM density profile only through their current mass distribution (and resulting gravitational potential) -- is adequate for explaining the simulated SIDM density profiles in the presence of baryons. In Fig.~\ref{fig:goodness_of_fit_hydro} we plot $\delta_\mathrm{rms}$ for our ensemble of haloes including baryons. While Fig.~\ref{fig:vdSIDM_fit_hydro} demonstrated that the isothermal Jeans model struggles to infer the input cross-section for intermediate-mass haloes, Fig.~\ref{fig:goodness_of_fit_hydro} shows that this is not because it struggles to find a good fit to the density profiles. As in the Fig.~\ref{fig:baryons_MCMC} example, the isothermal Jeans model density profiles are a good match to the simulated density profiles, they are just not very sensitive to the value of the cross-section.

\subsection{Adiabatic contraction of haloes}
\label{sect:adiabtic}

Although the isothermal Jeans modelling we perform here accounts for the effect of baryons within the isothermal region of the halo, it does not take the baryons into account in the outer (NFW) part of the halo. For haloes with low (or zero) cross-section, this is the bulk of the halo, and so it is worth considering how this might affect our results with low cross-sections. Simulated CDM density profiles are affected by baryons \citep[e.g.][]{2015MNRAS.451.1247S}, typically becoming denser in their centres due to a process known as adiabatic contraction \citep{1986ApJ...301...27B, 2004ApJ...616...16G, 2010MNRAS.405.2161D, 2020MNRAS.495...12C}. In low-mass haloes, feedback-driven winds can actually reduce the central CDM density \citep{2005MNRAS.356..107R, 2014Natur.506..171P, 2015MNRAS.454.2981C, 2016MNRAS.456.3542T}, but this does not happen in the hydrodynamical simulations we use in this paper due to the fairly low gas density at which star formation occurs \citep{2019MNRAS.488.2387B}.

\begin{figure}
        \centering
        \includegraphics[width=\columnwidth]{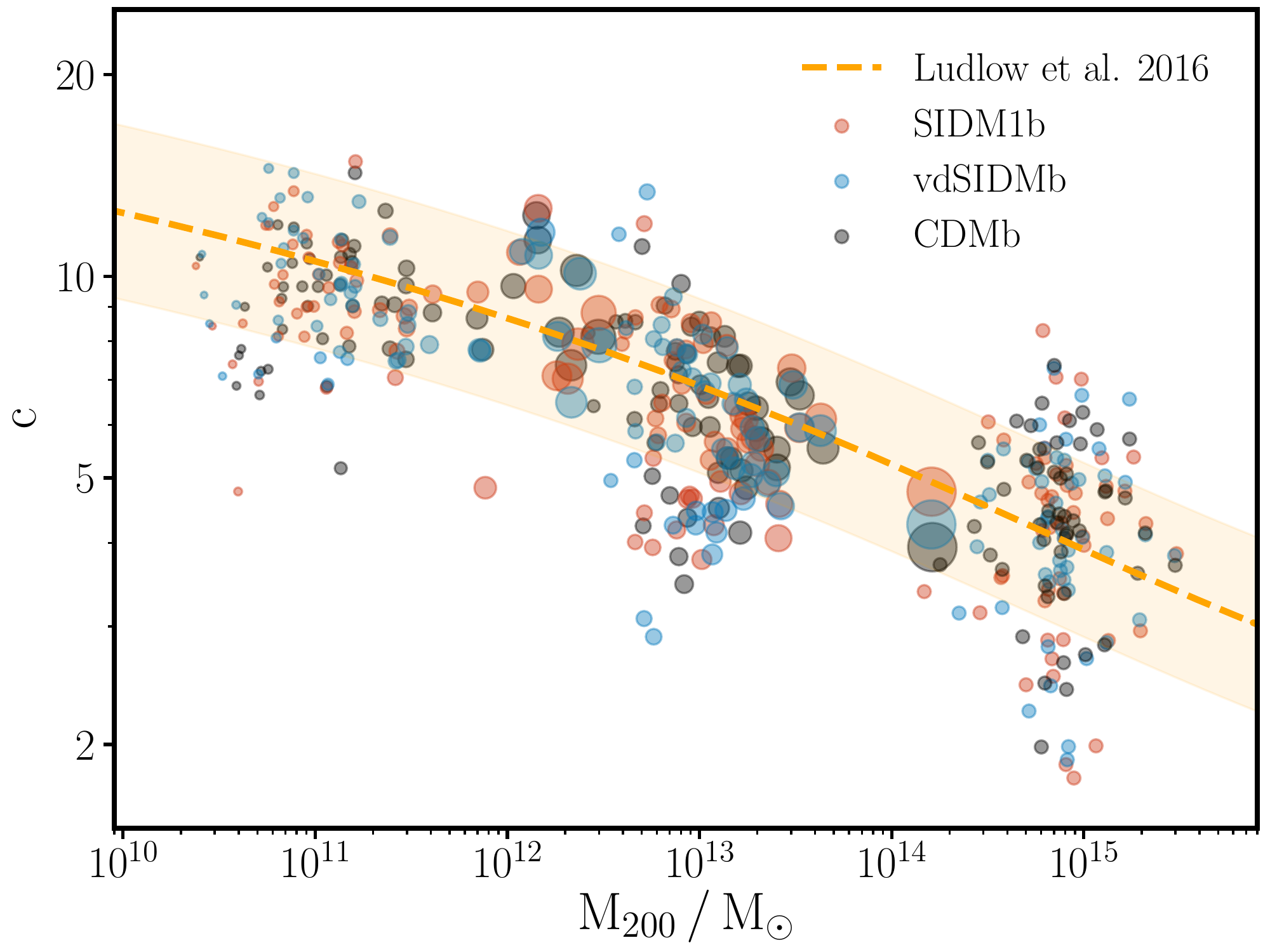}     
	\caption{The same as Fig.~\ref{fig:concentration_mass}, but for simulations including baryons. Note that the concentration is only for the DM, and is the concentration of the NFW profile that describes the outer part of the DM density profile. Despite the increased central density of DM haloes when including baryons (especially at intermediate halo masses), the halo concentrations end up being consistent with the CDM-only concentration-mass relation.}
	\label{fig:concentration_mass_hydro}
\end{figure}

\begin{figure}
        \centering
        \includegraphics[width=\columnwidth]{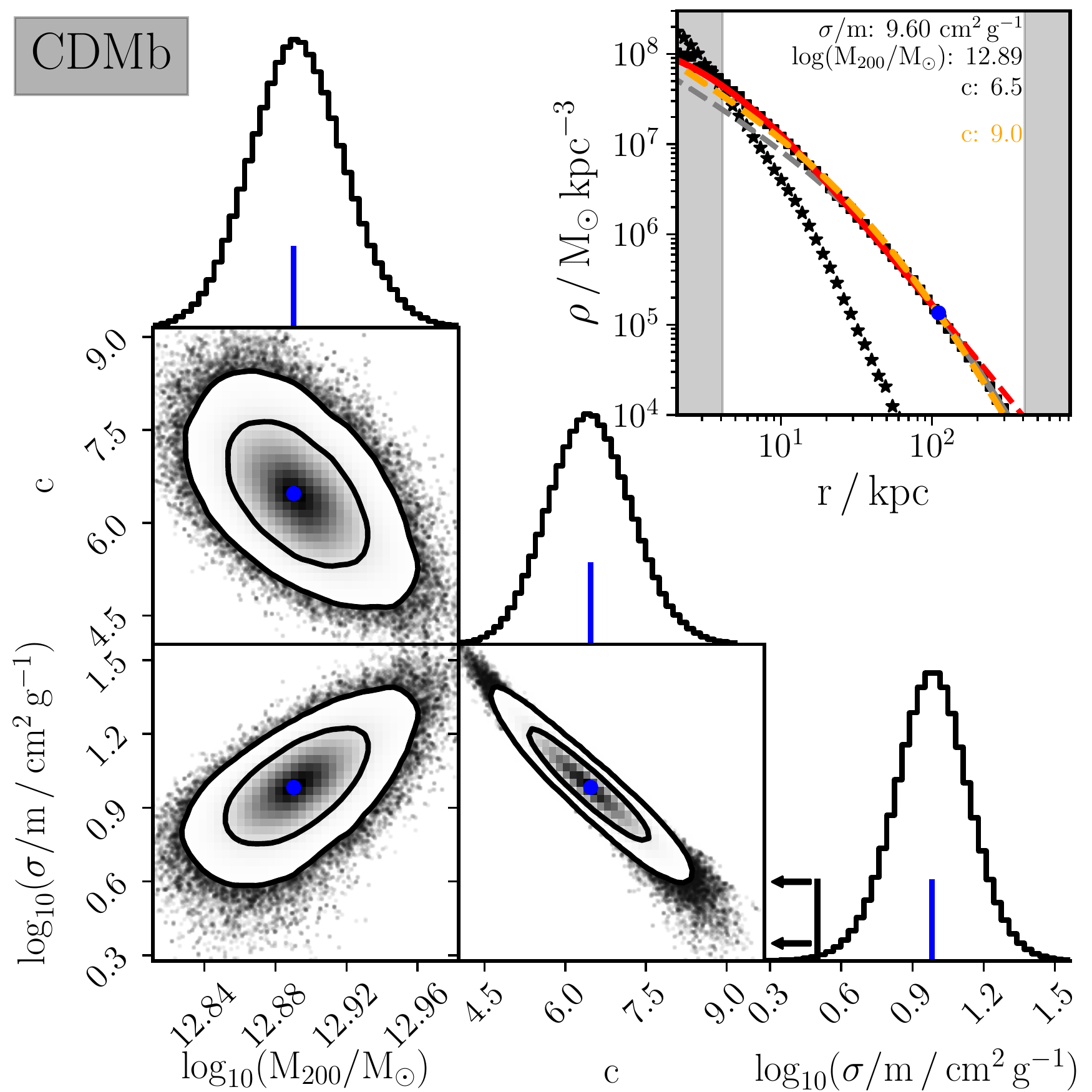}     
	\caption{An example isothermal Jeans model fit to a CDM+baryons halo. In addition to the features plotted in previous corner plots, there is an orange dashed line in the top-right panel, which shows the best-fit NFW profile (which has $c=9$) to the DM density profile. Despite the DM in the simulation being collisionless, the isothermal Jeans model prefers a cross-section of order $10 \cmsg$. This is because the isothermal Jeans model prediction with zero cross-section is an NFW profile, but the inner halo has been adiabatically contracted to be steeper than NFW, reflected in the orange line being below the simulated profile at low-radii.}
	\label{fig:CDMb_MCMC}
\end{figure}

If one considers the process of adiabatic contraction as increasing the NFW concentration of haloes \citep[e.g.][]{2008ApJ...672...19R}, then our model can actually account for this, because the halo concentrations are free to vary above those predicted by CDM-only simulations. In Fig.~\ref{fig:concentration_mass_hydro} we plot the isothermal Jeans model concentration-mass relation from our simulations including baryons, and find that the results are in fairly good agreement with the CDM-only $c(M)$ relation. This is surprising, given that for all DM models the haloes are significantly denser in their central regions than in the DM-only equivalents, especially for $M_{200}$ around $10^{12} - 10^{13} \msun$. The reason we do not see an increase in halo concentration in these intermediate-mass haloes with CDM+baryons is that the more centrally dense haloes are typically better-fit by a low $c$, high $\sigma/m$ solution, than an NFW profile with high $c$. We show an example CDM halo in Fig.~\ref{fig:CDMb_MCMC}, with a best-fit isothermal Jeans model concentration of 6.5. If just fitting an NFW profile to this same halo, the best-fit concentration is 9.0, but as can be seen in the top-right panel, this NFW profile cannot create an inner density profile with a slope as steep as the true one.

Adiabatic contraction leads to CDM density profiles that are not usually well fit by NFW profiles in their centres \citep{2014MNRAS.442.2641V}, typically having steeper inner density slopes \citep{2010MNRAS.405.2161D}. The isothermal parts of our model profiles can have steep inner slopes in the presence of significant baryon potentials, while NFW profiles always have $\rho \propto 1/r$ at small radii. If we just fit NFW profiles to the DM density in our CDM+baryons simulations then we find concentrations that typically lie above the CDM-only $c(M)$ relation, with a pronounced bump above the relation at $10^{12} - 10^{13} \msun$, consistent with the expectation that baryons increase halo concentrations. But when we are free to vary the cross-section, and hence include a central isothermal component, this can better match the steeper than $1/r$ density profiles at low radii.

In the future it would be good to include a model of adiabatic contraction that would affect the outer halo (which includes small radii for small cross-sections). Given that we do our matching inside-out, we currently benefit from having an analytical outer profile for which the density and enclosed mass at a given radius can be easily converted into the parameters describing the outer halo (i.e. $M_{200}$ and $c$). For this reason it is not simple for us to test the effects of including adiabatic contraction for the outer halo. For outside-in matching there is no such requirement, and so including adiabatic contraction would be better suited to an outside-in method, which would also circumvent the problems discussed in Appendix~\ref{App:priors} related to sampling from the inner-profile parameters disfavouring small $\sigma/m$. Including adiabatic contraction of the outer halo was recently done in \citet{2020arXiv200612515S}, who implemented the adiabatic contraction models from \citet{1986ApJ...301...27B} and \cite{2004ApJ...616...16G} into the isothermal Jeans model, and used it to analyse a sample of galaxy groups and clusters. They do the matching outside-in, using a relaxation method to find the inner profile, that removes the need to iteratively find the $\rho_0$ and $\sigma_0$ that match onto the outer profile. Implementing such a method is beyond the scope of this current work, but we hope to implement and test this method in the future, at which point we can also assess the impact of including adiabatic contraction of the outer halo on the quality of the fits as well as on the accuracy of the cross-section inference.

Irrespective of whether a more sophisticated treatment of baryons would improve the cross-section inferred from isothermal Jeans modelling, it is unlikely that the density profiles of intermediate-mass haloes will be particularly useful as probes of SIDM, for the simple reason that the density profiles themselves are very similar between different DM models in this mass range. As such, whatever technique is used to analyse them will struggle to tell SIDM and CDM apart. Returning to Fig.~\ref{fig:stacked_hydro_profiles}, we can see that both the low and high-mass bins show a clear trend of decreasing central density with increasing cross-section. It is therefore at dwarf galaxy and galaxy group/cluster scales where we expect to obtain the best measurements of (or limits on) the SIDM cross-section.

\section{Discussion and outlook}
\label{sect:discussion}

\citet{2018JCAP...12..038S} demonstrated that some of the assumptions made in the isothermal Jeans model are not satisfied by simulated SIDM halos. In particular, they showed that SIDM particle orbits in the inner regions of halos are not exactly isotropic, and that the radius $r_1$ does not correspond to the radius at which matched CDM and SIDM haloes enclose the same amount of mass. Our simulated haloes exhibit these same departures from the assumptions of the isothermal Jeans model, which should not come as a surprise given that the isothermal Jeans model is necessarily simplistic in assuming that fewer than one scattering per particle will have no effect on the DM distribution, while greater than one scattering per particle will fully thermalise the DM. Nevertheless, we find the model to provide a good description of simulated SIDM density profiles, and (importantly) find that the isothermal Jeans model can be used to infer the cross-section from a simulated halo's density profile, which works especially well for large cross-sections.

In light of the isothermal Jeans model's assumptions not holding exactly, \citet{2018JCAP...12..038S} advocate comparing simulated systems directly with observations \citep[as done in][for example]{2019MNRAS.488.3646R, 2020arXiv200606623B,  2020arXiv200608596V}. While this is certainly a worthwhile approach, it is worth mentioning the advantages of using something like the isothermal Jeans model in order to interpret observations. A large advantage is that the isothermal Jeans model is much less computationally expensive than simulations. This allows a scan over the SIDM parameter space (as was done when fitting to density profiles using MCMC in this paper), whereas when directly comparing with simulations one is typically comparing observations with (at most) a few different simulated cross-sections. A second advantage is that it can be used to model specific systems. This is especially important in the case of SIDM, where the distribution of baryons strongly influences the DM distribution. When comparing simulations directly with observations, this necessitates simulating many objects to find analogues of particular observed systems. With the isothermal Jeans model, the system can have the correct baryon distribution by construction.

The major downside of the isothermal Jeans model is that it is only approximate, requiring various assumptions that we know are not precisely true. We have already discussed in Section~\ref{sect:adiabtic} that it would be good to include a model for adiabatic contraction \citep[see][]{2020arXiv200612515S}, removing the (incorrect) assumption that collisionless DM is unaffected by baryons. It would also be beneficial to investigate removing the assumption of spherical symmetry -- which has already been done in the context of modelling observed disk galaxies \citep{2014PhRvL.113b1302K, 2017PhRvL.119k1102K, 2019PhRvX...9c1020R} -- and in future work we hope to test a non-spherical isothermal Jeans model using these simulations.

Another important aspect of the isothermal Jeans model that would benefit from testing with a large number of simulated systems across the full range of mass scales, is how well it works when density profiles are not known, but rather there are some observables (rotation curves, gravitational lensing, etc.) from which they are being inferred -- in other words, how well does the isothermal Jeans model work in practice? In general, it is non-trivial to infer the DM distribution from observations \citep[e.g.][]{2013MNRAS.431.2796K, 2018MNRAS.481L..89H, 2019MNRAS.482..821O, 2020MNRAS.498..144G, 2020MNRAS.496.4717H} and it is important to test how these difficulties impact the inferred SIDM cross-section when isothermal Jeans modelling is applied to observations. This was recently done in the context of galaxy cluster observations by \citet{2020arXiv200612515S}, who found that the inferred cross-section from isothermal Jeans modelling using mock observations of simulated clusters was in agreement with the cross-section used in the simulations. However, the small scales that need to be resolved for generating the relevant mock observables require high-resolution simulations, which meant that \citet{2020arXiv200612515S} only had two simulated systems on which to test their method.

While this paper has generally shown the isothermal Jeans model to be an effective way to describe SIDM density profiles, and to draw inference on the SIDM cross-section, this is less true for low cross-sections, particularly in lower mass haloes. Given that we have a log-uniform prior on the cross-section in the range $-2 <  \log_{10} \sigma/m / \cmsg < 2$, we could not have expected our CDM results to return $\sigma/m = 0$. However, our marginalised $\sigma/m$ posteriors for CDM haloes do not bunch up towards the lower edge of our prior. While this may be partly due to resolution-effects, $\sigma/m$ against $M_{200}$ does not show clear discontinuities when jumping between simulations with very different resolutions (Figures~\ref{fig:vdSIDM_fit_DMO} and \ref{fig:vdSIDM_fit_hydro}) so these are likely small. Instead, we attribute this primarily to a bias against small cross-sections that is inherent to the inside-out method we use for sampling the halo parameters (see Appendices~\ref{App:priors} and \ref{App:resolution}).

Sampling of the inner halo parameters has been used in a number of previous works that have presented positive evidence for SIDM \citep{2014PhRvL.113b1302K, 2018NatAs...2..907V, 2019PhRvX...9c1020R}, and so the fact that this method can lead to erroneously inferred non-zero cross-sections should be a cause for concern. That said, \citet{2014PhRvL.113b1302K} and \citet{2018NatAs...2..907V} used $r_1$ rather than $\sigma/m$ as a free parameter, and so will likely have different biases against certain regions of parameter space, while \citet{2019PhRvX...9c1020R} used a fixed cross-section throughout their work. Going forward, it would be good to demonstrate a method that can reliably infer cross-sections from simulated data, when the simulated cross-section is small (or zero). It may be that sampling from the outside-in is preferable for this, as \citet{2020arXiv200612515S} found that this method gives increased weight to low cross-section regions of parameter space.

\section{Conclusions}
\label{sect:conclusions}

Overall, we find that the isothermal Jeans model provides a good description of simulated SIDM density profiles, both DM-only and including baryons. Turning this around, we can use the isothermal Jeans model and a known DM density profile to determine the DM--DM scattering cross-section. This works especially well for large cross-sections, while with CDM our results tend to incorrectly favour non-zero cross-sections, driven by a bias against small cross-sections inherent in our adopted method of sampling the isothermal Jeans model parameter space (`inside-out').

We find that the quality of the fits are typically better than 20\% (0.08 dex) averaged across the  radial range from $0.01 \, r_{200}$ to $r_{200}$, similar to the level at which NFW profiles fit CDM-only simulated haloes. We also find that the NFW concentrations from isothermal Jeans modelling of SIDM haloes agree with the CDM concentration-mass relation, suggesting that the isothermal Jeans model works in a manner close to the way in which it was envisaged -- namely, it starts from the expected density profile for collisionless DM and predicts how self-interactions should change this. 

We find that (when assuming a constant age for all haloes) haloes that formed earlier have higher inferred cross-sections, in keeping with the isothermal Jeans model prediction that it is the product of cross-section and halo age that is important for the effects of SIDM on a density profile. That said, the correlation between halo age and inferred cross-section is not particularly tight, reflecting the fact that formation histories are complicated and are not captured by a single definition of age.

When modelling haloes simulated with a velocity dependent cross-section, we find that this leads to an inferred cross-section that varies as a function of halo mass, because the typical relative velocities between particles are higher in more massive haloes. Fits to individual systems have a strong covariance between the inner-halo's velocity dispersion and the cross-section, which makes properly extracting the best-fit velocity-dependent cross-section from an ensemble of haloes a statistically challenging problem. We did not attempt this here, but demonstrated that (with an appropriate method for assigning a typical relative velocity to a halo) the inferred cross-section from the isothermal Jeans model, as a function of velocity, scatters around the cross-section used to run the simulations.

In general the isothermal Jeans model works well for cases involving baryons, with the quality of the model fits comparable with the DM-only fits. However, at intermediate halo masses ($M_{200} = 10^{12} - 10^{13} \msun$) the inner halo is baryon dominated, and SIDM and CDM density profiles end up looking very similar. This makes it difficult to use these haloes to infer the cross-section, but this is at least reflected in broad $\sigma/m$ posteriors in such cases. As such, the best constraints on (or measurements of) the SIDM cross-section are likely to come from a mix of dwarf galaxies and galaxy groups and clusters.

\section*{Acknowledgments}

We thank Manoj Kaplinghat, Felix Kahlhoefer, Laura Sagunski and Sean Tulin for helpful discussions about isothermal Jeans modelling. Thanks also to Matthieu Schaller, Rob Crain and Richard Bower for providing the \eagle simulation code, Alejandro Ben\'itez-Llambay for providing the initial conditions and CDM \eagle-12 simulations, and Ian McCarthy for providing the \bahamas code and CDM simulations.

AR is supported by the European Research Council's Horizon2020 project `EWC' (award AMD-776247-6). RM acknowledges the support of a Royal Society University Research Fellowship. VRE acknowledges support from STFC grant ST/P000541/1. This work used the DiRAC Data Centric system at Durham University, operated by the Institute for Computational Cosmology on behalf of the STFC DiRAC HPC Facility (www.dirac.ac.uk). This equipment was funded by BIS National E-infrastructure capital grant ST/K00042X/1, STFC capital grants ST/H008519/1 and ST/K00087X/1, STFC DiRAC Operations grant  ST/K003267/1 and Durham University. DiRAC is part of the National E-Infrastructure.

This work made use of the following software: \texttt{Astropy} \citep{2013A&A...558A..33A}, \texttt{Colossus} \citep{2018ApJS..239...35D}, \texttt{corner} \citep{2016JOSS....1...24F}, \texttt{emcee} \citep{2013PASP..125..306F}, \texttt{IPython} \citep{2007CSE.....9c..21P}, \texttt{Matplotlib} \citep{2007CSE.....9...90H}, \texttt{NumPy} \citep{2011CSE....13b..22V} and \texttt{SciPy} \citep{2011CSE....13b..22V}. 

\section*{Data availability}

The data underlying this article will be shared on reasonable request to the corresponding author.

\bibliographystyle{mnras}

\bibliography{bibliography}

\appendix

\section{Outside-in matching without iteration}
\label{App:matching}

A helpful way to think about matching from an NFW profile to the inner (DM-only) isothermal profile is to consider both the isothermal and NFW density profiles as defined by a scale-density and scale-radius,
\begin{equation}
\rho_\mathrm{iso}(r) = \rho_0 \, f(r/r_0)
\end{equation}
\begin{equation}
\rho_\mathrm{NFW}(r) = \rho_\mathrm{s} \, g(r/r_\mathrm{s}),
\end{equation}
where $f(x)$ can be found numerically (see Section~\ref{sect:isothermal_profiles}) and $g(y) = 1 / \left( y (1+y)^2\right)$ (see equation~\ref{eq:NFWrho}).

\begin{figure}
        \centering
        \includegraphics[width=\columnwidth]{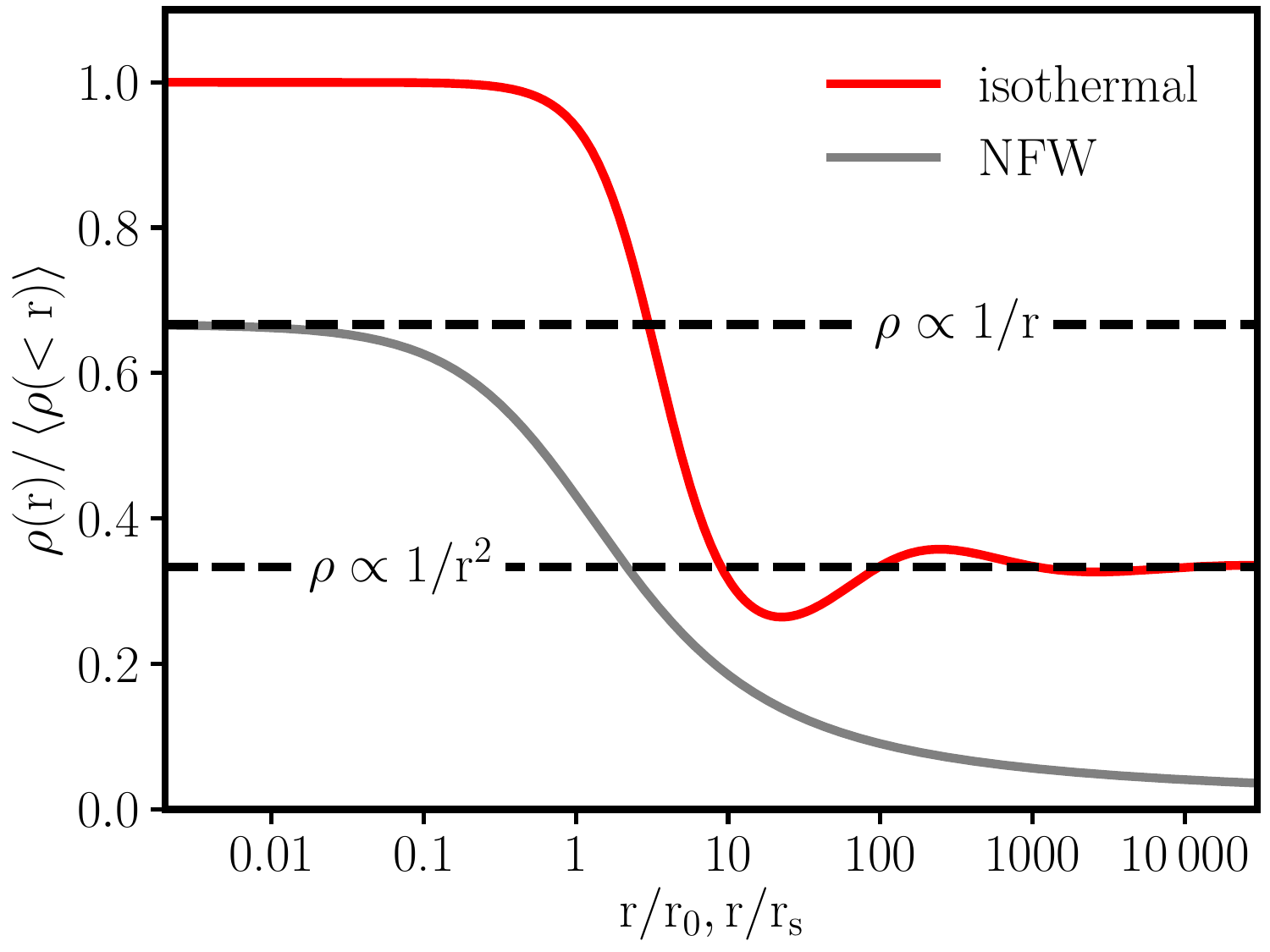}     
	\caption{The density at radius $r$ divided by the mean density within $r$, as a function of $r$, for an NFW profile and for a DM-only isothermal profile. The $x$-axis is a dimensionless radius, which is $r/r_\mathrm{s}$ for the NFW profile and $r/r_0$ for the isothermal profile. Power-law density profiles have constant values of $\rho / \langle \rho \rangle$, with the values for $\rho \propto 1/r$ and $\rho \propto 1/r^2$ indicated by the horizontal dashed lines.}
	\label{fig:analytical_matching}
\end{figure}

The isothermal profile can be scaled arbitrarily in density normalisation, which means that for a given set of $\rho_\mathrm{s}$, $r_\mathrm{s}$, $r_0$ and $r_1$ there is a $\ro$ for which $\rho_\mathrm{iso}(r_1) = \rho_\mathrm{NFW}(r_1)$, and there is also a $\ro$ for which $M_\mathrm{iso}(< r_1) = M_\mathrm{NFW}(<r_1)$. However, what we require is that these two matching criteria are simultaneously met, which means that $\rho_\mathrm{iso}(r_1) / M_\mathrm{iso}(< r_1) = \rho_\mathrm{NFW}(r_1) / M_\mathrm{NFW}(<r_1)$ or equivalently $\rho_\mathrm{iso}(r_1) / \langle \rho_\mathrm{iso}(< r_1) \rangle = \rho_\mathrm{NFW}(r_1) / \langle \rho_\mathrm{NFW}(< r_1) \rangle$. Given this, it is instructive to look at the form of $\rho / \langle \rho \rangle$ as a function of radius for both the NFW and DM-only isothermal profiles. These can be calculated from $f(x)$ and $g(y)$ as 
\begin{equation}
F(x=r/r_0) = \frac{\frac{4 \pi}{3} r^3 \rho(r)}{\int_0^r 4 \pi r^{\prime 2} \rho(r^\prime) \mathrm{d}r^\prime} = \frac{x^3 f(x)}{3 \int_0^x x^{\prime2} f(x^\prime) \, \mathrm{d}x^\prime},
\end{equation}
and analogously for $G(y)$.

Defining $x_1 = r_1 / r_0$ and $y_1 = r_1 / r_\mathrm{s}$ we can find $\rho_0$ and $r_0$ from $\rho_\mathrm{s}$, $r_\mathrm{s}$ and $r_1$ as follows:
$\bullet$ $y_1 = r_1 / r_\mathrm{s}$
$\bullet$ use $F(x_1) = G(y_1)$ to find $x_1$
$\bullet$ $r_0 = r_1 / x_1$
$\bullet$ $\rho_0 = \rho_1 / f(x_1)$.

The process of using $F(x_1) = G(y_1)$ to find $x_1$ is illustrated graphically in Fig.~\ref{fig:analytical_matching}, where the isothermal line is $F(x)$ and the NFW line is $G(y)$. One can immediately see that there are values for $x$ and $y$ for which a match cannot be found. In particular, the minimum value of F(x) is $\approx 0.26$, which is the value of $G(y)$ when $y \approx 4$. As such, when $y>4$, i.e. when $r_1 > 4 r_\mathrm{s}$, there is no isothermal solution that can match the NFW, as the NFW density has fallen off too quickly with radius to be reproduced by an isothermal profile.

One interesting feature in Fig.~\ref{fig:analytical_matching} is that $F(x)$ is not monotonically decreasing with increasing $x$, instead oscillating around the value of $1/3$ which corresponds to $\rho / \langle \rho \rangle$ for a $1/r^2$ density profile. This happens because at large $x$, $f(x) \propto 1/x^2$, but with a logarithmic slope that oscillates about the limiting value of -2. Turning to our procedure for matching, this would lead to multiple possible $r_0$, each with a corresponding $\rho_0$, that produce isothermal profiles that can match onto the NFW at $r_1$. These multiple solutions only arise when $G(y) < 0.36$ (the height of the positive peak in $F(x)$ at $x \approx 240$), which is when $r_1 > 1.78 r_\mathrm{s}$. As such, for $r_1 \lesssim r_\mathrm{s}$ there is always one and only one isothermal solution that matches, while for large cross-sections (leading to large $r_1$) there can be multiple or no matching solutions.

We stress that the results in this section apply only to DM-only haloes, as in cases involving baryons the isothermal profile is no longer a separable function of $\rho_0$ and $r_0$. In these cases it is not clear which NFW profiles and associated $r_1$ can be matched by an isothermal profile that is in hydrostatic equilibrium with the total potential (and therefore depends on the baryon distribution). This is one of the main reasons that we primarily used inside-out matching in this paper, as the NFW profile is not affected by baryons and so it is a simple procedure to find the NFW profile that matches onto a given isothermal profile, or to show that there is no such NFW profile.

\section{Effective priors}
\label{App:priors}

Using inside-out matching, we sample from the parameters of the inner halo. However, the outer-halo parameters are more natural for describing the halo as a whole. Given our priors on $\sigma/m$, $N_0$ and $\sigma_0$ it is interesting to see how these map on to \emph{effective priors} on the parameters $M_{200}$, $c$ and $\sigma/m$. Note that the effective $\sigma / m$ prior can differ from the actual $\sigma/m$ prior, because only certain regions of the prior parameter space map onto valid NFW profiles. Also, while our actual prior is a separable function of  $\sigma/m$, $N_0$ and $\sigma_0$, the effective priors can have a more complex structure.

\begin{figure*}
        \centering
        \includegraphics[width=\textwidth]{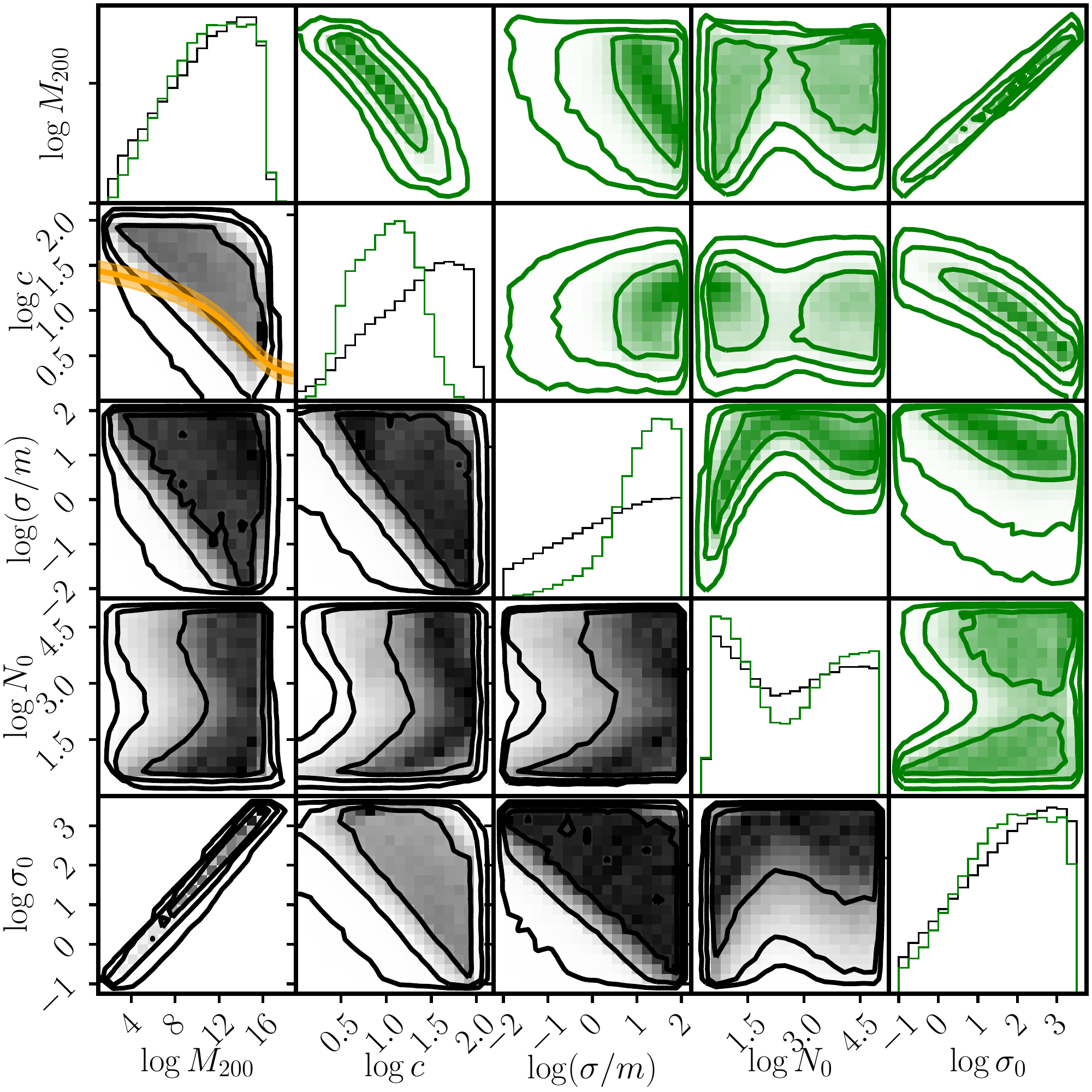}     
	\caption{A corner plot of the \emph{effective priors} described in Appendix~\ref{App:priors}. The lower-left (black) half of the plot shows points uniformly sampled from flat priors on $\log (\sigma/m)$, $\log N_0$ and $\log \sigma_0$, with those that don't map to a valid NFW profile removed. Cosmological haloes follow the concentration-mass relation, which we plot in orange on the $M_{200} - c$ panel, with the shaded region showing the $1\sigma$ scatter. The upper-right (green) half of the plot is a corner plot of the same points sampled from our priors shown in black, but downweighting points that lie far from the concentration-mass relation, following equation~\eqref{eq:cM_weights}.}
	\label{fig:effective_priors}
\end{figure*}

In order to assess the effective priors, we draw samples from our priors and find the corresponding NFW profiles, discarding points in parameter space that do not map onto a valid NFW. In practice, we do this by generating an MCMC chain in the same manner as when fitting to a density profile, but setting the likelihood to  a constant value (unless there is no matching NFW in which case the likelihood is zero). The resulting corner plot is shown in Fig.~\ref{fig:effective_priors}. Looking at the $\sigma/m$, $N_0$, $\sigma_0$ region of the plot, the effect of requiring a matching NFW profile can be seen, as the effective priors on these parameters do not match the priors actually imposed on these parameters. In particular, small $\sigma_0$ and/or small $\sigma/m$ are disfavoured. Mapping these same sets of points into the $M_{200}$ and $c$ parameter space, we can see that the marginalised $\log M_{200}$ and $\log c$ priors are both relatively flat, with the marginalised prior density for $\log M_{200}$ varying by less than 30\% over the range of halo masses for which we have simulations.

Given that the NFW parts of our isothermal Jeans model fits end up roughly following the CDM-only $c(M)$ relation (Fig.~\ref{fig:concentration_mass}), it is instructive to look at the subset of our effective prior volume that lies close to this $c(M)$ relation. To this end, we take the effective prior MCMC chain and re-weight each point by its closeness to the $c(M)$ relation. Specifically, we give each point a weight
\begin{equation}
w_i = \exp { - \left( \frac{\log_{10}\left[ c_i / c_\mathrm{L16}(M_i) \right] }{\sigma_{\log c}} \right)^2},
\label{eq:cM_weights}
\end{equation}
where $c_i$ and $M_i$ are the concentration and halo mass of the $i$th point in the MCMC chain, $c_\mathrm{L16}(M)$ is the median concentration-mass relation from \citet{2016MNRAS.460.1214L} and $\sigma_{\log c} = 0.13$~dex \citep{2014MNRAS.441.3359D}.

Having applied these weights, the resulting corner plot is shown in the top-right of Fig.~\ref{fig:effective_priors}. Of particular note is that the marginalised effective prior on $\sigma/m$ now strongly favours larger cross-sections, which we consider to be the driver of the moderately large cross-sections found when fitting the isothermal Jeans model to CDM simulated density profiles. From the $\sigma/m$--$M_{200}$ panel, one can see that this bias towards larger cross-sections increases with decreasing halo mass, which is reflected in the increasing $\sigma/m$ towards low halo-masses seen for CDM in Fig.~\ref{fig:vdSIDM_fit_DMO}. This happens because the mapping from inner to outer halo, which causes the bias, depends on the ratio of $r_1$ to the scale parameters of the inner and outer haloes, $r_0$ and $r_\mathrm{s}$. At fixed cross-section, $r_0 / r_\mathrm{s}$ decreases with decreasing halo mass, because the scattering rate is proportional to the velocity dispersion, which decreases with decreasing halo mass. This means that a larger cross-section is required in lower-mass haloes to achieve a particular $r_0 / r_\mathrm{s}$, and so the bias against low $\sigma/m$ is more pronounced in lower mass haloes.

\section{MCMC sampling}
\label{App:MCMC}

Our MCMC sampling was done using \texttt{emcee} \citep{2013PASP..125..306F}, which uses an ensemble of \emph{walkers}, with the next proposed jump in parameter space for a given walker being based upon the location of another of the walkers. As such, the proposal distribution is automatically tuned to the shape of the posterior distribution.

For each halo (both DM-only and those with baryons) the MCMC sampling of the posterior distribution for the isothermal Jeans model parameters was done with $K=56$ walkers. Walkers were started in a tight ball around $N_0 = 7$, $\sigma/m = 1 \cmsg$ and $\sigma_0 = (M^\mathrm{sim}_{200} / 10^{15} \msun)^{1/3} \, 700 \kms$, with an initial burn-in period of 1000 steps which was run and then discarded. The location of the walkers at the end of the burn-in was then their starting location for the MCMC chains that we use in our analysis, with a chain length of $N_\mathrm{chain} = 5000$ (for each walker).

We found that a small number of walkers would sometimes get stuck in local peaks of the posterior density. These peaks could have substantially worse posterior densities than those associated with points sampled about the true posterior density peaks (differences of 100s or 1000s in the natural logarithm of the posterior density), without covering a correspondingly larger volume of the parameter space that they should actually contribute a non-negligible amount to the probability mass. Following \citet{2012ApJ...745..198H} we adopted a procedure of \emph{pruning} the MCMC chains, removing these stuck walkers. Specifically, for each walker we calculated the mean value of the negative log likelihood 
\begin{equation}
L_k = \frac{1}{N_\mathrm{chain}} \sum_{t=1}^{N_\mathrm{chain}} - \ln \mathcal{L}(\vec{\theta_k} (t) | D),
\end{equation}
where $\vec{\theta_k} (t)$ is the set of parameter values for the $t$th step in walker $k$'s chain, $D$ is the data (the DM density profile of the simulated halo in question), and $\mathcal{L}$ is the likelihood function defined by equation~\eqref{eq:chi2}. We then rank all walkers based on $L_k$, such that $\{ \vec{\theta_1}, \vec{\theta_2},...,\vec{\theta_K} \}$ is in the order of increasing $L_k$. Starting from $k=1$ we find the difference in successive $L_k$ values, and stop when this difference is substantially larger than the average difference before. Specifically, we find the first successive pair of $L_k$ where
\begin{equation}
L_{j+1} - L_{j} > C \frac{L_{j} - L_{1}}{j-1}
\label{eq:prune_criteria}
\end{equation}
and throw away all $\vec{\theta_k}$ with $k>j$. We use $C=100$, but found that $C=10$ leads to almost identical results. Note that equation~\eqref{eq:prune_criteria} can lead to removing perfectly valid chains if, for example, $L_1$ and $L_2$ happen to be extremely close to another (such that the jump to $L_3$ looks large in comparison but is small in absolute terms). This was the case for one of our haloes, and so we additionally required that we keep at least the first $K/4$ walkers, checking the criteria in equation~\eqref{eq:prune_criteria} only for $j \geq K/4$.

As recommended by \citet{2018ApJS..236...11H} we use the \emph{integrated autocorrelation time} to test for convergence of our MCMC chains. We calculate the autocorrelation times using \texttt{emcee}'s built in functionality \citep{2019JOSS....4.1864F}. The chains mix quickly in terms of $M_{200}$, but the main quantity of interest, $\sigma/m$, typically has a longer autocorrelation time because of the more complex shape of the posterior as a function of $\sigma/m$ (in particular, the $\sigma/m$--$c$ degeneracy). For our fits to DM-only density profiles, the autocorrelation time for $\log \sigma/m$ was typically 30--80 steps, so that our chain length of 5000 corresponds to an effective number of independent samples per walker of order 100. For runs including baryons, the posteriors are often complicated, leading to increased autocorrelation times. Our longest set of autocorrelation times was for \eagle-12 with CDMb. These had estimated autocorrelation times of 150--400 steps, which is likely an underestimate given this was calculated from fewer than 50 autocorrelation times' worth of samples.\footnote{See \href{https://dfm.io/posts/autocorr/}{https://dfm.io/posts/autocorr/} for a discussion of how to calculate autocorrelation times with multi-walker chains, and what happens when estimating autocorrelation times from short chains.} While this leads to fairly low effective sample sizes, running longer chains is computationally infeasible due to the fact that systems including baryons are the ones in which likelihood evaluations take longest (finding the numerical solution to equations \eqref{eq:hydrostatic} and \eqref{eq:enclosed_mass} takes longer when $M_\mathrm{bar}$ is significant). Nevertheless, we had originally run chains of length 1000 (also with a shorter burn-in) and while a few systems had noticeable shifts in their posterior distributions between these early runs and the ones we use in this paper, the overall picture expressed in Fig.~\ref{fig:vdSIDM_fit_hydro} was very similar between these two sets of MCMC chains.

\section{Velocity averaging}
\label{App:vel_av}

For a velocity dependent cross-section, the particle physics specifies the momentum transfer cross-section as a function of the relative velocity between two DM particles, $\sigTt(v_\mathrm{rel})$. Meanwhile, an isothermal Jeans model fit provides a measurement of the 1D velocity dispersion in the isothermal region, $\sigma_0$, and a value for the cross-section, $\sigma/m$, which is the velocity-independent cross-section that leads to 1 scattering per particle over the age of the halo at the radius $r_1$.

In Fig.~\ref{fig:sigma_vs_vrel} we plot the median posterior values of $\sigma/m$ against $\sigma_0$ from isothermal fits to simulated haloes. While $\sigma_0$ is a single velocity, it represents a velocity distribution in which pairs of particles will have different relative velocities. There are then different ways in which a velocity-dependent cross-section can be plotted onto this, depending on how one takes a velocity-average of the velocity-dependent cross-section \citep[see][for further discussion]{2020JCAP...06..043C}. If we call the effective velocity-independent momentum-transfer cross-section in an isothermal region $\sigma_\mathrm{eff}$, then possibilities for how to calculate $\sigma_\mathrm{eff}$ as a function of $\sigma_0$ include:
\begin{equation}
\sigma^a_\mathrm{eff}(\sigma_0) = \sigTt(\langle v_\mathrm{rel}) \rangle) = \sigTt(4/\sqrt{\pi} \, \sigma_0)
\end{equation}
and
\begin{equation}
\sigma^b_\mathrm{eff}(\sigma_0) = \langle \sigTt(v_\mathrm{rel}) \rangle = \int_0^\infty \sigTt(v) f(v) \, \mathrm{d}v, 
\end{equation}
where
\begin{equation}
f(v) = \frac{1}{\sqrt{4 \pi}} \frac{v^2}{\sigma_0^3} \exp\left( - \frac{v^2}{4 \sigma_0^2} \right) 
\end{equation}
is the pairwise velocity distribution for a Maxwellian velocity distribution with 1D dispersion, $\sigma_0$,\footnote{The pairwise velocity distribution looks like a Maxwell-Boltzmann distribution (i.e. the distribution function of individual particles' speeds) but with a 1D velocity dispersion of $\sqrt{2} \, \sigma_0$.} and $\langle ... \rangle$ represents averaging over this pairwise velocity distribution. $\sigma^a_\mathrm{eff}$ is just the cross-section evaluated at the mean pairwise velocity of particles, while $\sigma^b_\mathrm{eff}$ is the mean cross-section over all particle pairs.

Another alternative is 
\begin{equation}
\sigma^c_\mathrm{eff}(\sigma_0) = \frac{\langle \sigTt(v_\mathrm{rel}) v_\mathrm{rel} \rangle}{\langle v_\mathrm{rel} \rangle} = \frac{\int_0^\infty \sigTt(v) v f(v) \, \mathrm{d}v}{\int_0^\infty v f(v) \, \mathrm{d}v}, 
\end{equation}
which is the pairwise-velocity-weighted mean cross-section over all particle pairs, where the $v_\mathrm{rel}$ weighting is motivated by the fact that the scattering probability for a pair of particles is proportional to $\sigma(v_\mathrm{rel}) v_\mathrm{rel}$. With this weighting scheme we would expect the total scattering rate of particles (with a Maxwellian velocity distribution) to be equal for the true velocity-dependent cross-section, and for a velocity-independent cross-section of $\sigma^c_\mathrm{eff}$.

A final velocity averaging scheme that we consider is 
\begin{equation}
\sigma^d_\mathrm{eff}(\sigma_0) = \frac{\langle \sigTt(v_\mathrm{rel}) v_\mathrm{rel}^2 \rangle}{\langle v_\mathrm{rel}^2 \rangle} = \frac{\int_0^\infty \sigTt(v) v^2 f(v) \, \mathrm{d}v}{\int_0^\infty v^2 f(v) \, \mathrm{d}v}, 
\end{equation}
which has an additional factor of the pairwise velocity in the weight function, to reflect the fact that for some fixed scattering angle, the amount of momentum transferred in a scattering event is proportional to the relative velocity of the scattering particles. Therefore, $\sigma^d_\mathrm{eff}$ would be the relevant quantity if we want to find the velocity-independent cross-section that would lead to the same rate of momentum exchange through particle scattering as the underlying velocity-dependent cross-section.

For the vdSIDM model that we simulate, the cross-section at high-velocities decreases with increasing velocity. This means that averaging schemes that weight more towards high-velocity pairs of particles (i.e. $\sigma^c_\mathrm{eff}$ and $\sigma^d_\mathrm{eff}$) predict lower effective cross-sections than if weighting each particle pair equally (i.e. $\sigma^b_\mathrm{eff}$). In Fig.~\ref{fig:sigma_vs_vrel} the dashed line is $\sigma^b_\mathrm{eff}$, the dot-dashed line $\sigma^c_\mathrm{eff}$ and the dotted line $\sigma^d_\mathrm{eff}$. It appears that $\sigma^b_\mathrm{eff}$ and $\sigma^c_\mathrm{eff}$ are better ways to calculate an effective cross-section than $\sigma^a_\mathrm{eff}$ or $\sigma^d_\mathrm{eff}$. Given that it is the transfer of energy and momentum between particles that allows self-interactions to establish thermal equilibrium, it is perhaps surprising that $\sigma^d_\mathrm{eff}$ does not provide the best description. It is possible that this is because haloes grow through time, so the velocity dispersions increase over time. This means that a system with a particular effective cross-section now will have had a larger effective cross-section in the past, and so (averaging over the age of the halo) schemes that over-predict the effective cross-section now may do better at describing the effective cross-section over the whole halo history. Alternatively, it may be that the inferred cross-sections from modelling vdSIDM galaxy clusters are artificially high for the same reasons that we typically find modest non-zero cross-sections with CDM (see Appendix~\ref{App:resolution}). It could then be the case that -- without a bias towards larger cross-sections -- the inferred cross-sections with vdSIDM in clusters would be lower, and so better described by $\sigma^d_\mathrm{eff}$.

\section{Simulation resolution and the cross-section posteriors for CDM}
\label{App:resolution}

It is well documented that $N$-body simulations are affected by resolution, and that the densities at the centre of CDM-only simulated haloes are systematically lower in low-resolution simulations compared with higher-resolution ones \citep{2003MNRAS.338...14P, 2004MNRAS.348..977D, 2008MNRAS.391.1685S, 2010MNRAS.402...21N, 2012MNRAS.425.2169G}. As a lowered central density is also the primary effect of DM self-interactions it is important that we understand what influence the resolution of our simulations might have on our inference of the cross-section from the simulated density profiles. In general, SIDM haloes are converged down to smaller radii than their CDM counterparts \citep[see][]{2012MNRAS.423.3740V, 2020arXiv200403872S}, owing to the fact that physical two body interactions between particles (i.e. the self-interactions) dominate over the gravitational two-body interactions that are the primary driver of numerical core formation with collisionless DM \citep{2019MNRAS.488.3663L}. Given this, we focus on the CDM simulations in this Appendix.

Recent work by \citet{2019MNRAS.488.3663L} has shown that a fairly simple expression can be derived for the minimum radius at which simulated CDM-only density profiles are converged, $r_\mathrm{conv}$. This depends primarily on the simulation particle mass, depending only weakly on the gravitational softening length as well as the halo mass and concentration. They find that for typical haloes -- and with a definition of `converged' that is convergence in the circular velocity to better than 10\% -- $r_\mathrm{conv} \approx 0.05 \, l$, where $l$ is the mean inter-particle spacing.

\begin{figure}
        \centering
        \includegraphics[width=\columnwidth]{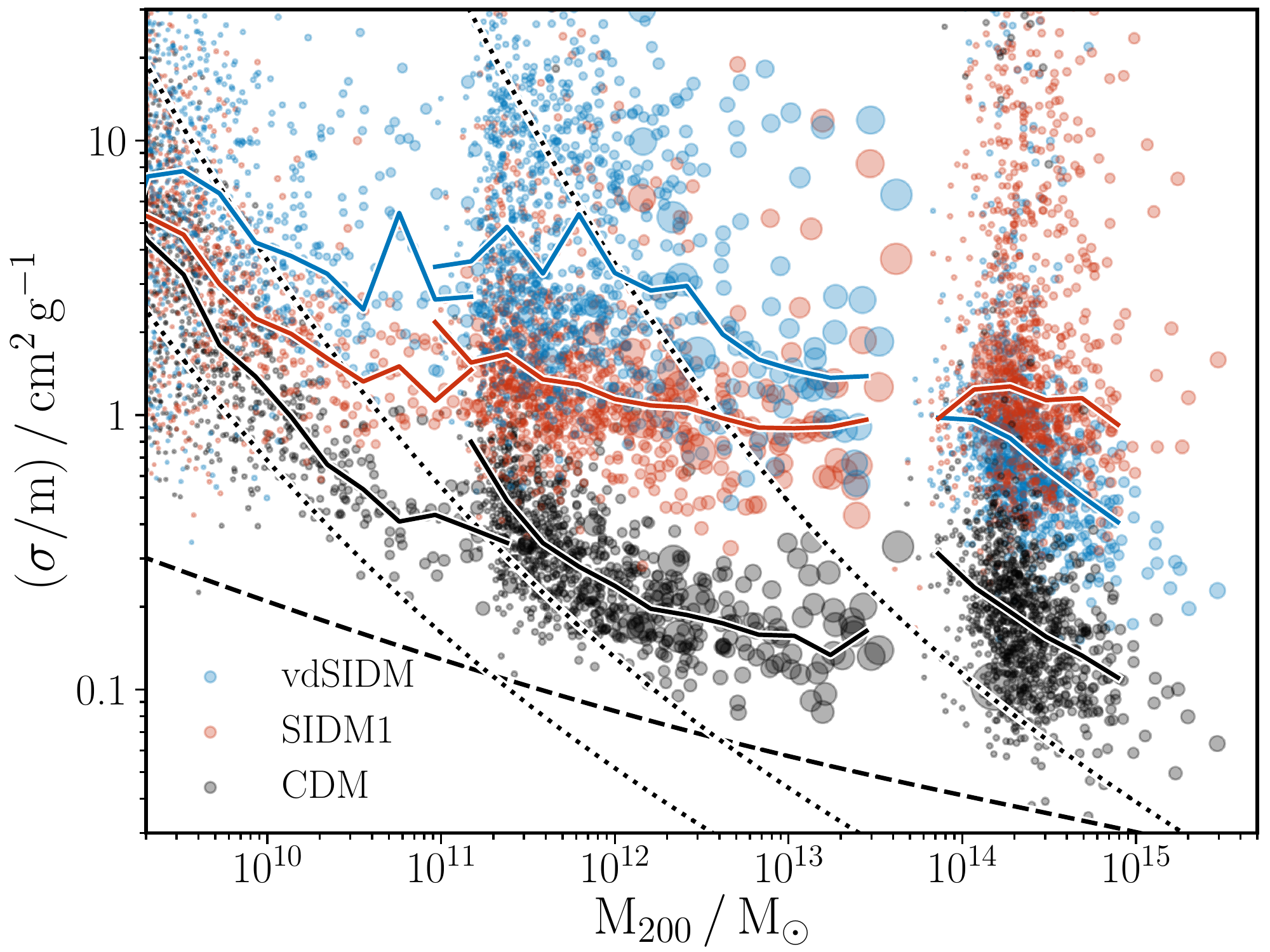}     
	\caption{The median cross-section as a function of median halo mass for isothermal Jeans model fits to the 1000 most massive FOF groups from each of our simulations. This includes haloes that are poorly resolved and so have density profiles that (in the region that we are fitting to) are strongly affected by resolution. The solid lines show running median cross-sections as a function of halo mass for the different simulations. The dotted lines are where $r_1 = r_\mathrm{conv}$ (assuming that haloes follow the median concentration-mass relation) for our three different simulation resolutions. The dashed line is where $r_1 = 0.01 \, r_{200}$.}
	\label{fig:M200_sigma_all_DMO}
\end{figure}

Given that simulated halo density profiles are significantly affected by resolution for $r \lesssim r_\mathrm{conv}$, and that within the isothermal Jeans model departures from an NFW profile happen for $r < r_1$, it is instructive to consider the cross-section required such that $r_1 = r_\mathrm{conv}$. This depends on the simulation resolution as well as the mass of the halo in question, and we plot three curves (one for each simulation resolution) in Fig.~\ref{fig:M200_sigma_all_DMO} that correspond to $r_1 = r_\mathrm{conv}$ for haloes following the \citet{2016MNRAS.460.1214L} concentration-mass relation. The low-mass CDM haloes within each of the three CDM-only simulations have inferred cross-sections that are close to this line, suggesting that cross-sections derived from fitting to the density profiles of these low-mass haloes are reflecting the numerical cores formed due to limited resolution. Note however, that the 50 most massive FOF groups from each simulation (which are the ones used throughout this paper) have $r_\mathrm{conv} \lesssim 0.01 \, r_{200}$. We only fit the density profiles down to 1\% of $r_{200}$ and therefore do not expect resolution to have a significant impact on the haloes used throughout this paper.

The highest-mass (and therefore best-resolved) haloes in each CDM-only simulation do not appear to follow the $r_1 = r_\mathrm{conv}$ lines, instead having larger inferred cross-sections. Given that we only fit density profiles at radii above 1\% of $r_{200}$, isothermal Jeans model fits with a cross-section for which $r_1 < 0.01 \, r_{200}$ are actually just fits of NFW profiles to the simulated density profiles. To illustrate the haloes for which this could be important, we include a line corresponding to $r_1 = 0.01 \, r_{200}$ in Fig.~\ref{fig:M200_sigma_all_DMO}. It is not so obvious in Fig.~\ref{fig:M200_sigma_all_DMO} due to the absence of error bars,\footnote{These are omitted for clarity, including them leads to a mess!} but if including error bars like those in Fig.~\ref{fig:vdSIDM_fit_DMO}, a significant fraction of the CDM-only haloes contain posterior probability at cross-sections below the $r_1 = 0.01 \, r_{200}$ line. Given that for all cross-sections below this line the same likelihood is achievable (by matching onto a common NFW profile), the isothermal Jeans model should not have a preference between cross-sections just below this line or at much lower cross-sections (down to the lower-limit of our prior at $0.01 \cmsg$). The reason this does not end up being the case is related to the `effective priors' discussed in Appendix~\ref{App:priors}, which (despite adopting a flat prior on $\log \sigma/m$) favour larger cross-sections, especially in lower-mass haloes. This happens because the prior area in $\log N_0$--$\log \sigma_0$ that maps into a given area in $M_{200}$--$c$ decreases with decreasing $\sigma/m$. Thus, larger cross-sections are favoured not because of an improvement to the likelihood, but because (for a given $M_{200}$ and $c$) a larger volume of prior space maps to this $M_{200}$ and $c$ when adopting a larger cross-section.

This behaviour is not particularly easy to understand, but inspection of Fig.~\ref{fig:analytical_matching} can go some way to explaining it. If we consider some fixed NFW profile and then ask what isothermal profiles match onto it, we see that as the cross-section decreases (and so $r_1$ decreases) $\rho(r_1) / \langle \rho(<r_1) \rangle \to 2/3$. This means that small cross-sections require a fine-tuned value of $r_1 / r_0 \approx 3$, which can be seen from where the red isothermal line crosses a value of 2/3 in Fig.~\ref{fig:analytical_matching}. In terms of our MCMC analysis, our log-uniform priors were on the quantities $N_0$ and $\sigma_0$. This fine-tuned value of $r_1 / r_0$ translates into a fine-tuned value of $N_0 \approx 2.91$, and it is this required fine-tuning that produces the bias against small cross-sections. 

We explicitly verified that changing our priors to boost the prior volume associated with profiles that map onto an NFW profile with $r_1 \ll r_\mathrm{s}$ shifts the CDM-only $\sigma/m$ posteriors downwards, by re-running our MCMC analyses using a reparameterised model. We parameterised the isothermal profile in terms of $\sigma_0$ and $\tilde{N}_0 = N_0 - 2.91$, and then adopted a flat prior on $\log_{10} \tilde{N}_0$ between -2 and 5. This does indeed shift down the $\sigma/m$ posteriors for CDM, leaving fits to the simulations with non-zero cross-sections relatively unchanged. However, this procedure is fairly arbitrary\footnote{If our prior on $\log_{10} \tilde{N}_0$ extended to even lower values than -2 then this would further boost the prior volume associated with small cross-sections.} and, importantly, only works for DM-only haloes. In the presence of baryons the isothermal profiles are altered and it is no longer true that $N_0 \approx 2.91$ corresponds to an isothermal profile that can match an NFW at $r_1 \ll r_\mathrm{s}$.

To conclude on the effects of resolution, we do not believe that the results for the 50 most massive haloes in each of our simulations should be affected much by the spatial resolution of our simulations, given that we only fit the density profiles down to 1\% of $r_{200}$. However, there are complications associated with prior volumes and inside-out matching that lead to a bias towards larger cross-sections when $r_1 \ll r_\mathrm{s}$.

\bsp
\label{lastpage}

\end{document}